\documentclass{aa}
\usepackage{natbib}
\usepackage{graphicx}
\usepackage{txfonts}

\begin{document}

   \title{FIRST-based survey of Compact Steep Spectrum sources}

   \subtitle{III. MERLIN and VLBI observations of subarcsecond-scale objects}

   \author{A. Marecki\inst{1}
          \and M. Kunert-Bajraszewska\inst{1}
          \and R. E. Spencer\inst{2}
          }
    \offprints{Andrzej Marecki \email{amr@astro.uni.torun.pl}}
    \institute{Toru\'n Centre for Astronomy, N. Copernicus University,
              87-100 Toru\'n, Poland
    \and
              Jodrell Bank Observatory, The University of Manchester,
              Macclesfield, Cheshire, SK11 9DL, UK
    }

    \date{Received 27 June 2005 /Accepted 25 November 2005}

\abstract
{According to a generally accepted paradigm, small intrinsic sizes of Compact
Steep Spectrum (CSS) radio sources are a direct consequence of their youth,
but in later stages of their evolution they are believed to become
large-scale sources. However, this notion was established mainly for strong
CSS sources.}
{In this series of papers we test this paradigm on 60 weaker objects selected
from the VLA FIRST survey. They have 5-GHz flux densities in the range
$150 < S_{5{\rm GHz}} < 550$\,mJy and steep spectra in the range
$0.365 \le \nu \le 5$\,GHz. The present paper is focused on sources that
fulfill the above criteria and have angular sizes in the range
$\sim$$0\farcs2$ -- $1\arcsec$.}
{Observations of 19~such sources were obtained using MERLIN in ``snapshot''
mode at 5\,GHz. They are presented along with 1.7-GHz VLBA and 5-GHz
EVN follow-up snapshot observations made for the majority of them.
For one of the sources in this subsample, 1123+340, a full-track 5-GHz
EVN observation was also carried out.}
{This study provides an important element to the standard theory of CSS
sources, namely that in a number of them the activity of their host galaxies
probably switched off quite recently and their further growth has been stopped
because of that. In the case of 1123+340, the relic of a compact ``dead
source'' is particularly well preserved by the presence of intracluster medium
of the putative cluster of galaxies surrounding it.}
{The observed overabundance of compact sources can readily be explained in the
framework of the scenario of ``premature'' cessation of the activity of the host
galaxy nucleus. It could also explain the relatively low radio flux densities of
many such sources and, in a few cases, their peculiar, asymmetric morphologies.
We propose a new interpretation of such asymmetries based on the light-travel
time argument.}


\maketitle

\section{Introduction}

Following the first paper defining the class of Compact Steep Spectrum (CSS)
sources \citep{pw82}, two almost identical surveys of CSS sources
\citep{spencer89,fanti90} based on the 3CR catalogue and the \citet{pw82}
list have been constructed. CSS sources collected in these samples --
they are often labelled jointly as the so-called 3CRPW sample -- are as
powerful as large-scale Fanaroff-Riley type\,II \citep[FR\,II,][]{fr74}
radio sources, yet their angular sizes are of the order of a few arcseconds.
Their interpretation was given by \citet{fanti90} and has become a
paradigm \citep[see][for a review]{odea98}. According to this paradigm, the
large majority of CSS sources are intrinsically small objects
-- their true linear sizes are $\la 20h^{-1}$\,kpc\footnote{For consistency
with earlier papers in this field, the following cosmological parameters have
been adopted throughout this paper: $H_0$=100${\rm\,km\,s^{-1}\,Mpc^{-1}}$ and
$q_0$=0.5. Wherever in the text we refer to linear sizes we introduce $h^{-1}$.
Whenever the redshift for a particular source described in this paper is not
available, a default value $z=1.25$ providing the maximum linear size for a
given angular size ($1\arcsec$ corresponds then to $4.3h^{-1}$\,kpc) and the
adopted cosmological model is assumed.} -- so that their small angular
sizes do not result from projection.

As their linear sizes are subgalactic, CSS sources are immersed in their host
galaxies and it is obvious that the environment responsible for the observable
effects, such as strong and
often asymmetric depolarisation of the radio emission caused by Faraday
rotation \citep{saikia85, saikia87, ag95, ludke98, f2004}, must have an
influence on morphologies and the evolution of CSS sources. A strong
interaction with the host galaxy interstellar medium (ISM) is a basis of the
so-called ``frustration scenario''.
According to this model, the small size of CSS sources may
be attributed to the presence of dense clouds of ISM \citep{bmh84}. However,
to date there is no proof that such clouds are dense {\em enough} to impede
further growth of the sources, so it is very plausible that the internal
pressure of a CSS source is sufficient to allow expansion in ram pressure
balance with the ambient ISM. Moreover, the infrared properties of GHz-Peaked
Spectrum (GPS) and CSS sources are consistent with those of large-scale radio
galaxies \citep{f2000}. Hence, the existence of a dense medium that enshrouds
galaxies hosting CSS sources and hampers the ``normal'' growth of a radio
source remains unproven.

The notion that CSS sources are not frustrated but instead young has been
present in the literature since the publication of the papers by \citet{pm82}
and \citet{c85}. \citet{f95} provided a comprehensive theoretical background
of the ``youth scenario'' which has subsequently been supported observationally.
In particular, \citet{murgia99} showed that the spectral ages of CSS sources
can be up to $10^5$ years. Consequently, the youth
scenario currently prevails over the frustration scenario; see the review by
\citet{fanti00}. Support for this view has been brought
by \citet{siem05} who observed diffuse X-ray emission around 3C186, a CSS
identified with a QSO, using the {\it Chandra} X-ray Observatory. The discovery
of a distant ($z=1.063$) cluster associated with this CSS source provides
direct observational evidence that at least this CSS source is not thermally
confined, as required by the frustration scenario. Instead, it appears
that 3C186 may be young and is observed at an early stage of its evolution.

If a CSS source lies close to the sky plane, it is perceived as a
Medium-sized Symmetric Object (MSO), which is
a scaled-down version of an FR\,II. Given that CSS sources are thought to be
young, this similarity can be explained by an evolutionary scenario.
Indeed, it has been argued \citep{r96} that Compact Symmetric Objects
(CSOs), MSOs and Large Symmetric Objects (LSOs) make up an evolutionary
sequence. Double-lobed radio sources continually increase their size with
time and LSOs begin their active phase as very compact sources and
must have passed the CSO and MSO phases. However, the converse may not be
true: it is not obvious that every CSO must become an LSO.

It also appears that there is no compelling reason to confine the CSS
class only to the most powerful objects fulfilling the compactness and
spectrum steepness criteria. In other words, the high radio power, typical
for bright CSS sources, is not an essential ingredient of the
definition of the CSS class. Indeed, the existence of weak CSS sources has
been confirmed observationally; see Sect.~\ref{s_new_samp}. Three
questions relating to these sources arise:

\begin{enumerate}
\item Are there any fundamental morphological differences between strong
and weak CSS sources analogous to e.g. the well-known FR\,I/FR\,II division
for LSOs which is clearly correlated with the radio luminosity \citep{fr74}?
\item Do weak CSS sources fit into the CSO--MSO--LSO evolutionary scheme
developed by \citet{r96}?
\item What are the actual causes of the low power output?
\end{enumerate}

This paper, along with its companion papers of the series, attempts to address
the above questions based on new, high resolution observations of weak CSS
sources selected from the {\it Faint Images of Radio Sky at Twenty} (FIRST)
survey \citep{wbhg97}\footnote{Website: http://sundog.stsci.edu}. It is
organised as follows. In Sect.~\ref{s_new_samp} the selection procedure of
the sample is presented. The technical details of the VLBI
observations are given in Sect.~\ref{s_obs} and the results of these
observations supplemented with the earlier results of MERLIN survey are
discussed in Sect.~\ref{s_notes}. Their astrophysical implications are
discussed in Sect.~\ref{s_discuss} and summarised in Sect.~\ref{s_concl}.

\begin{table*}[t]
\caption[]{FIRST coordinates, flux densities and spectral indices of the
sources in Subsample Two.}
\begin{tabular}{l|c c|c c c|r r|c}
\hline
& RA & Dec & $F_{365\mathrm{MHz}}$& $F_{1.4\,\mathrm{GHz}}$&
$F_{4.85\,\mathrm{GHz}}$& $\alpha_{365\,\mathrm{MHz}}^{1.4\,\mathrm{GHz}}$&
$\alpha_{1.4\,\mathrm{GHz}}^{4.85\,\mathrm{GHz}}$&
$F_{1.655\,\mathrm{GHz}}$\\
\multicolumn{1}{c|} {Source name} &
\multicolumn{2}{c|} {(J2000)} &
\multicolumn{3}{c|} {[Jy]} & & &
\multicolumn{1}{c} {[Jy]} \\
\multicolumn{1}{c|} {(1)} & (2) & (3) & {(4)} & {(5)} & {(6)} &
\multicolumn{1}{c} {(7)} &
\multicolumn{1}{c|} {(8)} & {(9)}\\
\hline
\hline
0744+291$^\ast$ & 07 48 05.332 & 29 03 22.54 & 1.25 & 0.48 & 0.17 & $-$0.71 & $-$0.83 &  0.41\\
0747+314 &        07 50 12.318 & 31 19 47.52 & 2.62 & 1.05 & 0.35 & $-$0.68 & $-$0.89 &  0.90\\
0811+360 &        08 14 49.068 & 35 53 49.70 & 1.71 & 0.58 & 0.17 & $-$0.80 & $-$0.99 &  0.49\\
0853+291 &        08 56 01.169 & 28 58 35.06 & 2.28 & 0.63 & 0.19 & $-$0.95 & $-$0.97 &  0.54\\
0902+416 &        09 05 22.197 & 41 28 39.65 & 1.10 & 0.48 & 0.17 & $-$0.61 & $-$0.86 &  0.42\\
0922+322$^\ast$ & 09 25 32.727 & 31 59 52.87 & 1.80 & 0.53 & 0.20 & $-$0.91 & $-$0.79 &  0.47\\
1123+340 &        11 26 23.674 & 33 45 26.64 & 3.80 & 1.32 & 0.38 & $-$0.79 & $-$1.01 &  1.11\\
1232+295 &        12 34 54.387 & 29 17 43.79 & 1.10 & 0.46 & 0.16 & $-$0.65 & $-$0.84 &  0.40\\
1242+364 &        12 44 49.679 & 36 09 25.53 & 2.75 & 0.78 & 0.21 & $-$0.94 & $-$1.07 &  0.65\\
1251+308 &        12 53 25.750 & 30 36 35.03 & 1.03 & 0.45 & 0.20 & $-$0.61 & $-$0.66 &  0.41\\
1343+386 &        13 45 36.948 & 38 23 12.62 & 1.81 & 0.85 & 0.44 & $-$0.56 & $-$0.53 &  0.78\\
1401+353$^\ast$ & 14 03 19.238 & 35 08 11.88 & 2.27 & 0.63 & 0.18 & $-$0.96 & $-$1.01 &  0.53\\
1441+409 &        14 42 59.307 & 40 44 28.79 & 1.51 & 0.97 & 0.30 & $-$0.33 & $-$0.94 &  0.83\\
1601+382$^\ast$ & 16 03 35.150 & 38 06 42.93 & 0.99 & 0.43 & 0.20 & $-$0.62 & $-$0.61 &  0.39\\
1619+378 &        16 21 11.290 & 37 46 04.94 & 1.62 & 0.64 & 0.20 & $-$0.69 & $-$0.94 &  0.55\\
1632+391 &        16 34 02.910 & 39 00 00.18 & 1.98 & 0.93 & 0.37 & $-$0.56 & $-$0.74 &  0.82\\
1656+391 &        16 58 22.172 & 39 06 25.55 & 1.34 & 0.65 & 0.24 & $-$0.53 & $-$0.80 &  0.57\\
1709+303 &        17 11 19.939 & 30 19 17.67 & 1.83 & 1.03 & 0.37 & $-$0.43 & $-$0.83 &  0.90\\
1717+315 &        17 19 30.062 & 31 28 48.12 & 1.06 & 0.45 & 0.15 & $-$0.63 & $-$0.87 &  0.39\\
\hline
\end{tabular}
\vspace{0.2cm}

$^\ast$ Sources with names marked with an asterisk were not followed-up using VLBI.

Description of the columns:
Col.~~(1): Source name in the IAU format;
Col.~~(2): Source right ascension (J2000) extracted from FIRST;
Col.~~(3): Source declination (J2000) extracted from FIRST;
Col.~~(4): Total flux density at 365\,MHz extracted from Texas Catalogue;
Col.~~(5): Total flux density at 1.4\,GHz extracted from FIRST;
Col.~~(6): Total flux density at 4.85\,GHz extracted from GB6;
Col.~~(7): Spectral index ($S\propto\nu^{\alpha}$) between 365 and 1400\,MHz
calculated using flux densities in cols. (4) and (5);
Col.~~(8): Spectral index between 1.4 and 4.85\,GHz calculated using flux
densities in cols. (5) and (6);
Col.~~(9): Estimated total flux density at 1.655\,GHz interpolated from data
in cols. (5) and (6).

\label{t_radio}
\end{table*}

\begin{table*}[t]
\caption[]{Redshifts and photometric data of the sources in Subsample Two.}
\begin{tabular}{l l c l|c c|c c c c c}
\hline
\multicolumn{1}{c|} {Source} &
\multicolumn{1}{c} {$z$} & {Ref.} & {Opt.} &
\multicolumn{2}{c|} {APM} &
\multicolumn{5}{c} {SDSS/DR4} \\
\multicolumn{1}{c|} {name} & & & \multicolumn{1}{c|} {ID} & $R$ & $B$ & $u$ & $g$ & $r$ & $i$ & $z$\\
\multicolumn{1}{c|} {(1)} & \multicolumn{1}{c} {(2)} &
(3) & \multicolumn{1}{c|} {(4)} & (5) & (6) & (7) & (8) & (9) & (10) & (11) \\
\hline
\hline
0744+291 &       &     & q    & 19.70 & 19.79 & 21.04 & 20.91 & 20.98 & 20.64 & 20.14\\
0747+314 &       &     & G    &       &       & 22.55 & 23.55 & 20.87 & 19.02 & 19.24\\
0811+360 &       &     & EF   &       &       &       &       &       &       &      \\
0853+291 & 1.085 & (5) & Q    & 18.45 & 18.79 & 18.99 & 18.91 & 18.67 & 18.69 & 18.79\\
0902+416 &       &     & G    &       &       & 22.55 & 22.67 & 21.77 & 21.07 & 20.27\\
0922+322 &       &     & G    &       &       & 23.55 & 22.71 & 22.32 & 22.08 & 21.70\\
1123+340 & 1.247 & (1) & G    &       &       &       &       &       &       &      \\
1242+364 &       &     & G    &       &       & 23.87 & 21.94 & 21.57 & 21.23 & 20.30\\
1251+308 &       &     & --   & 19.67 & 21.14 &       &       &       &       &      \\
1343+386 & 1.844 & (2) & Q    & 17.69 & --    & 18.42 & 18.24 & 18.05 & 17.60 & 17.52\\
1401+353 & 0.45$\dagger$ & (3) & q    & 19.40 & 20.32 & 21.09 & 20.35 & 19.86 & 19.54 & 19.02\\
1441+409 &       &     & EF   &       &       &       &       &       &       &      \\
1601+382 &       &     & G    & 18.45 & 20.96 & 21.42 & 19.60 & 18.23 & 17.66 & 17.24\\
1619+378 & 1.271 & (4) & Q    & 18.89 & 21.29 & 21.29 & 19.90 & 18.81 & 18.29 & 18.00\\
1632+391 & 1.085 & (5) & Q    & 17.93 & 18.33 & 18.95 & 18.68 & 18.34 & 18.21 & 18.04\\
1656+391 &       &     & G    &       &       & 24.08 & 22.04 & 20.26 & 19.56 & 19.13\\
1709+303 &       &     & EF   &       &       &       &       &       &       &      \\
\hline
\end{tabular}
\vspace{0.2cm}

Description of the columns:
Col.~~(1): Source name in the IAU format;
Col.~~(2): Redshift;
Col.~~(3): Reference for the redshift given in col. 2;
Col.~~(4): Optical identification: G -- galaxy, Q -- QSO, EF -- ``empty field'',
q -- a star-like object with no known redshift;
Cols.~(5-6): Magnitudes for the two POSS/APM colours;
Cols.~(7-11): Magnitudes for the five SDSS colours.\\
$\dagger$Photometric redshift.\\
References for redshifts:
(1) -- \citet{rel01}, (2) -- \citet{hb89}, (3) -- \citet{mach98},
(4) -- \citet{kock96}, (5) -- SDSS.\\
Sources 1232+295 and 1717+315 are not listed since there is no data
available for them to date and their possible
optical counterparts are too faint to be measured using APM.\\
Note for optical identifications: ``G'' should be treated with caution as these
designations mostly come from the ``automated classifier'' of SDSS.

\label{t_optical}
\end{table*}

\section{The FIRST-based CSS sources sample}\label{s_new_samp}

A number of samples of CSS sources other than 3CRPW have been generated so far.
In particular, \citet{fanti01} derived a sample of 87~CSS sources from the
B3-VLA sample \citep{vig89} and the parsec-scale structures of many
of them were studied using VLBI. Depending on the degree of the
compactness of the sources they were followed up either with the EVN combined
with MERLIN at 1.7\,GHz \citep{dallaca02b} -- hereafter D02b -- or with the
VLBA at three frequencies: 1.7\,GHz \citep{dallaca02a} -- hereafter D02a --
and 5/8.4\,GHz \citep{orienti04}.

The present study is based on a sample consisting of 60~sources extracted from
FIRST. The details of the adopted selection procedure have been given in
\citet{kmskn02} -- hereafter Paper~I. Our selection criteria and those used by
\citet{fanti01} were different, as were the RA and Dec limits of both samples,
although the two samples are slightly overlapping. Therefore, a few sources
described in this paper also belong to the B3-VLA CSS sample and so their
independent VLBI observations have been reported in D02a, D02b and
\citet{orienti04}. Unlike the sources in the B3-VLA CSS sample, the ones dealt
with here have steep spectra in a broad range of frequencies: $0.365 \le \nu
\le 5$\,GHz. This means that the GPS sources, which are sometimes treated as a
separate subclass of the CSS class, have been deliberately excluded.

The sample was surveyed with MERLIN at 5\,GHz by means of snapshot observations
(a ``pilot'' survey) carried out in two sessions in 1997 and 1999; see Paper~I
for the technical details of those observations. Based on the angular sizes of
radio structures detected by MERLIN, the whole sample was then divided into
three groups. The first one consists of 11~objects with angular sizes in the
range: $\sim$$1\arcsec$ -- $5\arcsec$. Five of them were presented in Paper~I.
With one exception (1201+394) they have asymmetric structures. The remaining
six objects are double or triple MSOs and were presented in \citet{kb05} --
hereafter Paper~II.

The second group -- hereafter Subsample Two -- consists of 19~sources with
angular sizes in the range $\sim$$0\farcs2$ -- $1\arcsec$ and the present paper
(Paper~III) is devoted to them. The basic parameters of their radio emission
extracted from Texas \citep{doug96}, FIRST and GB6 \citep{bec91} catalogues are
listed in Table~\ref{t_radio} and the redshifts, if available, and photometric
data are given in Table~\ref{t_optical}. The magnitudes shown in cols. 5 and 6
are extracted from the POSS plates using the Automatic Plate Measuring (APM)
machine whereas those in cols. 7 to 11 are taken from the Release~4 of the
Sloan Digital Sky Survey (SDSS/DR4) which is the up-to-date version at the time
of writing.

Objects in the third group -- they are labelled ultra-compact steep spectrum
objects -- have angular sizes below $\sim$$0\farcs2$ and were barely resolved
with MERLIN at 5\,GHz (resolution $\leq$ 50\,milliarcseconds) in the course of
the pilot survey. We found 16~such sources. Their milliarcsecond scale
structures have been investigated by means of multifrequency VLBA observations
and the results of that study are presented in Paper~IV \citep{kmt06} and
Paper~V (Kunert-Bajraszewska in~prep.).

\section{MERLIN and VLBI observations}\label{s_obs}

MERLIN images of all 19~objects of Subsample Two are shown in
Figs.~\ref{fig:0744+291} to~\ref{fig:1717+315} and they are discussed in
Sect.~\ref{s_notes}. The maps are plotted to scale: their sizes are
$3\farcs2 \times 3\farcs2$. Contours increase by a factor of 2 and the first
contour is at $\sim 3\sigma$ level. Table~5 
gives the flux densities of the main components of the structures derived from
these images using AIPS task {\tt JMFIT}.

Four sources (0744+314, 0922+322, 1401+353 and 1601+382) were not followed up
using the VLBI technique as the MERLIN images did not show noticeable compact
structures that would still be visible at milliarcsecond resolution.
For the remaining 15 targets, two-frequency VLBI observations were carried
out and their results are shown in Figs.~\ref{fig:0747+314}
to~\ref{fig:1717+315} except Figs.~\ref{fig:0922+322}, \ref{fig:1401+353}
and~\ref{fig:1601+382}. As in the case of MERLIN maps, contours increase by a
factor of 2 and the first contour is at $\sim 3\sigma$ level.

VLBI follow-up observations were carried out at two frequencies:
1.7\,GHz with the VLBA and 5\,GHz with the EVN. The angular resolutions of
the two instruments are comparable at the quoted frequencies with typical
beam sizes of $\simeq 5\times8$ milliarcseconds (VLBA) and $\simeq
7\times8$ milliarcseconds (EVN).

The VLBA observations were made in snapshot mode at 1655.4\,MHz with
128\,Mb/s recording (i.e. $2\times$32\,MHz bandwidth) during three observing
sessions as shown in Table~\ref{t_journal}. Each source was observed
for 30\,--\,40\,min in several scans of a typical length of 5\,min. The
data were correlated with the NRAO processor at Socorro. After system
temperature calibration and antenna gain and correlator corrections had been
applied, the data were further processed using standard AIPS procedures. The
positions of target sources were established from interleaved observations
of phase-reference sources. The locations of the maps' centres were then
formally shifted to the strongest features using AIPS task {\tt UVFIX}.
(This procedure does not change the absolute positions of the target sources.)

\begin{figure}[t]
\resizebox{\hsize}{!}{\includegraphics{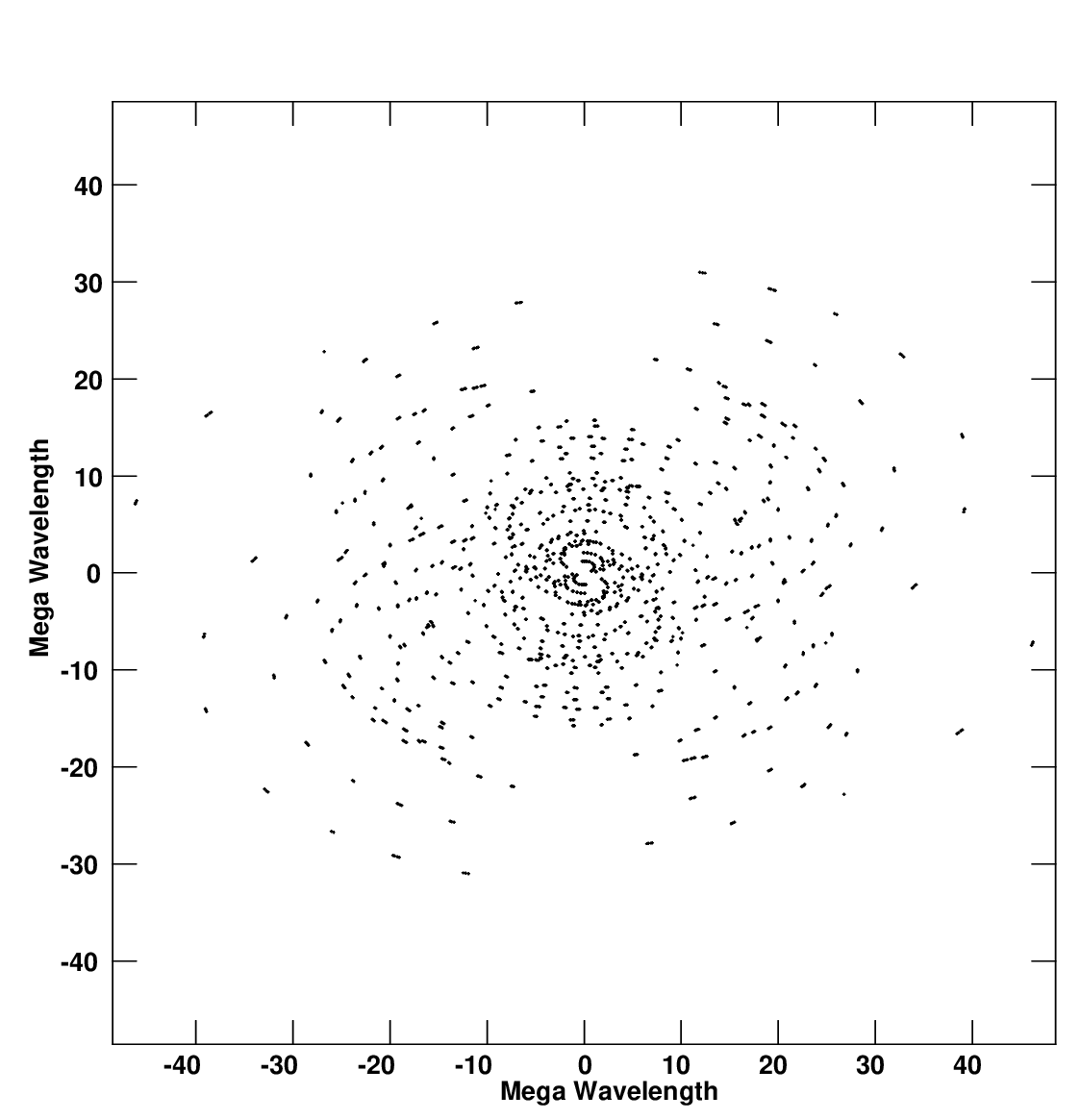}}
\caption{Typical $u$ -- $v$ coverage during VLBA 1.7-GHz observations.}
\label{f_uvVLBA}
\end{figure}

The noise levels in the images resulting from the
above observations are typically of $\sim$$30\mu$Jy/beam. A typical
$u$ -- $v$ coverage achieved (for 1123+340) is shown in Fig.~\ref{f_uvVLBA}.

The EVN snapshot observations were carried out in three observing sessions;
see Table~\ref{t_journal}. In 1998 the array comprised the following
telescopes: Cambridge, Effelsberg, Jodrell Bank~Mk2, Medicina, Onsala (26\,m),
Toru\'n and Westerbork (tied array). Left-hand circular polarisation with a
total bandwidth of 28\,MHz (Mk\,IIIB standard) at 4961\,MHz 
was recorded and the correlation was carried out using the Bonn correlator.
In the second EVN session the telescopes were as listed above, but excluding
Cambridge and including Noto. The observations were made at a frequency of
4971.5\,MHz and with a bandwidth of 32\,MHz. The two-bit sampled left-hand
circular polarisation data were correlated at JIVE. Each source was
observed for 30\,--\,40\,min in several scans of a typical length of 5\,min.

Calibration and further processing of the data were carried out using a
standard procedure in AIPS. However, the absolute positions of the sources were
established by using our 1.7-GHz VLBA images as initial models in the
fringe-fitting algorithm (AIPS task {\tt FRING}).
Such a procedure does not change the absolute positions of the target
sources more than a fraction of the beam size. The noise levels in the
resulting images are typically $\sim$$50\mu$Jy/beam. The flux densities
of the main components of the sources at the two frequencies derived using
AIPS task {\tt JMFIT} are given in Table~\ref{t_VLBI}.

\begin{figure}[t]
\resizebox{\hsize}{!}{\includegraphics{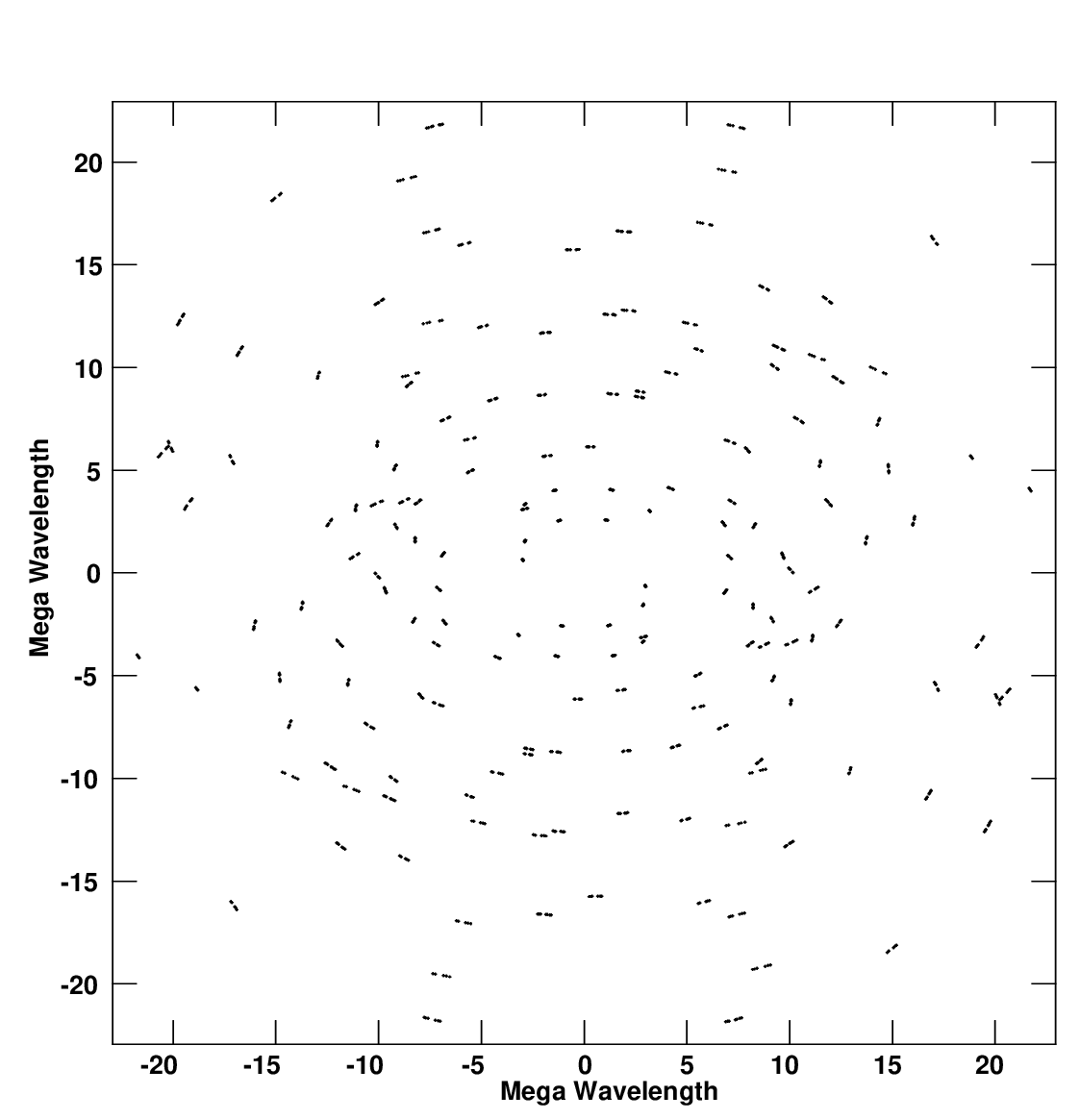}}
\caption{Typical $u$ -- $v$ coverage during EVN 5-GHz observations.}
\label{f_uvEVN}
\end{figure}

A typical $u$ -- $v$ coverage during the EVN observations (for 1123+340) is
shown in Fig.~\ref{f_uvEVN}. The shortest VLBA spacings are
a few times smaller than those available with the EVN so that the sensitivity
to the more extended elements of the images is much better in the case of the
VLBA observations. Consequently, there is often an appreciable fraction
of the ``missing flux'' in the EVN maps. In principle, this drawback could
be removed by combining the EVN data with the MERLIN data
from the pilot survey. Such data processing was attempted but it
was unsuccessful mainly because of the snapshot mode of MERLIN observations.

\begin{figure}[t]
\resizebox{\hsize}{!}{\includegraphics{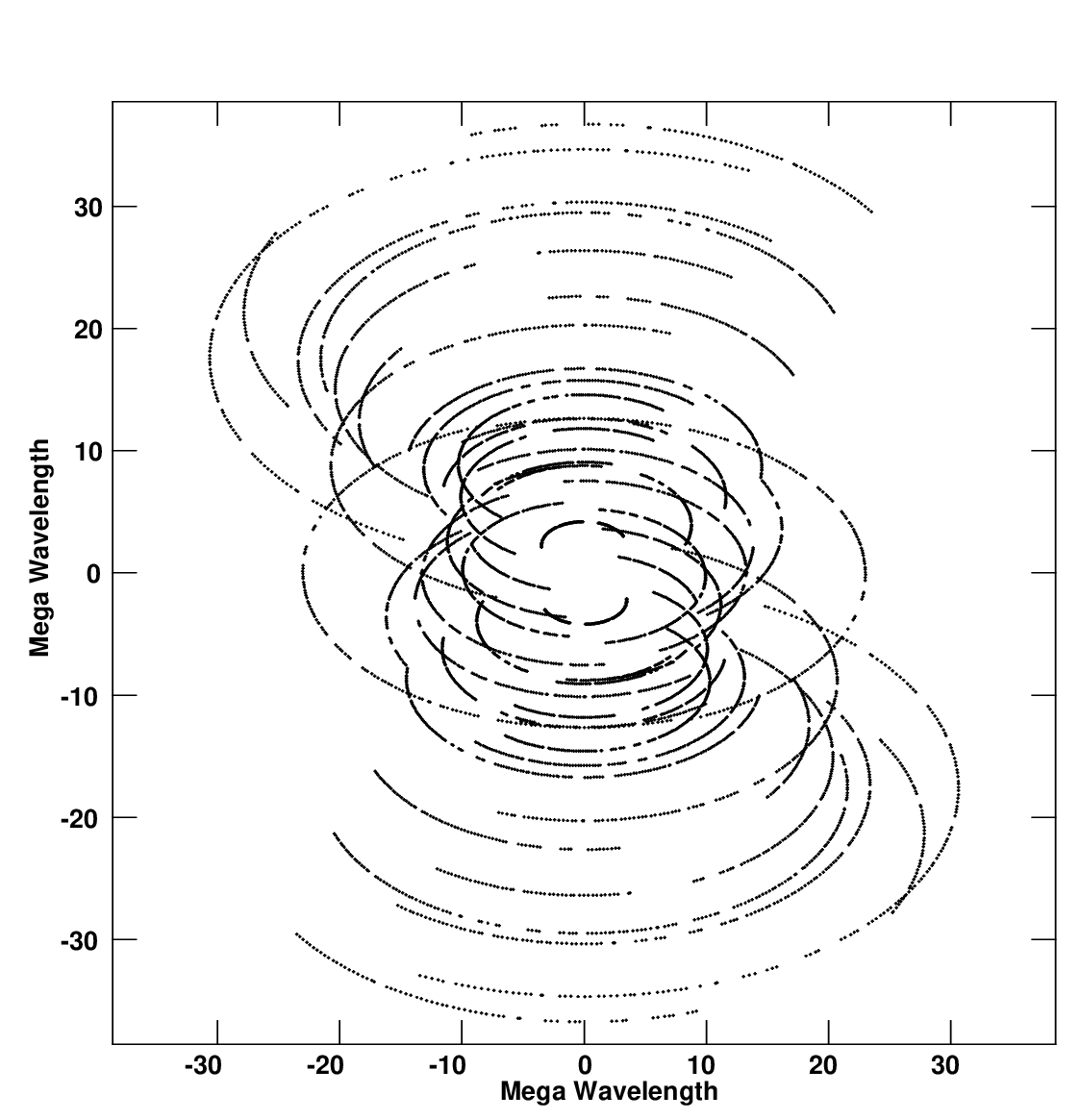}}
\caption{The $u$ -- $v$ coverage during the 12-hour EVN observation of 1123+340
at 5\,GHz.}
\label{f_uvEVN12}
\end{figure}

Initial analysis of the observations of 1123+340 indicated that the source
might be particularly interesting but the snapshot observations were
insufficient for a definitive interpretation. Consequently, 1123+340 was
re-observed in a dedicated EVN experiment at 5\,GHz. The observation took place
on 8~June~2001 with the participation of the Effelsberg, Jodrell Bank~Mk2,
Medicina, Noto, Onsala, Toru\'n and Westerbork telescopes. The resulting
$u$ -- $v$ coverage is shown in Fig.~\ref{f_uvEVN12}. The observational
configuration, the correlation and further data reduction were the same as for
the EVN observations made on 11~Feb.~2001. Since the image of 1123+340
resulting from this observation is a considerable improvement upon that
obtained from the observation of 16~Feb.~1998 -- the noise level in the
resulting image is $\sim$$10\mu$Jy/beam -- only the more recent image is
included in this paper. Also, the data given for 1123+340 in Table~\ref{t_VLBI}
and discussion related to the EVN-based results for this source refer to the
observation of 8~June~2001.

\begin{table}[ht]
\caption[]{Journal and modes of VLBI observations.}
\begin{tabular}{c|c|c|l}
\hline
\multicolumn{1} {c|} {Source} &
\multicolumn{3} {c} {Networks, frequencies and recording systems} \\
\cline{2-4}
& VLBA, 1.7\,GHz &
\multicolumn{2} {l} {EVN, 5\,GHz / recording} \\
\hline
\hline
0747+314 & 19 Sep 1998 & 16 Feb 1998 & Mk\,IIIB\\
0811+360 & '' & ''\\
0902+416 & '' & ''\\
1123+340 & '' & ''\\
1232+295 & '' & ''\\
1242+364 & '' & ''\\
1251+308 & '' & ''\\
\hline
1619+378 & 30 Mar 2001 & 11 Feb 2001 & Mk\,IV\\
1632+391 & '' & ''\\
1656+391 & '' & ''\\
1709+303 & '' & ''\\
1717+315 & '' & ''\\
\hline
0853+291 & 06 May 2001 & 11 Feb 2001 & Mk\,IV\\
1343+386 & '' & ''\\
1441+409 & '' & ''\\
\hline
1123+340 & ---- & 8 Jun 2001 & Mk\,IV\\
\hline
\end{tabular}

\label{t_journal}
\end{table}

\begin{table}[ht]
\caption[]{Morphological classification of the sources in Sample Two.}
\begin{tabular}{l|l}
\hline
\multicolumn{1}{c|} {Type} & \multicolumn{1}{|c} {Sources}\\
\hline
\hline
 & \\
Possible faders	&	0744+291, 0922+322, 1123+340$\dagger$,\\
    & 1656+391\,(?)\\
Double-lobed	&	0747+314, 0811+360, 0902+416$^\ast$,\\
    & 1123+340$\ddagger$,	1242+364, 1251+308\,(?),\\
    & 1441+409, 1656+391, 1709+303, 1717+315\\
Core-jet	&	1232+295, 1251+308\,(?), 1343+386,\\
    & 1601+382, 1619+378\\
Other	&	0853+291, 1401+353, 1632+391\\
 & \\
\hline
\end{tabular}
\vspace{0.5cm}

$^\ast$Classified as ``double-lobed'' based upon D02b.\\
$\dagger$Applies to `E' and `S' components.\\
$\ddagger$Applies to `N1/2' and `W1/2' components.\\

Some sources are listed more than once if they fulfill more than one criterion
or if the classification is uncertain. Sources in the latter category are
denoted with ``?''.

\label{t_types}
\end{table}

\section{Notes on individual sources}\label{s_notes}

The observational results for all 19~objects listed in Table~\ref{t_radio}
are commented on in this section. The morphological classification is
summarised in Table~\ref{t_types}.

\begin{figure*}
\centering
\includegraphics[scale=0.5]{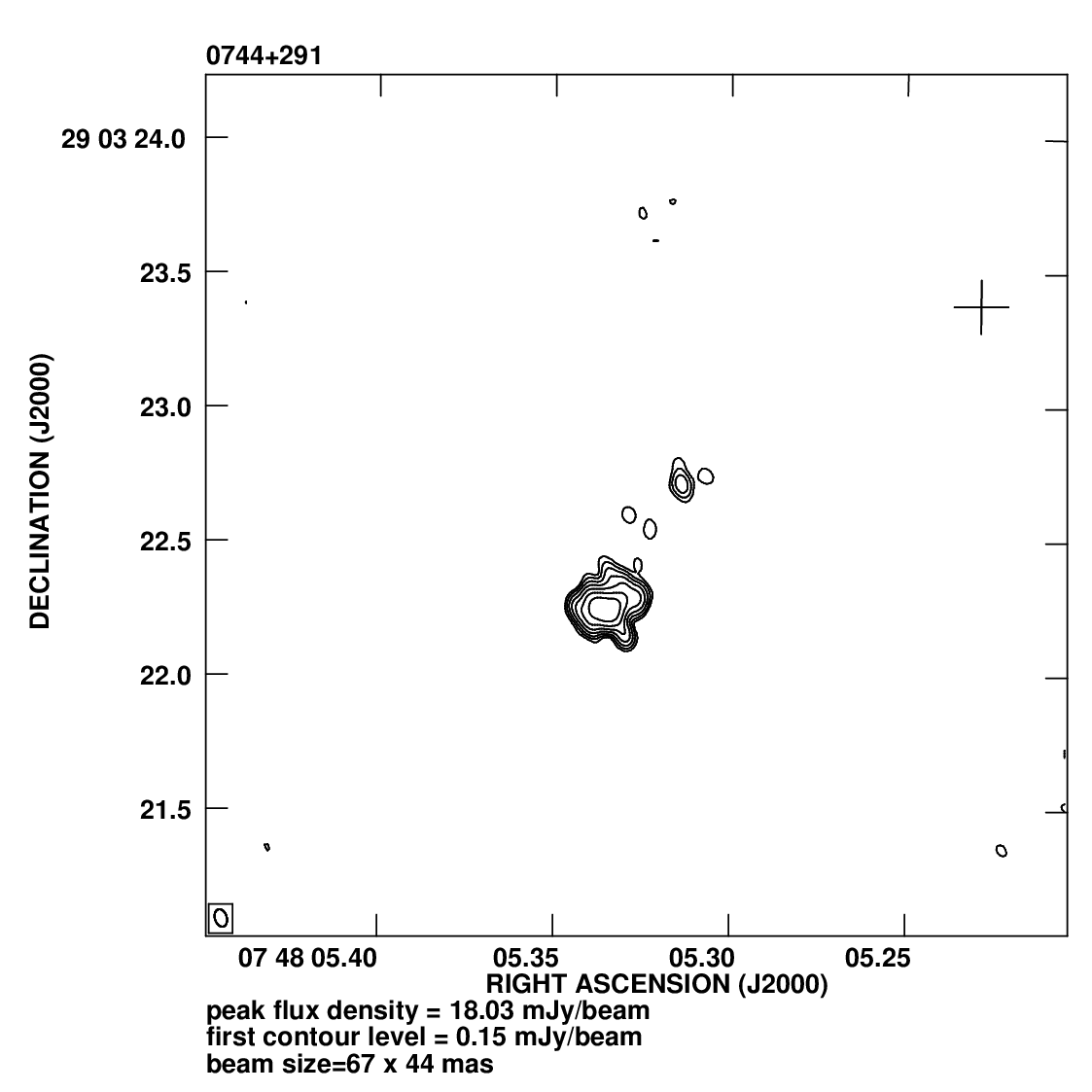}
\caption{MERLIN map of 0744+291 at 5\,GHz.
The cross indicates the
position of an optical object in SDSS/DR4.}
\label{fig:0744+291}
\end{figure*}

\noindent {\bf \object{0744+291}} (Fig.~\ref{fig:0744+291}). The northern
component visible in the MERLIN 5-GHz image is very weak (2.4\,mJy) so in
principle this object consists of only one diffuse element. Only 47\% of the
GB6 flux density is present in the MERLIN image, indicating that there may
exist an even more diffuse emission which is not reproduced there. The optical
counterpart of this source -- a $m_R=20.98$ star-like object -- is included in
SDSS/DR4 although no redshift is available from there. It is located at
RA=\,$7^{\rm h}48^{\rm m}05\fs229$, Dec=\,$+29\degr 03\arcmin 23\farcs38$ i.e.
$\sim$$2\arcsec$ north-west of the position of the main radio feature. There is
about 1-mag. difference between the APM and the SDSS measurements in both the
red and the blue parts of the spectrum (Table~\ref{t_optical}) which may
suggest variability.

This source can tentatively be interpreted as a relic lobe of an AGN located
in the vicinity of this faint optical object and perhaps it ended
its activity quite recently. The coordinates of the main component in the
MERLIN image coincide with those of the point-like object visible in the FIRST
map. This means that there is no hint of the existence of the second lobe in
FIRST, so even if such a lobe actually exists, it must be very faint.
To make this interpretation viable it has to be assumed that the whole radio
structure does not lie in the sky plane and that its observable part is more
distant than the optical object which, in turn, is more distant than the
putative second lobe. We can use the light-travel time argument, in a way
similar to that in \citet{ocp98}, but substantially simplified; the 
``current activity period'' -- see their Fig.~3 -- does not take place in the
sources investigated here and we do not strictly require a restarted
activity for them. It thus may be concluded that what is observed is the radio
lobe of a switched-off AGN as seen in an epoch earlier than that of the optical
object. An even later epoch emission from the other lobe is absent in the
image either because the lobe has decayed completely or its spectrum has
become very steep and so it is very weak at 5\,GHz. There is another object
in Subsample Two (0922+322) which could be interpreted in a similar
manner -- see below. In Sect.~\ref{dis_asym} the plausibility of the
hypothesis used to interpret these two sources is discussed.

As the radio structure visible in the MERLIN image does not have a conspicuous
compact component that might be an interesting target for follow-up
observations at high resolution, 0744+291 was not imaged using VLBI.

\begin{figure*}
\begin{flushleft}
\centering
\includegraphics[scale=0.5]{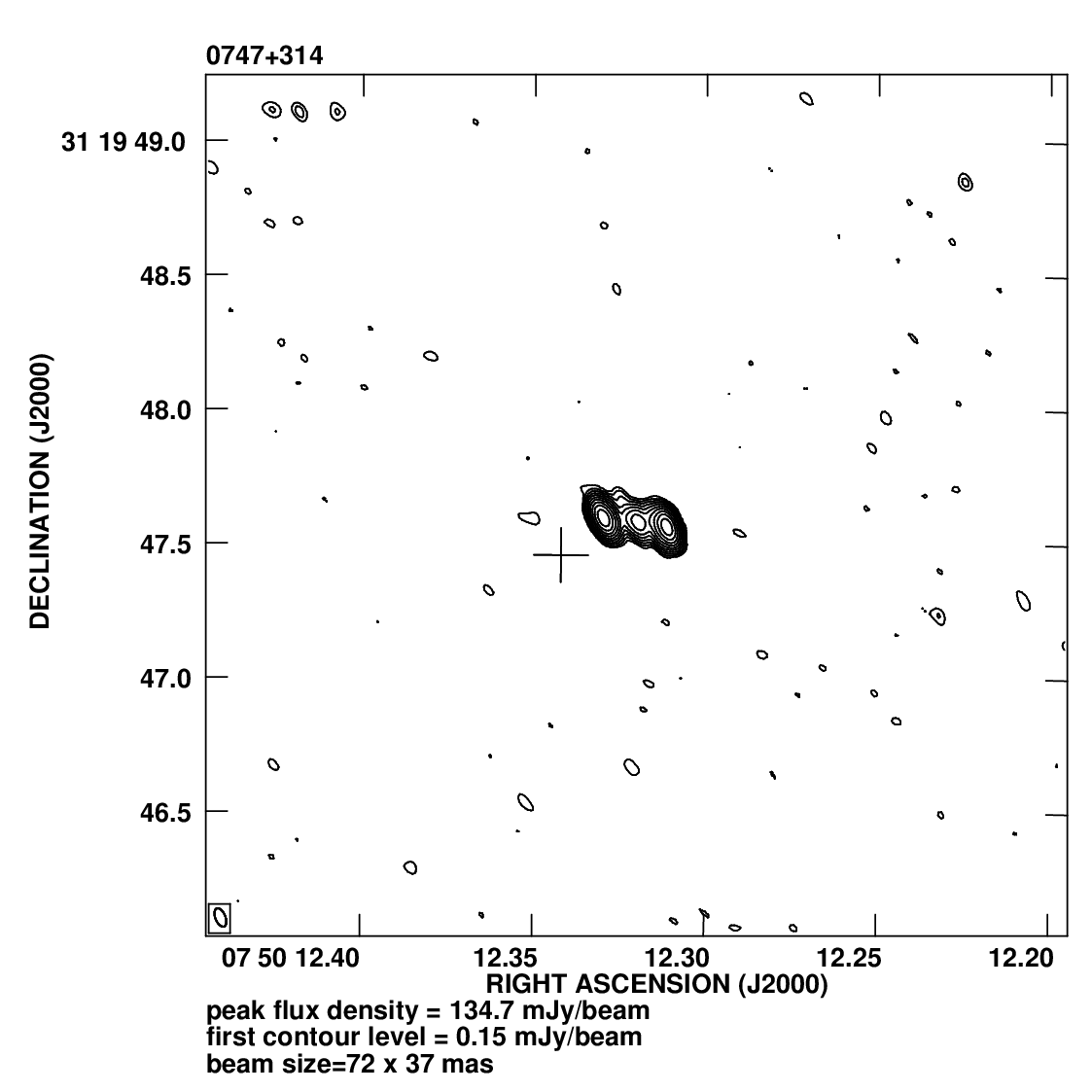}
\end{flushleft}
\begin{flushleft}
\centering
\includegraphics[scale=0.4]{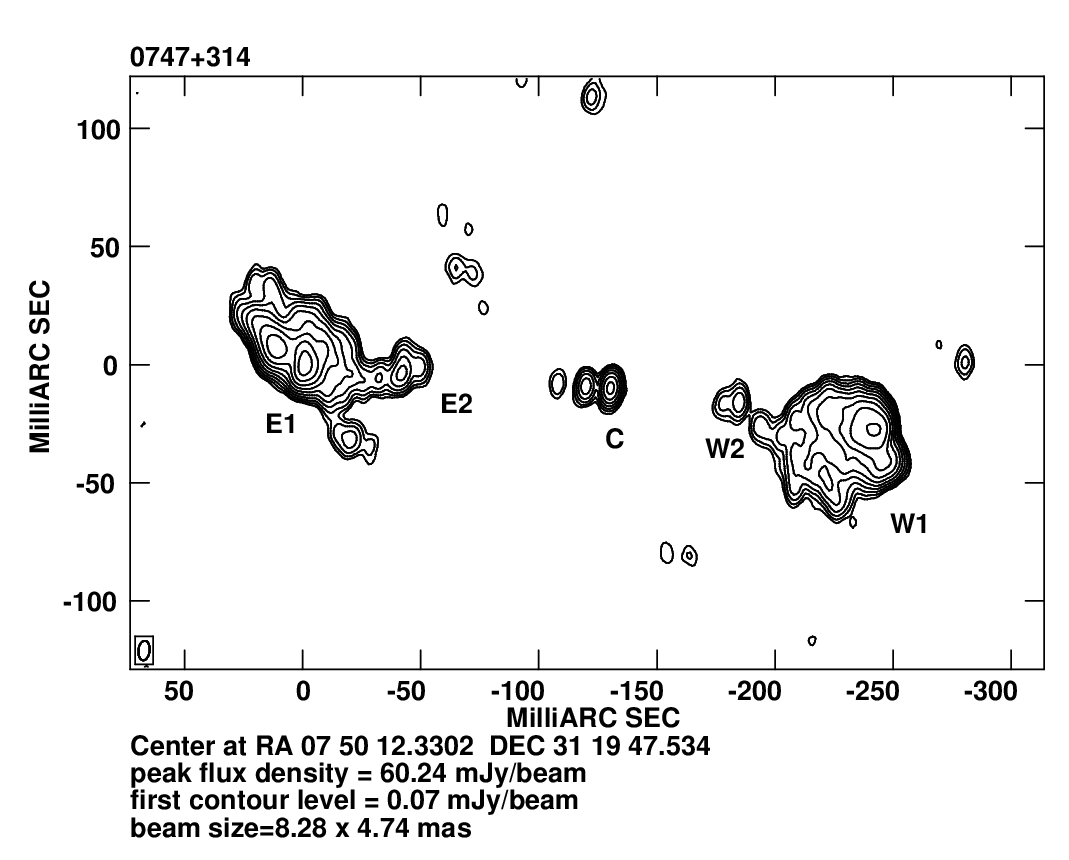}
\includegraphics[scale=0.4]{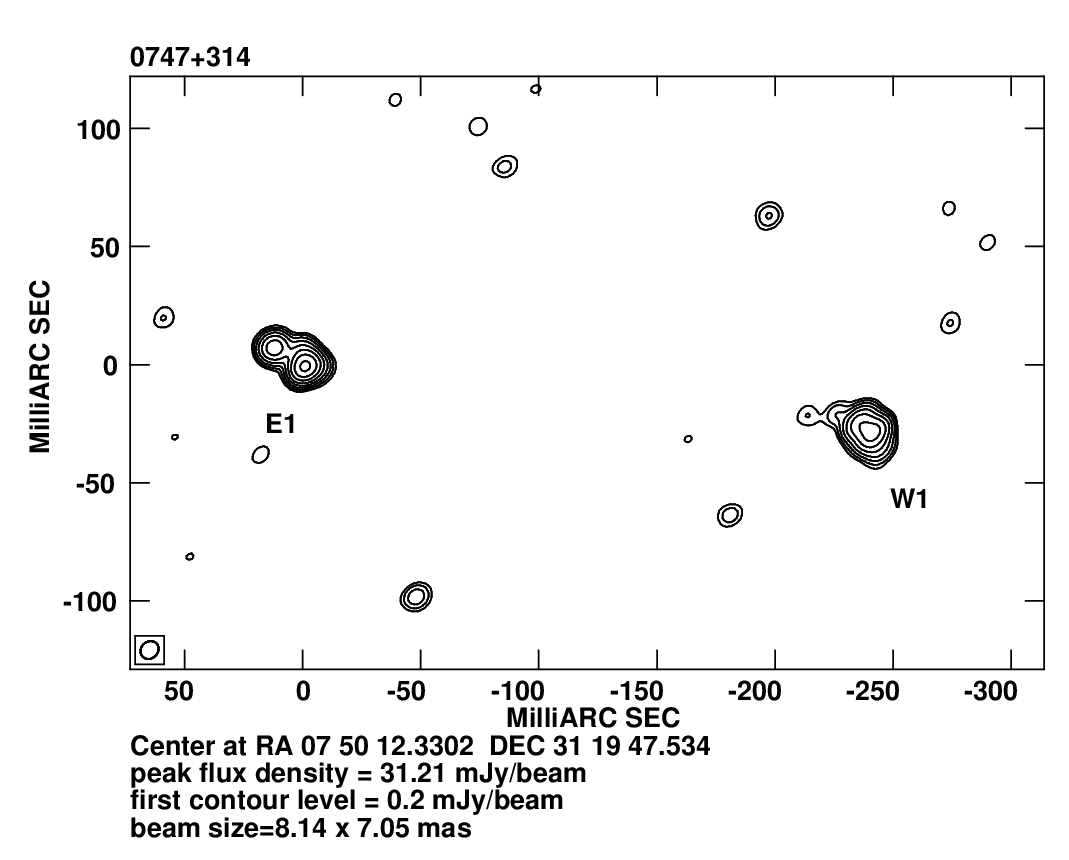}
\end{flushleft}
\caption{0747+314 --- MERLIN map at 5\,GHz (upper panel), VLBA map at 1.7\,GHz
(lower left panel) and EVN map at 5\,GHz (lower right panel).
The cross on the MERLIN
map indicates the position of an optical object in SDSS/DR4.}
\label{fig:0747+314}
\end{figure*}

\begin{figure*}
\begin{flushleft}
\centering
\includegraphics[scale=0.5]{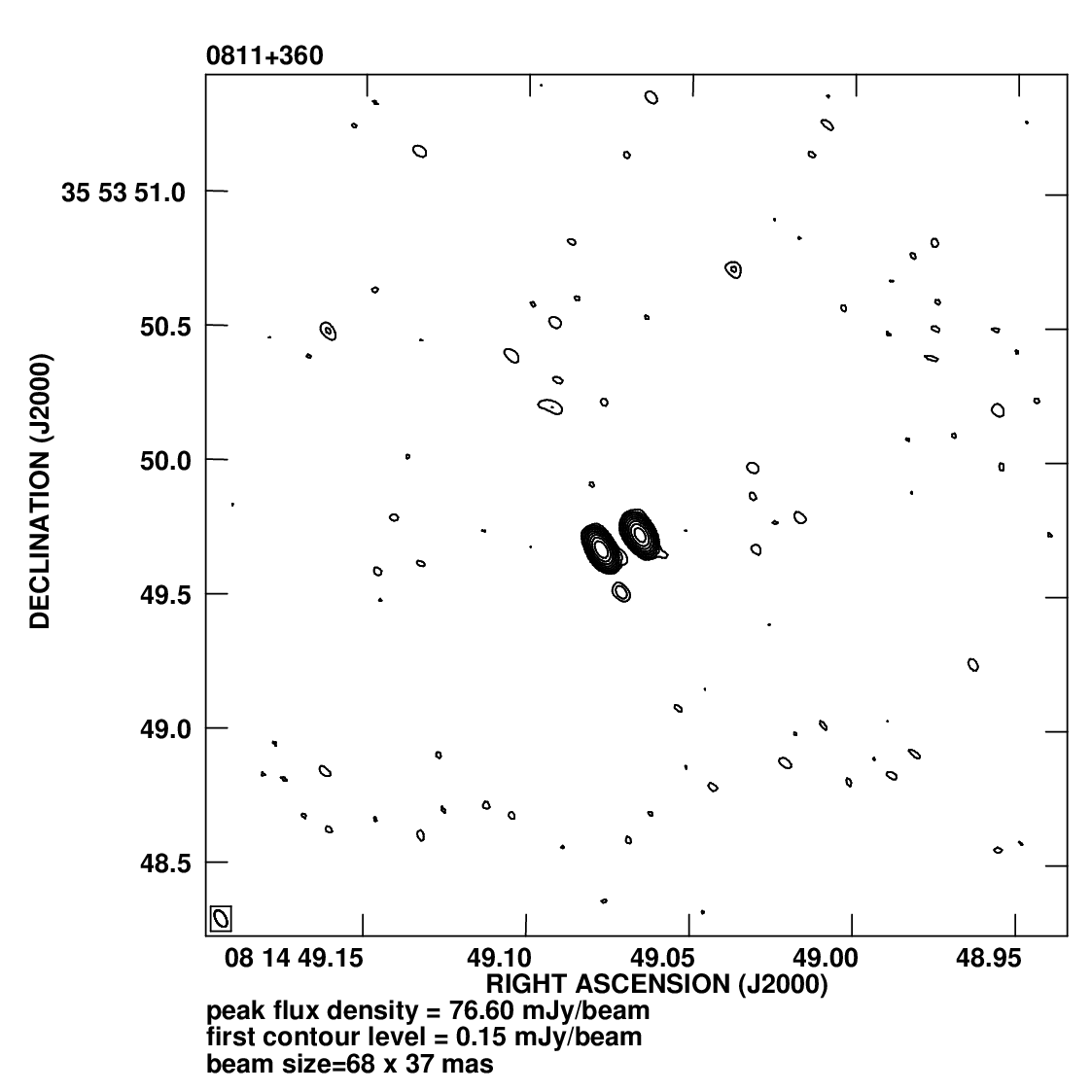}
\end{flushleft}
\begin{flushleft}
\centering
\includegraphics[scale=0.5]{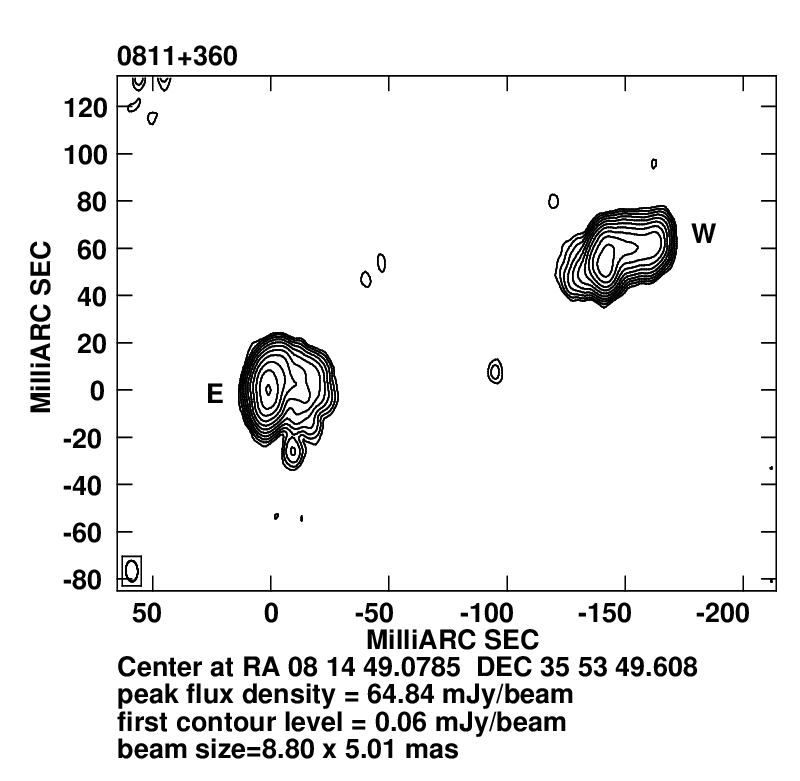}
\includegraphics[scale=0.5]{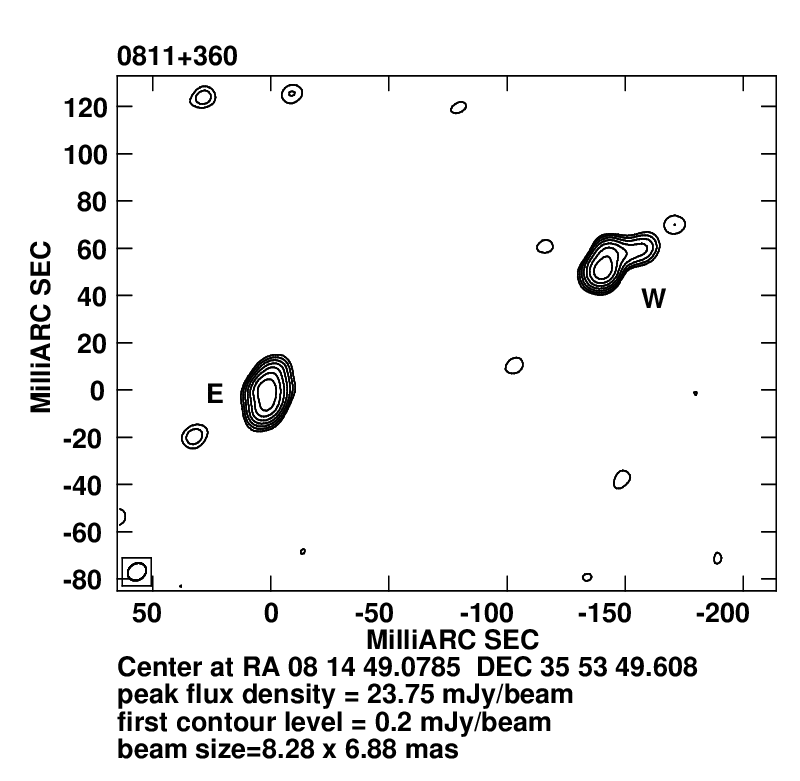}
\end{flushleft}
\caption{0811+360 --- MERLIN map at 5\,GHz (upper panel), VLBA map at 1.7\,GHz
(lower left panel) and EVN map at 5\,GHz (lower right panel).
}
\label{fig:0811+360}
\end{figure*}

\begin{figure*}
\begin{flushleft}
\centering
\includegraphics[scale=0.5]{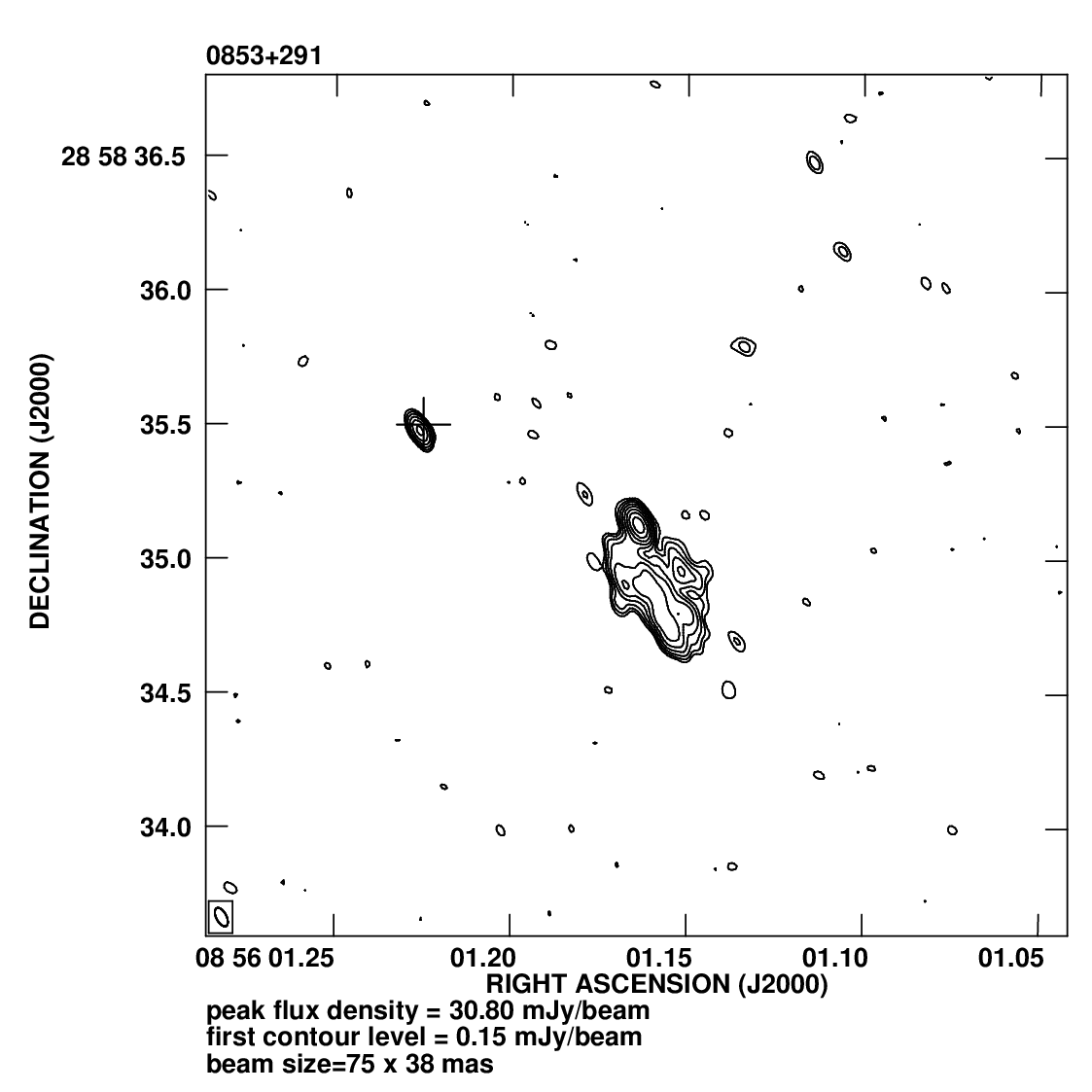}
\includegraphics[scale=0.5]{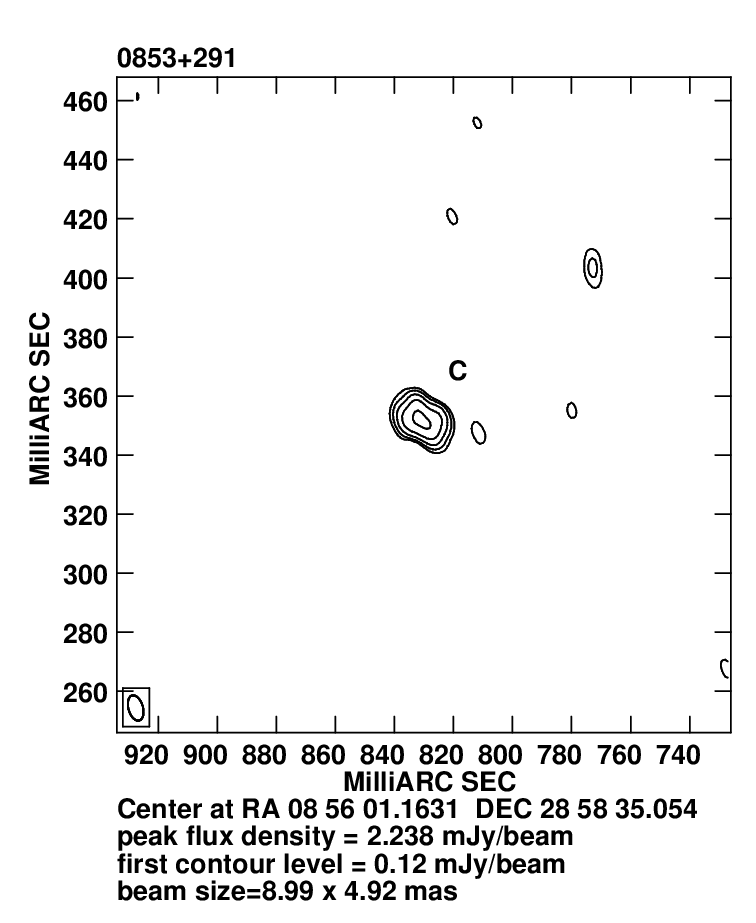}
\end{flushleft}
\begin{flushleft}
\centering
\includegraphics[scale=0.5]{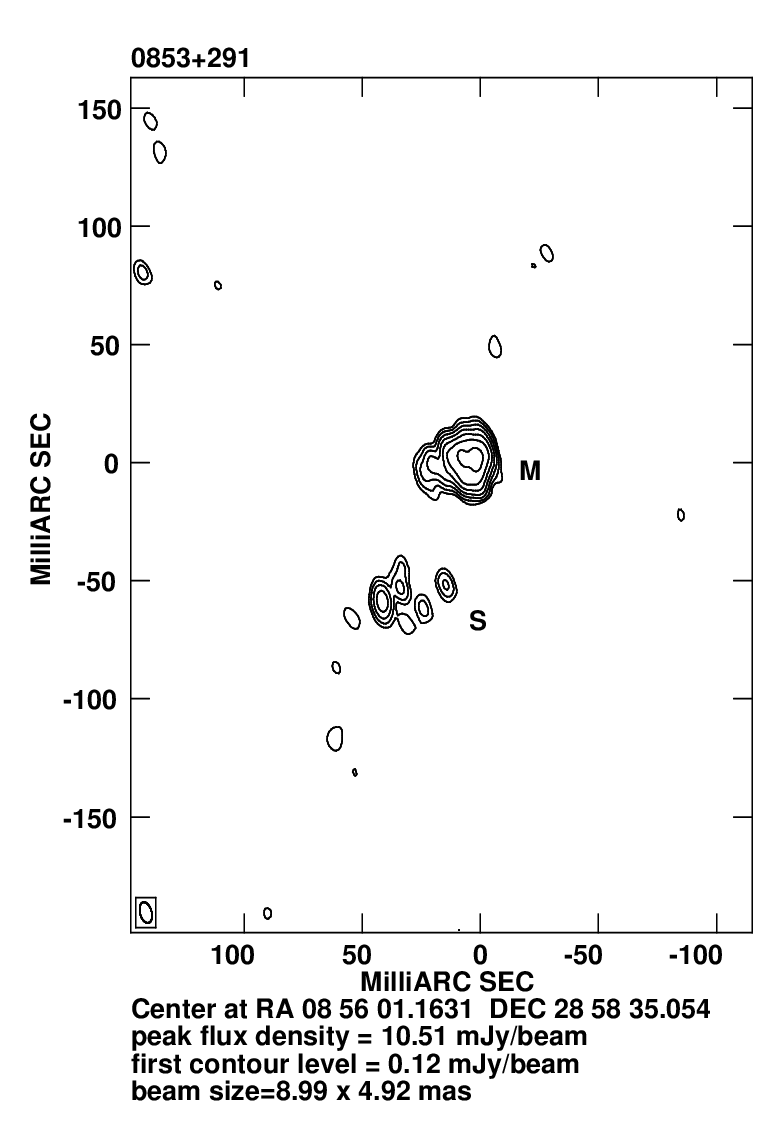}
\includegraphics[scale=0.5]{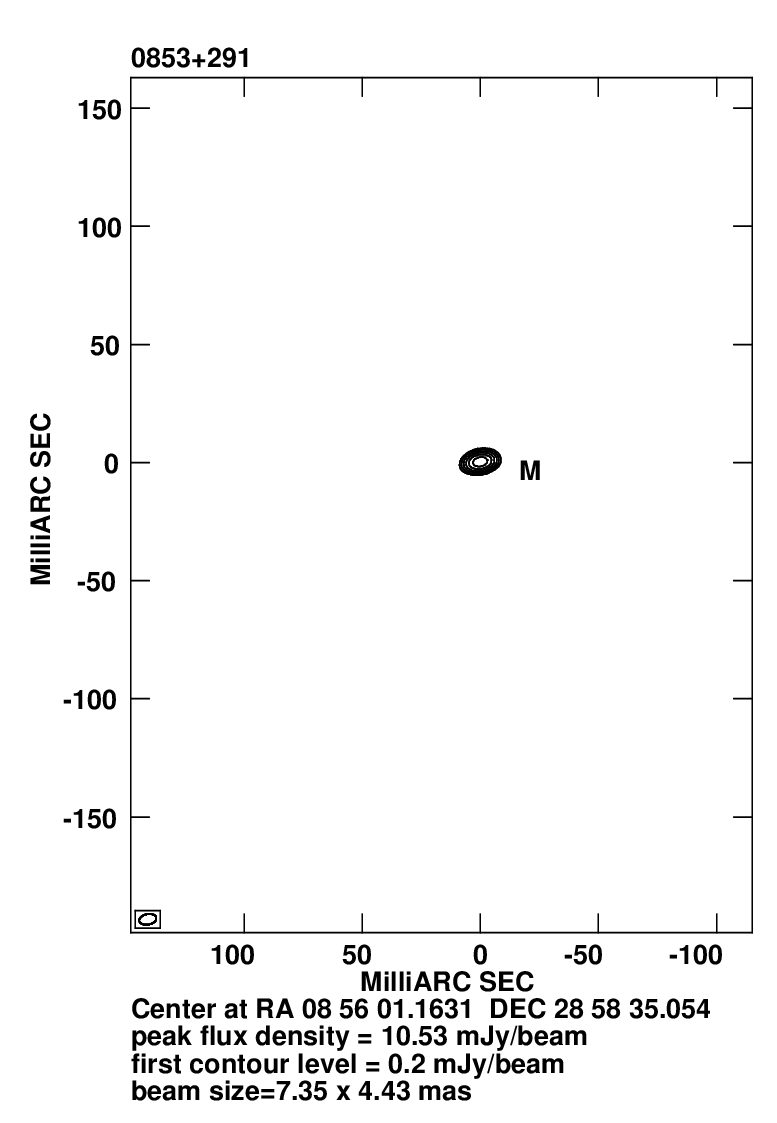}
\end{flushleft}
\caption{0853+291 --- MERLIN map at 5\,GHz (upper left panel), VLBA map at 1.7\,GHz
of the north-eastern component (upper right panel), VLBA map at 1.7\,GHz
(lower left panel) of the central component and EVN map at 5\,GHz of the central
component (lower right panel).
The cross on the MERLIN map indicates the position of an optical object in SDSS/DR4.}
\label{fig:0853+291}
\end{figure*}

\begin{figure*}
\begin{flushleft}
\centering
\includegraphics[scale=0.5]{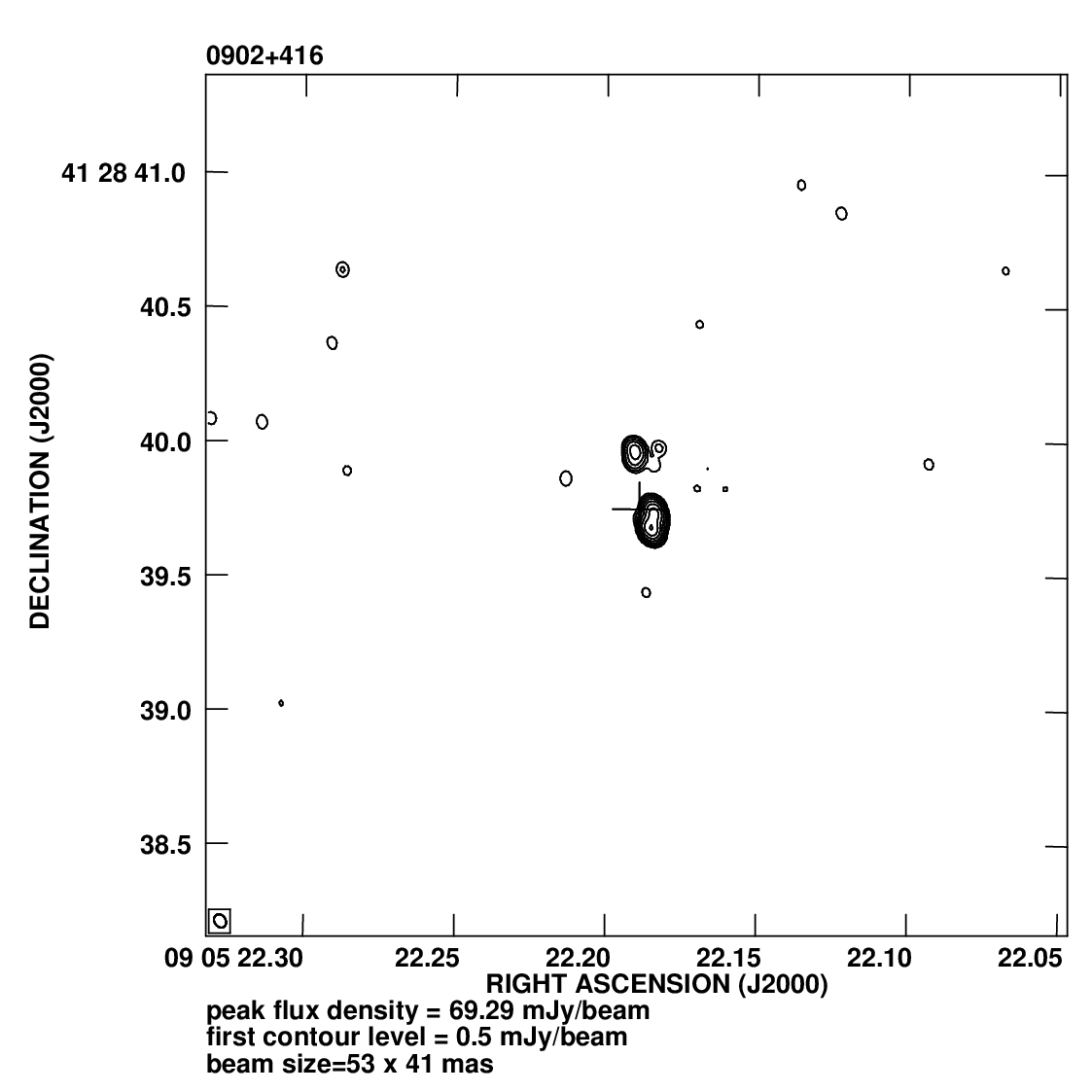}
\end{flushleft}
\begin{flushleft}
\centering
\includegraphics[scale=0.5]{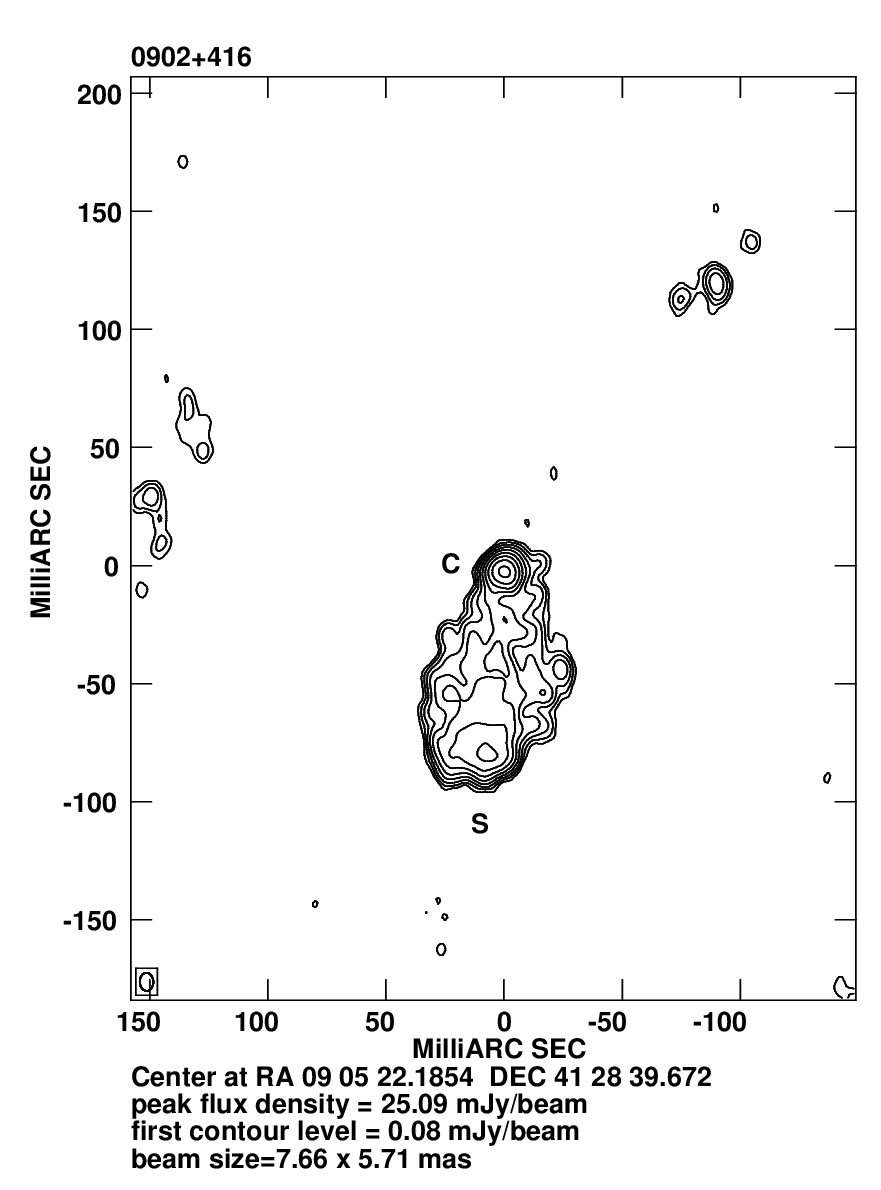}
\includegraphics[scale=0.5]{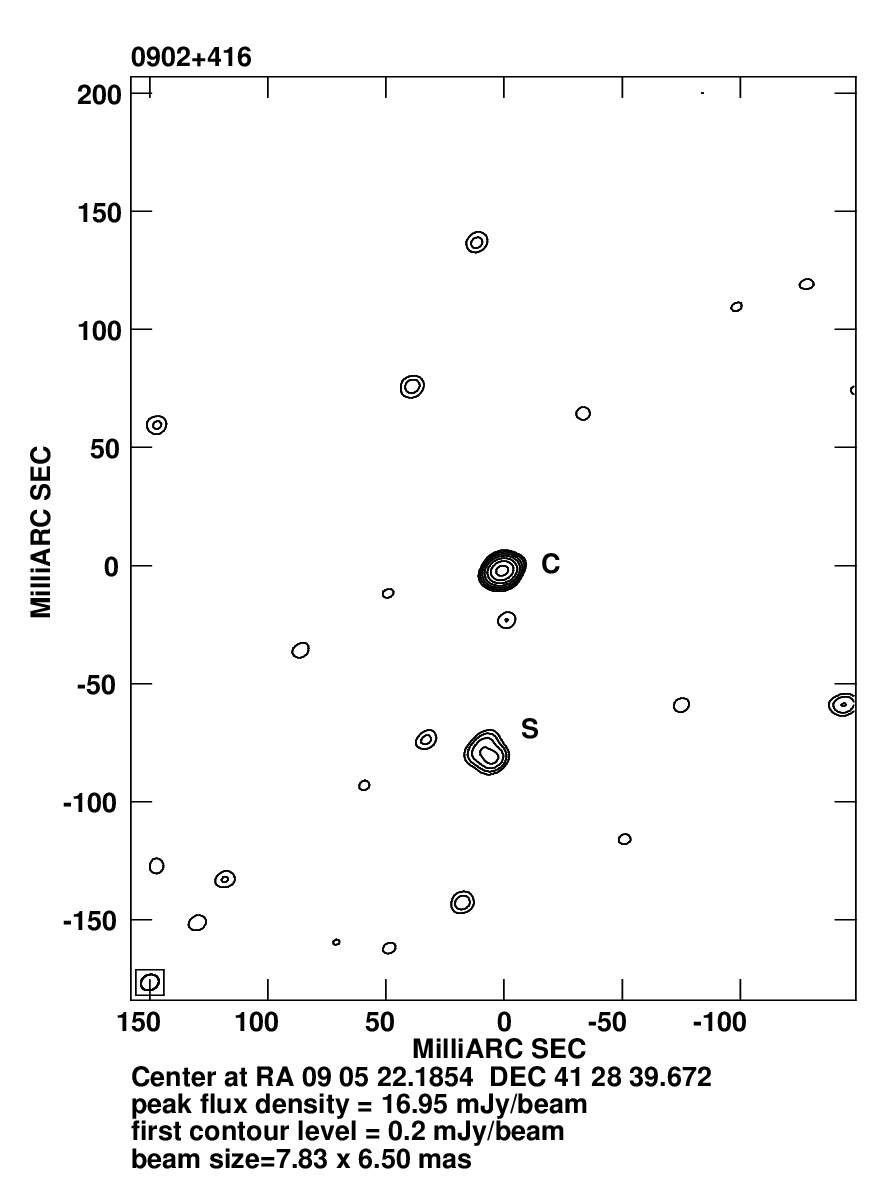}
\end{flushleft}
\caption{0902+416 --- MERLIN map at 5\,GHz (upper panel), VLBA map at 1.7\,GHz
(lower left panel) and EVN map at 5\,GHz (lower right panel).
The cross on the MERLIN
map indicates the position of an optical object in SDSS/DR4.}
\label{fig:0902+416}
\end{figure*}

\begin{figure*}
\begin{flushleft}
\centering
\includegraphics[scale=0.5]{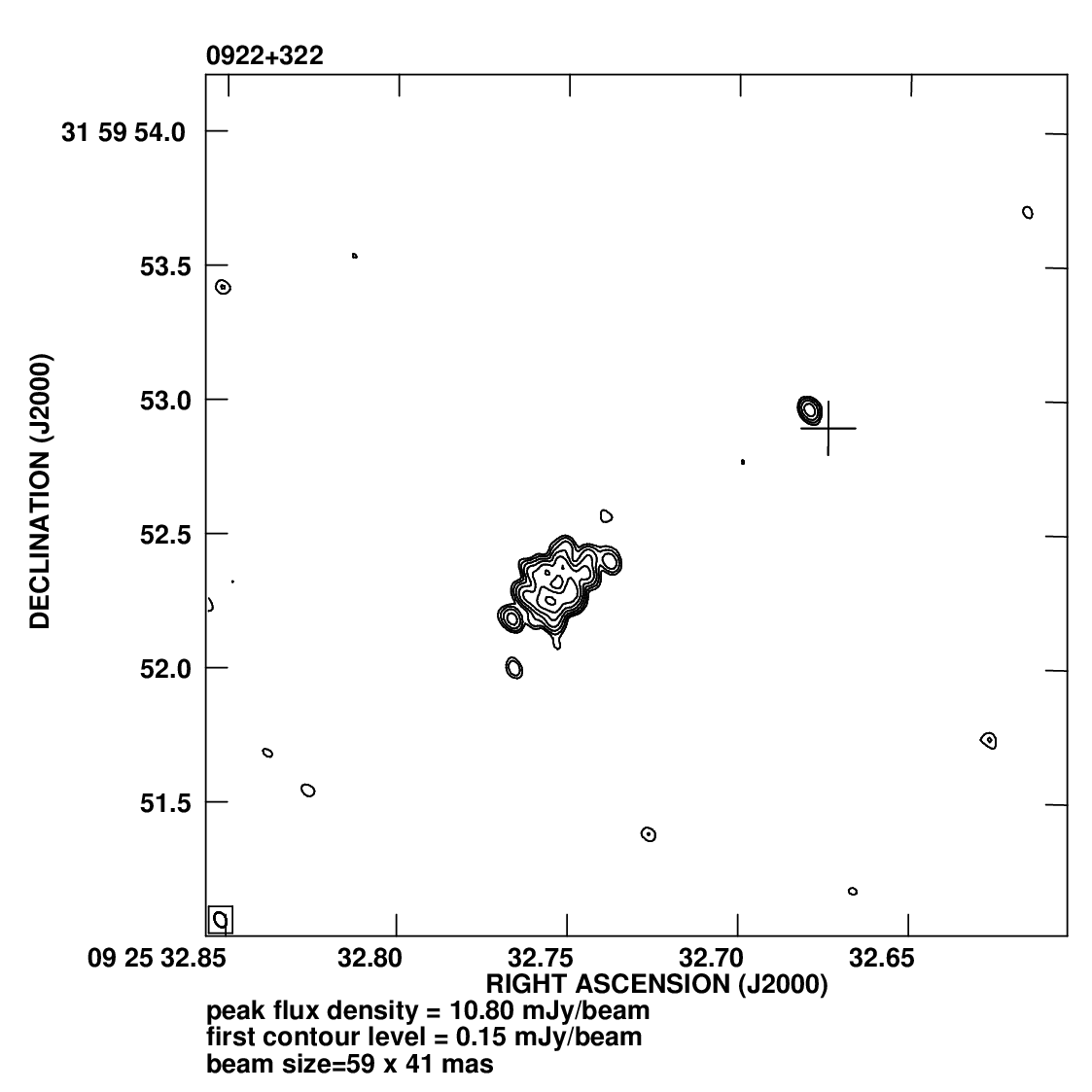}
\end{flushleft}
\caption{MERLIN map of 0922+322 at 5\,GHz.
The cross on the MERLIN
map indicates the position of an optical object in SDSS/DR4.}
\label{fig:0922+322}
\end{figure*}

\begin{figure*}
\begin{flushleft}
\centering
\includegraphics[scale=0.5]{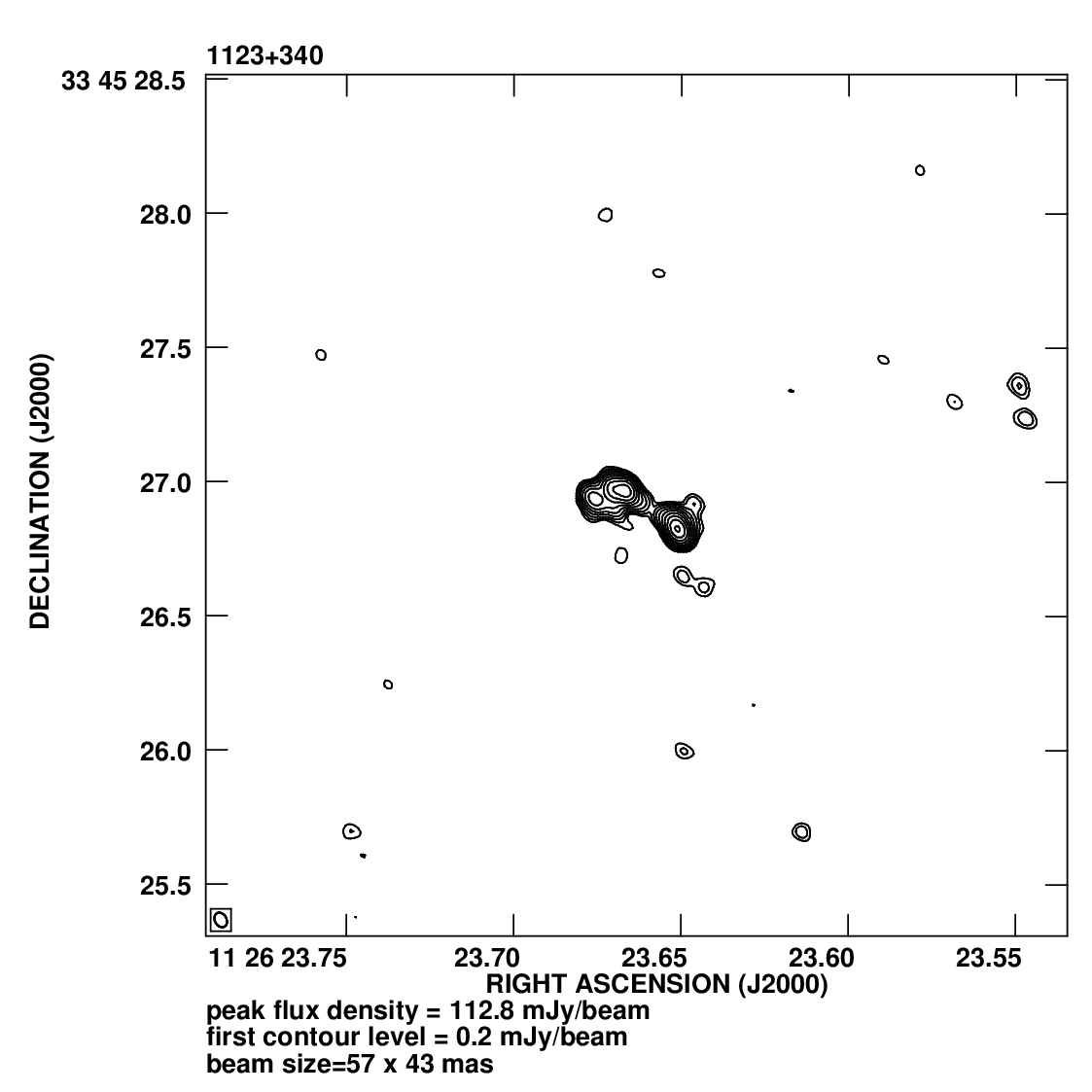}
\end{flushleft}
\caption{1123+340 --- MERLIN map at 5\,GHz.}
\label{fig:1123+340}
\end{figure*}

\noindent {\bf \object{0747+314}} (Fig.~\ref{fig:0747+314}). The 1.7-GHz
milliarcsecond structure, particularly as seen in the VLBA image, leaves no
doubt that the source is a classical, double-lobed triple with the central
component being the core. A possible optical counterpart of this source is
present in the SDSS/DR4 and identified with a $m_R=20.87$ galaxy. A relatively
bright ($\sim$$13^m$) object (a field star?) is located $5\farcs78$ north of
the position of the galaxy so only with SDSS is a correct optical
identification possible. No redshift for that galaxy is available at the
time of writing. The position of the optical object
(RA=\,$7^{\rm h}50^{\rm m}12\fs342$, Dec=\,$+31\degr 19\arcmin 47\farcs46$)
does not coincide well with the position of the radio source (see
Fig.~\ref{fig:0747+314}) but given the diffuse shape of the optical object, the
centroiding errors can be larger than usual and so it is a likely optical
counterpart of the radio source.

The MERLIN image accounts for the entire GB6 flux density of this object. It
was not possible to reproduce the core in the EVN image. Based on the value of
the spectral index between 1.4 and 5\,GHz and the flux densities at these
frequencies, the interpolated total flux density of the source at 1.655\,GHz
(i.e. the VLBA observing frequency) has been calculated\footnote{The same
calculation has been carried out for the remaining sources in Subsample Two.}
and listed in col.~9 of Table~\ref{t_radio}. It follows that the VLBA image
accounts for 63\% of the interpolated total flux density.

\noindent {\bf \object{0811+360}} (Fig.~\ref{fig:0811+360}). This is a simple
double source with both parts of roughly equal flux densities. The MERLIN image
accounts for almost entire GB6 flux density of this object whereas the
VLBA image accounts for 84\% of the interpolated total flux density. No optical
identification is available for this object in the literature. Based on the
SDSS/DR4 image, which covers the field of 0811+360, this source has no
optical identification.

\noindent {\bf \object{0853+291}} (Fig.~\ref{fig:0853+291}).
According to SDSS/DR4, the northeastern unresolved component seen in
the MERLIN image is coincident with a QSO at a redshift of $z=1.085$
located at RA=\,$8^{\rm h}56^{\rm m}01\fs225$, Dec=\,$+28\degr 58\arcmin
35\farcs50$ but it is the western part that is a dominating radio component.
The 1.7-GHz VLBA image of that part, which accounts for only 10\% of the
interpolated total flux density, shows that there is a jet emerging from 
there towards east. To match it with the fuzzy structure seen in the MERLIN
image it has to be assumed that the jet moves along a complicated, perhaps
helical path. The 1.7-GHz VLBA image of the northeastern component 
identified with the optical object is also presented. It is slightly resolved
and shows an extension oriented exactly towards the dominating component.
The MERLIN image accounts for 76\% of the GB6 flux density.

\noindent {\bf \object{0902+416}} (Fig.~\ref{fig:0902+416}).
This source belongs to the B3-VLA CSS sample and as such has been thoroughly
investigated by D02b. The MERLIN image is fully consistent
with the MERLIN 1.7-GHz image shown in D02b. The source has a
double structure and almost entire GB6 flux density is accounted for in our
image whereas there is a large discrepancy between our 1.7-GHz VLBA image and
the EVN 1.7-GHz image presented in D02b; it was not possible to reproduce the
northern lobe from the VLBA data. Our image accounts for only 47\% of the
interpolated total 1.7-GHz flux density.

This object is present in SDSS/DR4 and identified with a $m_R=21.77$ galaxy.
Its redshift is currently unknown. Its position
(RA=\,$9^{\rm h}05^{\rm m} 22\fs189$, Dec=\,$+41\degr 28\arcmin 39\farcs75$)
is coincident with the brightest component in our VLBI images which is labelled
``S2'' in the D02b EVN image. This means that the ``C'' component in the D02b
EVN image is less likely to be the core.

\noindent {\bf \object{0922+322}} (Fig.~\ref{fig:0922+322}). According to
SDSS/DR4, the northwestern unresolved component seen in the MERLIN image is
coincident with a $m_R=22.32$ galaxy of unknown redshift located at
RA=\,$9^{\rm h}25^{\rm m}32\fs674$, Dec=\,$+31\degr 59\arcmin 52\farcs90$. The
other component, because of its diffuse shape and a lack of a hotspot, could be
interpreted as a relic lobe somewhat similar to that in 0744+291. Only 38\% of
the GB6 flux density is accounted for in the MERLIN image. The northwestern
component is a likely core and so the other lobe, which is unobservable at
5\,GHz, should be located even farther north-west. It follows that a similar
assumption as for 0744+291 could be made in this case, namely that the
invisible northwestern lobe has completely dispersed. The combination of the
source's possible linear size and the timescale of the dispersion make the
light-travel delay an essential factor responsible for the apparent shape of
this source. Further discussion and interpretation of 0922+322 is given in
Sect.~\ref{dis_asym}.

0922+322 is a member of 7C and 9C surveys
\citep{7C,9C} and its 151-MHz and 15-GHz flux densities are 3410 and 68\,mJy,
respectively. It follows from these values supplemented with the measurements
at three intermediate frequencies quoted in Table~\ref{t_radio} that the source
has had a steep spectrum throughout two decades of frequency.

0922+322 was not imaged using VLBI.

\noindent {\bf \object{1123+340}} (Fig.~\ref{fig:1123+340}). In the 15-GHz VLA
map this object appears as a compact double source \citep{naun92}. The MERLIN
image confirms that apart from the two equally bright components a weaker one
east of these two is also seen. (However, the
restoring beam of the 15-GHz map by Naundorf et al. is five times larger
so the eastern protrusion might not be resolved.) The MERLIN image accounts
for 90\% of the GB6 flux density so it seems to be fairly complete. The 1.7-GHz
VLBA image that accounts for 75\% of the interpolated total flux density at
1.7\,GHz shows a very complicated structure.

The suggested interpretation is as follows.
The field of 1123+340 encompasses {\em two} separate radio sources.
One of them consists of the components labelled ``E'' and ``S''. Based on
the morphology alone, the ``E'' component resembles a lobe that is not edge
brightened whereas the ``S'' component appears in the VLBA image as
a collection of more than ten faint, compact subcomponents located in the area
$\sim$$50 \times 50$\,milliarcseconds. We regard them as remnants
of a lobe being at an advanced stage of decay. The ``S'' component does not
appear in any of the 5-GHz images so that an assessment of its spectral index
is impossible but it is obvious that the spectrum must be very steep.

It appears, therefore, that the components ``E'' and ``S'' are the lobes of
a double source which is ``dying'': there is no core or jet visible and the
lobes are in the coasting phase. The projected linear size of this source is
$1.9h^{-1}$\,kpc. Assuming that the source does not lie exactly in the sky
plane but the ``S'' lobe is closer to the observer, a simple interpretation
of the difference between ``E'' and ``S'' is that ``S'' is seen at a later
epoch, i.e. at a more advanced stage of the ageing process, and thus is more
diffuse than ``E''. The likely steepness of the spectrum of the ``S''
component explains why it is not visible at 5\,GHz or at 15\,GHz. The
remaining components make a double source, ``N2'' being the putative core
given it has the flattest spectrum.

Although this source is reminiscent of a lens such
a possibility is unlikely (Browne, Rawlings priv. comm.).

The properties of 1123+340 will be further discussed in
Sect.~\ref{s_1123}.

\noindent {\bf \object{1232+295}} (Fig.~\ref{fig:1232+295}). The MERLIN image
accounts for entire GB6 flux density. Its brightest central component most
likely harbours the core and so the source has a triple structure. The 1.7-GHz
VLBA image that accounts for 46\% of the interpolated total flux density at
1.7\,GHz shows quite a complicated core-jet structure. The core is the only
component reproduced in the EVN image.

The most unusual feature of this source is that according to NVSS its
fractional polarisation is 9.4\%. Given that such high polarisation is not
normally observed in CSS sources, particularly at low frequencies, 1232+295 may
be intrinsically larger than normally attributed to CSS sources and substantial
parts of its radio structure may lie outside of the inner 2-3\,kpc region
responsible for Faraday depolarisation \citep{f2004}. Thus, this source may
predominantly be distorted by beaming effects.

No optical counterpart of this object is known to date. 

\noindent {\bf \object{1242+364}} (Fig.~\ref{fig:1242+364}).
In all the images shown, the source has a classic triple design with a core and
two lobes. Component ``C'' in the VLBI images is most likely the core
given that it is coincident with a $m_R=21.57$ galaxy located at
RA=\,$12^{\rm h}44^{\rm m}49\fs679$, Dec=\,$+36\degr 09\arcmin 25\farcs50$ (The
identification usually given in the literature -- a $\sim$20\,mag. object --
is in fact 6\farcs5 off the position of the radio source towards the
north-east.) It follows that the three compact components located between the
core and the northern lobe are parts of a jet. The possibility that the source
might be a gravitational lens system has been ruled out by \citet{king99}. The
MERLIN image accounts for almost entire GB6 flux density of this object whereas
the VLBA image accounts for 60\% of the interpolated total flux density at
1.7\,GHz. The spectrum of this source is one of the steepest among the sources
in Subsample Two: $\alpha\approx -1.0$. The measurement at 10\,GHz by
\citet{machin90} confirms this. At 151\,MHz the source is slightly variable
\citep{mr00}, however even assuming the maximum value of the flux density
measured by Minns \& Riley, the spectrum flattens below 365\,MHz
($\alpha=-0.42$).

\noindent {\bf \object{1251+308}} (Fig.~\ref{fig:1251+308}).
The MERLIN image, which accounts for 66\% of the GB6 flux density, suggests it
might be a double source and the bright components at its extremities are
connected by a bridge. This picture is roughly confirmed by the VLBA 1.7-GHz
image but the extended low surface brightness ``bridge'' is poorly reproduced
because of the lack of short spacings in the $u$ -- $v$ coverage. This is also
indicated by the fact that the VLBA image accounts for only 30\% of the
interpolated total flux density at 1.7\,GHz. Given that, at present, there is
no known optical counterpart of this source, it is not possible to make a
reliable classification -- it can be either a simple core-jet (the northern
feature being the core) or a double source.

\noindent {\bf \object{1343+386}} (Fig.~\ref{fig:1343+386}).
In the FIRST image there are two sources in this field: the one which is
referred to in Table~\ref{t_radio} and a 19~times weaker companion source
12\farcs8 to the north. Whether the northern weak companion is related to the
main one is unclear. The 1.46-GHz VLA map of 1343+386 by \citet{mc83} also
shows that the brighter one seems to be a compact double. The MERLIN and
EVN images confirm this but the 1.7-GHz VLBA image shows much more detail of
the emission coming from the area between to two strongest components and fully
confirms the image shown by D02a. The VLBA image accounts for 83\% of the
interpolated total flux density at 1.7\,GHz. Similarly, the MERLIN image
accounts for 79\% of the GB6 flux density. Having made a provision for a
missing part of the 1.7-GHz flux and comparing it with MERLIN 5-GHz flux
density we conclude that the component ``N'' (D02a denote it as ``N1'') has a
flat spectrum and so the source as a whole is of the core-jet type. The same
conclusion has been reached by \citet{orienti04}.

The optical identification of the source is a QSO at the redshift $z=1.844$
located at
RA=\,$13^{\rm h}45^{\rm m}36\fs942$, Dec=\,$+38\degr 23\arcmin 12\farcs51$.
Given that CSOs are usually identified with galaxies but not quasars, this fact
provides an additional constraint that 1343+386 is not a CSO.

\noindent {\bf \object{1401+353}} (Fig.~\ref{fig:1401+353}).
The 1.46-GHz VLA map presented by \citet{mc83} features a bright peak and
a $5\arcsec$ extension to the north-east. 1401+353 is present in SDSS/DR4
and identified with a $m_R=19.86$ star-like object located at
RA=\,$14^{\rm h}03^{\rm m}19\fs309$, Dec=\,$+35\degr 08\arcmin 13\farcs28$
so that it is coincident with the northeastern compact component seen in our
MERLIN image. Thus, 1401+353 is an asymmetric triple where the northeastern
extension visible in the \citet{mc83} map is a lobe while our MERLIN map shows
the core and the other lobe. The MERLIN image accounts for all the GB6 flux
density so that the northeastern lobe must have a steep spectrum.

According to NVSS, the fractional polarisation of 1401+353 is 5.3\%
which is unusual for CSS sources. To explain this it has to be assumed
that, similarly to 1232+295, 1401+353 lies predominantly outside the inner
region responsible for Faraday depolarisation. It follows that its intrinsic
size must be larger than that of {\em bona fide} CSS sources and it is beamed
to the observer so that it is foreshortened by the projection. A further
consequence of such an assumption is that it is most likely a switched-off
source and because of that, a core-jet structure, which is typical for beamed
RLAGNs, is not observed here. The light-travel scenario works particularly well
for such an orientation and so the conspicuous asymmetry of the whole object,
as seen in the map by \citet{mc83}, is explained.

1401+353 was not imaged using VLBI.

\noindent {\bf \object{1441+409}} (Fig.~\ref{fig:1441+409}). The MERLIN image
accounts for the entire GB6 flux density and it shows a simple core-jet
structure whereas at least four components are present in the VLBI images. All
of them have very steep spectra calculated based on the flux densities
extracted from 1.7-GHz VLBA and 5-GHz EVN images. This is in an agreement with
the fact that the source as a whole has a very steep spectrum between 1.4\,GHz
and 5\,GHz ($\alpha=-0.94$) but between 365\,MHz and 1.4\,GHz the spectrum is
relatively flat ($\alpha=-0.33$). In general, the source has a
CSO layout. However, based on the VLBA and EVN images it is not possible to
localize the core. 1441+409 belongs to the B3-VLA sample and its VLBA
images are shown in D02a and \citet{orienti04}. Only the 5-GHz
VLBA image shown by \citet{orienti04} has a sufficient resolution and overall
quality to pinpoint the core. Based on the SDSS/DR4 image, which covers
the field of 1441+409, this source has no optical identification.

The VLBA image accounts for 84\% of the interpolated total flux density at
1.7\,GHz.

\noindent {\bf \object{1601+382}} (Fig.~\ref{fig:1601+382}).
This object is listed as a $m_R=18.23$ galaxy in SDSS/DR4 and its optical
position
(RA=\,$16^{\rm h}03^{\rm m}35\fs171$, Dec=\,$+38\degr 06\arcmin 42\farcs92$)
coincides with the strongest radio feature. It can be tentatively assumed
that this is a core-jet source but spectral index information is required
to confirm this. The MERLIN image accounts for 66\% of the GB6
flux density. No VLBI follow-up has been made for this object. No redshift is
known at the time of writing.

\noindent {\bf \object{1619+378}} (Fig.~\ref{fig:1619+378}).
The MERLIN image accounts for almost entire GB6 flux density and is made up
of two main components. They are also present in the VLBI images; however, a
large fraction of the flux is missing there mainly because the
VLBI arrays could not reproduce the low surface brightness ``bridge'' seen in
the MERLIN image. In the 1.7-GHz VLBA image which accounts for 43\%
of the interpolated total flux density at 1.7\,GHz both components have
roughly equal flux densities whereas in the EVN 5-GHz image the northeastern
one is much weaker. This means it either has a steep spectrum or, because of
its diffuseness, it could not be reproduced with the EVN snapshot
observation. This is a clear hint that the northeastern feature is a
lobe and so the south-west compact feature could be identified as a core. Thus,
1619+378 is asymmetric and could be a core-jet-lobe source.

1619+378 has an optical identification (a QSO) and its optical position given
by SDSS/DR4
(RA=\,$16^{\rm h}21^{\rm m}11\fs276$, Dec=\,$+37\degr 46\arcmin 04\farcs92$)
coincides with the feature labelled here as a core. The redshift of 1619+378 is
$z=1.271$ \citep{kock96} and SDSS confirms this.

\noindent {\bf \object{1632+391}} (Fig.~\ref{fig:1632+391}). This a QSO at
$z=1.085$. It has been included in the SDSS/DR4 and its optical position is
RA=\,$16^{\rm h}34^{\rm m}02\fs954$, Dec=\,$+39\degr 00\arcmin 00\farcs56$.
The MERLIN image accounts for 82\% of the GB6 flux density whereas the VLBI
observations account only for a minute fraction of the full flux density: at
1.7\,GHz it is only 8\% of the interpolated total flux density at 1.7\,GHz,
whereas at 5\,GHz is it extremely low so that it was impossible to produce a
map at all. Therefore, the bulk of the low frequency flux must come from a
diffuse parts of the structure. The two southern diffuse components seen in the
1.7-GHz VLBA image seem to support this conjecture.

\noindent {\bf \object{1656+391}} (Fig.~\ref{fig:1656+391}).
According to the SDSS/DR4, this is a $m_R=20.26$ galaxy at the position
RA=\,$16^{\rm h}58^{\rm m}22\fs185$, Dec=\,$+39\degr 06\arcmin 25\farcs58$.
In the radio domain this object is clearly a double. The MERLIN image accounts
only for 54\% of the GB6 flux density. The most striking feature of this source
is that according to the 1.7-GHz VLBA image -- it accounts for 63\%
of the interpolated total flux density at 1.7\,GHz -- both putative lobes of
the source are not edge brightened as is normally expected in FR\,II-like
objects. Moreover, since they are diffuse and contain no compact features
(hotspots), the source is very poorly visible in the 5-GHz EVN image.
It appears that 1656+391 may be a relic of a switched-off object.

\noindent {\bf \object{1709+303}} (Fig.~\ref{fig:1709+303}).
This object is clearly an edge brightened double and as such can be safely
classified as a standard MSO. The MERLIN image accounts only for 47\% of the
GB6 flux density whereas the VLBA image accounts for 55\% of the interpolated
total flux density at 1.7\,GHz. According to SDSS/DR4, there is a
$m_R=21.27$ galaxy $\sim$$0\farcs4$ south-west of the radio structure.
It is not likely to be related to the radio source.

\noindent {\bf \object{1717+315}} (Fig.~\ref{fig:1717+315}).
This double source seems to have hotspots at
the lobes' extremities -- see the MERLIN image -- but the VLBI observations
poorly reproduce this. In particular, the source remains undetected with
the EVN at 5\,GHz. Most of the flux is also missing in the 1.7-GHz VLBA image
(it accounts only for 24\% of the interpolated total flux density at 1.7\,GHz)
so it is likely that the lobes are much more diffuse than might be expected
based on the MERLIN image alone, which, in turn, accounts only for 30\% of
the GB6 flux density. There is no known optical counterpart of this source.

\section{Discussion}\label{s_discuss}

\subsection{The intermittent activity scenario}

\citet{rb97} -- hereafter RB97 -- proposed a model in which extragalactic
radio sources with modest linear sizes can be intermittent on timescales of
$\sim$$10^4$--$10^5$ years. They assume that an initially small source
undergoes $10^4$ year long bursts that recur every $10^5$ years. When the
power supply cuts off, the cocoon/shocked-shell system enters a coasting phase
which is still pressure driven. Because of the drop in the source pressure, the
radio luminosity of the cocoon will fall rapidly once the ``central engine'' has
turned off. However, the shocked matter continues to expand supersonically and
keeps the basic source structure unchanged. It is to be noted that the
radiative losses were not included in the model so all of the fading is due to
expansion. This model predicts that there should be a considerable number of
MSOs which are weaker than those known so far because of the power cut-off.

RB97 do not investigate whether such short outbursts re-occurring relatively
often can actually take place in radio-loud AGNs and if so, what sort
of physical mechanism drives them. As suggested by \citet{b99}, it is possible
to combine the RB97 model with the mechanism of thermal-viscous instabilities
in the accretion disks of AGNs presented by \citet{siem96} and \citet{se97}.
That theory has been further developed by \citet{hse2001} and
\citet{janiuk04}. Further discussion of possible explanation of intermittency
in radio sources by means of this theory is presented in Sect.~5 of Paper~II.

It has been shown in this paper (but also in Paper~II) that switched-off
sources can have arbitrary linear sizes. It follows that the activity in a
radio-loud AGN {\em can} cease at early stages
of its evolution. The existence of such objects was predicted by RB97 and they
may be responsible for the overabundance of small-scale radio sources
\citep{b99}. This ``population problem'' was first noticed by \citet{ob97}
who pointed out that for the sources with linear sizes below a few kpc their
number is constant regardless of the size. This cannot be easily solved in
the framework of the ``youth scenario''.

\citet{b96} developed a simple model of the evolution of a radio source
which treats it as a bubble expanding into a radially stratified medium. It
is assumed that the external density is governed by the power law: $n\propto
r^{-\beta}$ and the luminosity is proportional to:
\begin{equation}
l^{(-\beta+4)/12},
\end{equation}
where $l$ is the source's linear size. The assumption of $\beta$ between 1.5
and 2.0 is usually made and it is reasonable for the interstellar medium in
giant elliptical galaxies. The number of sources per octave of size selected
from a flux density limited sample obeys:
\begin{equation}
N(l)\propto l^{(28-11\beta)/24}.
\end{equation}
To make the model consistent with the relationship $N\propto l^{0.4}$ observed
for larger sources \citep{ob97}, $\beta\approx 1.7$ is to be assumed.
For smaller sources $N={\rm const.}$ and so the parameter $\beta$ has to be
increased considerably. \citet{ob97} put $\beta=2.6$ within the inner few kpc
which is effectively a variant of the frustration scenario, and -- as they
admit -- it is hard to understand within the paradigm of the King-type
model of the ISM. Following \citet{b99}, we suggest that the evolution of
small sources is not necessarily different, and there is no need to adopt
extreme values for $\beta$. Instead, the intermittency which manifests
itself as a ``premature'' cease of the activity in particular phases of the
source evolution can well account for the excess of small-scale sources.

\subsection{The case of 1123+340: a switched-off object in a
cluster}\label{s_1123}

At the time of writing, this object is not included in SDSS, but
according to Rawlings (priv. comm.) the radio source 1123+340 is identified
with a near-IR giant elliptical galaxy located at
RA=\,$11^{\rm h}26^{\rm m}23\fs654$, Dec=\,$+33\degr 45\arcmin 26\farcs99$.
That galaxy is most likely the N1/2 component of the 1.7-GHz VLBA image.
There is only one definite emission line in the spectrum of this object so
its redshift $z=1.247$ is not certain \citep{rel01}. The near-IR image shows
an appreciable number of red objects close to that galaxy -- see the figure in
\citet{eales97} -- but the spectra show no evidence of any second redshift
system (Rawlings priv. comm.). It is plausible that the 1123+340
field encompasses a  cluster of galaxies and the two independent, very compact
radio sources seen in the VLBI images are hosted by two galaxies of the
cluster. Circumstantial evidence for the existence of a cluster is provided
by the fact that N1/2 and W1/2 components form a double source with
Wide-Angle-Tailed morphology which is a good tracer of the existence of the
surrounding cluster.

\citet{murgia04} reported the discovery of three ``dying'' radio galaxies (they
are sometimes termed ``faders''; see Paper~II and references therein) located
at the centre of an X-ray emitting cluster. They argue that the association of
faders with clusters implies that the pressure of the dense intracluster
medium, perhaps a cooling flow, prevents a quick dispersion of a relic radio
lobe through adiabatic expansion. Since the source expansion is reduced or
even stopped, the chance of detecting relic radio lobes is higher than without
the presence of intracluster gas. As shown in Sect.~\ref{s_notes}, the
eastern double source seen in the 1.7-GHz image (Fig.~\ref{fig:1123+340})
is fader-like. Hence, our result is similar to that of
\citet{murgia04} except that the fader seen in 1123+340 field is two orders
of magnitude more compact than those presented by Murgia et~al.

\subsection{The asymmetries caused by light-travel time effect}\label{dis_asym}

It has been suggested in Sect.~\ref{s_notes} that a number of sources
dealt with here are ``dying''; see Table~\ref{t_types}. The major problem
with accepting the above scenario is that according to the notion well
established for the LSOs \citep{kg94} -- hereafter KG94:

\begin{itemize}
\item faders should have ultra steep spectra which is not the case for any of
these sources,
\item the ``coasting phase'' of the lobes of a formerly radio-loud AGN may last
up to $10^8$~years which is three orders of magnitude more than the (spectral)
ages of the CSS sources \citep{murgia99}.
\end{itemize}

However, while the findings of KG94 are valid
for LSOs with the linear sizes of the order of 1\,Mpc, an important difference
between LSOs and MSOs exists, namely that the latter ones have overpressured
hotspots and probably the whole sources are overpressured\footnote{For
example, \citet{siem05} provided direct evidence that 3C186 is, indeed,
overpressured.} due to their subgalactic sizes. Thus, when the jets
switch off, the radio structure will still expand and suffer dramatic adiabatic
losses which are explicitly neglected by KG94 but are likely to be much more
effective in dimming a compact source than radiative losses taken into account
by KG94 (Leahy, priv. comm.). Consequently, the theory developed by KG94
predicting long decay timescales that take place in coasting LSOs is not
applicable to relics of CSS sources. As the expansion losses dominate in
subgalactic-scale faders and they could occur in a comparatively short period
of time -- $10^5$\,years (RB97) -- the lobes would quickly take the typical
form of a fader without their spectra showing signs of ageing for frequencies
below 5\,GHz. This issue was raised in Paper~II for the case of
1542+323 but more recently \citet{gir05} provided compelling evidence that,
indeed, an MSO can take the form of a fader whereas its spectrum is not
steeper than typical for CSS sources. One of the low-power compact sources
they present, 1855+37, clearly shows the features of a fader: low
power of the core, the lack of visible jets and diffuse lobes without hotspots.
However, the spectral index between 1.4 and 5\,GHz estimated from their plot is
only $\alpha \approx -0.8$.

If a double-lobed radio source does not lie close to the sky plane then a
significant time lag between the apparent stages of the evolution of the two
lobes that are no longer fuelled should be observed. However, for the LSOs,
even with linear sizes $\ga$1\,Mpc, the light-travel time, i.e. the time lag
between the images of the lobes as the observer perceives them, is about two
orders of magnitude less than the timescale of the relic lobes decay time as
given by KG94 and so in practice it is not observable. Contrary to that, in a
(sub)galactic-scale source with a linear size of a typical galaxy diameter i.e.
$\approx10^5$ light years, the time lag between the images of both lobes starts
to play a role. This is because the lobes' decay time is of a similar order
of magnitude ($10^5$ years, RB97) as the time lag which, albeit shorter
than the source linear size expressed in the light-travel time due to
inclination of the source axis with respect to the sky plane, still has an
order of magnitude of $\sim$$10^5$ years provided the inclination angle is not
very small. As a result, differences in the luminosities and morphologies of
the lobes might become conspicuous.

In the case of an appreciable alignment of the source's axis with respect
to the line of sight, polarisation asymmetry is also expected, given that
the larger parts of the source would have already been out of the ``Faraday
fog'' found in the vicinity of the centre of the host galaxy. Moreover, an
assumption that the source is in a switched-off state is necessary here to
explain the lack of dominating core-jet structure coincident with the optical
counterpart of the radio source as expected for a switched-on source.
1401+353 is perhaps a good example of a switched-off source beamed towards
the observer.

The issue of luminosity asymmetries in CSS sources was thoroughly investigated
both observationally \citep{saikia01, saikia02} and theoretically
\citep{jey05}. They noticed that MSOs in general exhibit much
larger asymmetries than LSOs and that this could be attributed to asymmetries
in the distribution of gas on the opposite sides of the nucleus of the radio
source. While this explanation remains valid both for ``normal'' sources where
energy is currently transported to the lobes and ``dying'' sources, we
suggest that the time-lag scenario outlined above may also play an
important role in the case of the latter class of the sources. High dynamic
range observations of the sources like 0744+291 and 0922+322 leading to a
possible discovery of ``missing'' lobes and estimation of the spectral indices
of both lobes are necessary to confirm the validity of the interpretation of
the asymmetries by means of the rapid decay of relic lobes in compact sources
and time lags due to the source orientation. Additionally, this would provide
firm proof that the timescales of the decay of relic lobes in compact sources
are indeed short, in accordance with RB97.

Finally, while we argue above that compact sizes of sources make the asymmetry
caused by light-travel time effect more likely than in the case of LSOs,
asymmetries of this kind {\em are} observed in objects appreciably larger than
CSS sources. In TXS\,1602+324 \citep{mc83} the southern lobe is devoid of a
hotspot whereas the northern one is clearly of FR\,II type. It may be
speculated that at a later stage of the evolution the southern lobe of
TXS\,1602+324 could fade out and become almost completely invisible while the
northern one would remain observable. If this scenario actually develops,
TXS\,1602+324 might transform into a source similar to PKS\,1400$-$33
\citep{subra03} where, indeed, only one lobe is observed.

\subsection{Where are young FR\,Is?}

For LSOs there are two distinct morphological Fanaroff--Riley classes and the
FR\,I objects are less luminous at 178\,MHz \citep{fr74}. Given that
according to the evolutionary scenario, 3CRPW CSS sources evolve towards
FR\,II radio galaxies, it can be speculated that there might be a
relationship between weak CSS sources and FR\,Is. In the simplest
form such a relationship would mean a morphological similarity of a
weak CSS source to an FR\,I. However, among the sources investigated here
the FR\,I-like morphology is never observed. This not difficult to interpret
taking into account that CSS sources have subgalactic linear sizes. The
propagation in the atmosphere of the host galaxy will tend to keep the jets
overpressured, and so to maintain sources as FR\,IIs until they exit the
galaxy atmosphere \citep[see e.g.][]{co02}. This would suggest that a possible
transition from ``post-CSS'' FR\,II phase to FR\,I would only occur once the
radio source has propagated well beyond the host galaxy.

\citet{orienti04} claim they have found two FR\,I-like
CSS sources in their sample -- 1242+410 and 2358+406 -- but in our opinion
this interpretation is uncertain. 1242+410 is a QSO so the radio source is
likely to be distorted by beaming and 2358+406, although it
resembles an FR\,I in the 1.7-GHz image (D02a), has a quite
complicated, multiple structure in 5 and 8.4-GHz images \citep{orienti04}.

Nevertheless, a relationship between weak CSS sources and FR\,Is might
exist. According to the ``bubble model'' \citep{b96}, a
$\sim$100\,pc-sized source after having evolved into 100\,kpc source would
decrease its luminosity by a factor of 20. This means that only
bright CSS sources would evolve into FR\,IIs. Thus, it cannot be ruled
out that weak CSS sources, even if they resemble FR\,IIs at the MSO stage,
are progenitors of large-scale FR\,Is \citep{snellen00}. However, the
transition from an FR\,II to FR\,I phase is still not well understood,
although it has been predicted by \citet{gc2001}.

\section{Conclusions}\label{s_concl}

The conclusions of the study of 19~weak CSS sources with sizes below
$\approx 1\arcsec$ presented here are as follows.

\begin{enumerate}

\item The sources in Subsample Two have a variety of morphological types as
shown in Table~\ref{t_types}. Such a variety remains in agreement with
the general picture drawn by \citet{fanti90} that CSS sources are intrinsically
small and can be randomly oriented with regard to the sky plane. However,
there are only 3 quasars among the objects investigated. Therefore, distortions
caused by beaming perhaps play a secondary role compared to those
resulting from the asymmetries in the distribution of host galaxy ISM (but see
also point 3. below).

\item The well established youth scenario remains valid. However, an important
factor has been added to it: the activity of AGNs hosting young radio sources
can be intermittent and so they can ``die early''. At least three sources
with clear signs of ceased activity have been found. The relic object in
1123+340 field is of a particular importance. It is most likely a member of a
cluster of galaxies and as such it is immersed in the intracluster gas which
slows down the dispersion of the lobes and so helps to preserve their shape.

\item According to RB97, the timescale of the drop in the radio luminosity
of ``dying'' CSS sources is of the order of $10^5$ years. It could therefore be
comparable to the light-travel time across the source itself. Unless the source
does not lie in the sky plane, the apparent differences in the evolutionary
stages of coasting lobes can be visible and take the form of high asymmetries
in the images of its lobes which, intrinsically, can be quite symmetric.

\item None of the sources has an FR\,I morphology, which is in agreement with
the notion that the FR\,I-like structures develop only above the atmosphere of
the host galaxy. It is not excluded, however, that weak CSS sources {\em can}
eventually become FR\,Is  but the transition from FR\,II-type morphology
typical for CSS sources to FR\,I-type morphology remains unclear.

\end{enumerate}

\begin{acknowledgements}

\item MERLIN is operated by the University of Manchester as a National Facility
on behalf of the Particle Physics \& Astronomy Research Council (PPARC).

\item The VLBA is operated by the National Radio Astronomy Observatory (NRAO),
a facility of the National Science Foundation (NSF) operated under
cooperative agreement by Associated Universities, Inc. (AUI).

\item JIVE is the Joint Institute for Very Long Baseline
Interferometry in Europe. It was created by the European Consortium for VLBI
and is a member of the European VLBI Network (EVN). Its primary task is to
operate the EVN Mk\,IV VLBI Data Processor.

\item The Automatic Plate Measuring (APM) machine is a National Astro\-nomy
Facility run by the Institute of Astronomy in Cambridge (UK). Official website:
http://www.ast.cam.ac.uk/$\sim$apmcat.

\item Use has been made of the fourth release of the Sloan Digital Sky
Survey (SDSS) Archive. Funding for the creation and distribution of the
SDSS Archive has been provided by the Alfred P. Sloan Foundation, the
Participating Institutions, the National Aeronautics and Space
Administration, the National Science Foundation, the U.S. Department of
Energy, the Japanese Monbukagakusho, and the Max Planck Society. The SDSS
Web site is http://www.sdss.org/. The SDSS is managed by the Astrophysical
Research Consortium (ARC) for the Participating Institutions. The
Participating Insti\-tutions are The University of Chicago, Fermilab, the
Insti\-tute for Advanced Study, the Japan Participation Group, The Johns
Hopkins University, Los Alamos National Labora\-tory, the
Max-Planck-Institute for Astronomy (MPIA), the Max-Planck-Institute for
Astrophysics (MPA), New Mexico State University, University of Pittsburgh,
Princeton University, the United States Naval Observatory, and the
University of Washington.

\item This research has made use of the NASA/IPAC Extragalactic Database
(NED) which is operated by the Jet Propulsion Laboratory, California
Institute of Technology, under contract with the National Aeronautics and
Space Administration.

\item Part of this research was made when MK-B stayed at Jodrell Bank
Observatory and received a scholarship provided by the EU under the
Marie Curie Training Site scheme.

\item Interesting discussions on thermal-viscous instabilities with
Bo\.zena Czerny and Agnieszka Janiuk made our findings on the role
of that mechanism clearer.

\item We are very grateful to Peter Thomasson for suggestions leading to a
significant improvement of this paper.

\item We thank the referee for a comprehensive report and discussion.

\end{acknowledgements}

\Online

\begin{table*}[t]
\caption[]{Fitted parameters of the MERLIN map components.}
\begin{flushleft}
\begin{tabular}{c|c|c|r|r|r|r|c}
\hline
Source &      &     & ${\rm S_{5GHz}}$& $\theta_{1}$ & $\theta_{2}$ & P.A.
& Opt.\\
name   & R.A. & Dec & [mJy] & [mas] & [mas] & [deg.] & ID \\
(1)    & (2) & (3)   & (4)    & (5)        & (6)      & (7) & (8) \\
\hline
\hline
0744+291 & 07 48 05.314 & 29 03 22.72 &   2.4 & 121 &  95 & 114 & \\
         & 07 48 05.335 & 29 03 22.25 &  77.5 & 124 &  80 & 107 & \\
\hline                                                
0747+314 & 07 50 12.311 & 31 19 47.57 & 166.6 &  36 &  23 &  74 & \\
         & 07 50 12.316 & 31 19 47.58 &  72.0 &  19 &  14 &  84 & \\
         & 07 50 12.330 & 31 19 47.60 & 163.7 &  32 &  17 &  73 & \\
\hline                                                
0811+360 & 08 14 49.066 & 35 53 49.72 &  71.4 &  22 &   6 & 114 & \\
         & 08 14 49.077 & 35 53 49.67 &  81.9 &  14 &   7 & 130 & \\
\hline                                                
0853+291 & 08 56 01.153 & 28 58 34.81 &  20.4 & 292 &  65 & 171 & \\
         & 08 56 01.161 & 28 58 34.87 &  72.9 & 299 &  91 &  36 & \\
         & 08 56 01.164 & 28 58 35.12 &  44.7 &  83 &  43 & 167 & \\
         & 08 56 01.226 & 28 58 35.48 &   5.6 &   0 &   0 & --- & $\bullet$\\
\hline                                                
0902+416 & 09 05 22.185 & 41 28 39.70 & 129.6 &  74 &  12 & 171 & $\bullet$\\
         & 09 05 22.190 & 41 28 39.96 &  19.6 &  47 &  39 & 174 & \\
\hline                                                
0922+322 & 09 25 32.679 & 31 59 52.98 &   2.1 &   0 &   0 & --- & $\bullet$\\
         & 09 25 32.753 & 31 59 52.29 &  73.5 & 185 & 118 & 133 & \\
\hline                                                
1123+340 & 11 26 23.651 & 33 45 26.83 & 141.0 &  36 &  16 &  25 & \\
         & 11 26 23.667 & 33 45 26.97 & 147.9 &  67 &  24 &  90 & \\
         & 11 26 23.674 & 33 45 26.95 &  46.7 &  75 &  34 & 118 & \\
\hline                                                
1232+295 & 12 34 54.373 & 29 17 43.96 &  56.5 &  93 &  63 & 119 & \\
         & 12 34 54.380 & 29 17 43.87 & 106.7 & 149 &  47 & 131 & \\
         & 12 34 54.387 & 29 17 43.80 &  10.0 &  16 &   5 & 137 & \\
         & 12 34 54.379 & 29 17 43.95 &  18.4 &  20 &   7 &  52 & \\
\hline                                                
1242+364 & 12 44 49.672 & 36 09 25.35 &  51.3 &  36 &  33 &  90 & \\
         & 12 44 49.685 & 36 09 25.53 &  20.2 &  71 &  38 &  32 & $\bullet$\\
         & 12 44 49.700 & 36 09 25.77 & 115.0 &  63 &  17 &  53 & \\
\hline                                                
1251+308 & 12 53 25.736 & 30 36 35.28 &  36.0 &  17 &   6 & 113 & \\
         & 12 53 25.741 & 30 36 35.18 &   4.5 &  26 &  12 & 167 & \\
         & 12 53 25.745 & 30 36 34.99 &  17.3 &  88 &  78 & 123 & \\
         & 12 53 25.747 & 30 36 34.91 &  15.2 &  76 &  47 &   3 & \\
         & 12 53 25.750 & 30 36 34.74 &  57.8 &  35 &  14 & 146 & \\
\hline                                                
1343+386 & 13 45 36.949 & 38 23 12.59 &  88.5 &  25 &  13 & 167 & \\
         & 13 45 36.951 & 38 23 12.49 & 260.9 &  17 &  12 & 135 & \\
\hline                                                
1401+353 & 14 03 19.231 & 35 08 11.81 &  36.9 &  81 &  53 &   8 & \\
         & 14 03 19.240 & 35 08 11.70 &  57.4 &  31 &   0 & 150 & \\
         & 14 03 19.244 & 35 08 11.97 &  73.0 & 151 &  61 &  39 & \\ 
         & 14 03 19.319 & 35 08 13.34 &   3.5 &  35 &  18 &  83 & $\bullet$\\
\hline
\end{tabular}
\end{flushleft}
\end{table*}
\begin{table*}
{\small{\bf Table \thetable{}.} continued}
\begin{flushleft}
\begin{tabular}{c|c|c|r|r|r|r|c}
\hline
Source &      &     & ${\rm S_{5GHz}}$& $\theta_{1}$ & $\theta_{2}$ &P.A.
& Opt.\\
name   & R.A. & Dec & [mJy] & [mas] & [mas] & [deg.]  & ID \\
(1)    & (2) & (3)   & (4)    & (5)        & (6)      & (7)  & (8) \\
\hline                                                
1441+409 & 14 42 59.327 & 40 44 29.01 & 317.4 &  90 &  14 &  63 & \\
\hline
1601+382 & 16 03 35.182 & 38 06 42.83 &  64.3 &  70 &  30 &   1 & $\bullet$\\
         & 16 03 35.185 & 38 06 43.06 &  37.1 & 222 &  98 &  41 & \\
         & 16 03 35.187 & 38 06 43.04 &  10.1 & 153 &  37 & 121 & \\
         & 16 03 35.190 & 38 06 43.28 &  21.4 &  86 &  32 & 175 & \\
\hline                                                
1619+378 & 16 21 11.278 & 37 46 04.69 &   5.7 &  69 &  63 &  30 & \\
         & 16 21 11.286 & 37 46 04.89 &  68.9 &  31 &  12 &  53 & $\bullet$\\
         & 16 21 11.299 & 37 46 05.05 &   9.7 & 209 &  99 &  64 & \\
         & 16 21 11.316 & 37 46 05.26 & 101.2 &  62 &  50 &  60 & \\
\hline                                                
1632+391 & 16 34 02.925 & 39 00 00.20 & 113.5 & 142 &  94 &  47 & \\
         & 16 34 02.940 & 39 00 00.29 & 122.4 & 114 &  74 &  25 & \\
         & 16 34 02.952 & 39 00 00.38 &  57.7 &  55 &  53 &  32 & \\
         & 16 34 02.963 & 39 00 00.63 &  13.5 &  34 &  22 & 164 & $\bullet$\\
\hline                                                
1656+391 & 16 58 22.176 & 39 06 25.63 &  90.1 &  21 &  12 & 112 & \\
         & 16 58 22.185 & 39 06 25.56 &  42.1 &  24 &  13 & 115 & \\
\hline                                                
1709+303 & 17 11 19.930 & 30 19 17.92 &  71.0 &  46 &  34 & 164 & \\
         & 17 11 19.959 & 30 19 17.73 & 102.6 &  49 &  19 & 114 & $\bullet$\\
\hline                                                
1717+315 & 17 19 30.062 & 31 28 48.34 &  33.2 & 100 &  45 & 114 & \\
         & 17 19 30.095 & 31 28 48.28 &   5.0 & 130 &  77 &  78 & \\
         & 17 19 30.104 & 31 28 48.30 &   8.3 &  39 &  31 & 102 & \\
\hline
\end{tabular}
\vspace{0.2cm}

In column (8) indication is given which component is uniquely identified with
SDSS/DR4 optical object.
\end{flushleft}
\label{t_MERLIN}
\end{table*}

\clearpage

\begin{table*}[ht]
\caption[]{Fitted parameters of the 1.7-GHz VLBA and 5-GHz EVN maps components.}
\begin{flushleft}
\begin{tabular}{c|c|r|r|r|r|r|r|r|r}
\hline
Source & Compo- & ${\rm S_{1.7GHz}}$& $\theta_{1}$ & $\theta_{2}$ &P.A. & ${\rm S_{5GHz}}$& $\theta_{1}$ & $\theta_{2}$ &P.A.\\
name   & nent(s)& [mJy] & [mas] & [mas] & [deg.] & [mJy] & [mas] & [mas] & [deg.]\\
(1)    & (2) & (3)   & (4)    & (5)        & (6)& (7)   & (8)    & (9)        & (10)\\
\hline
\hline
0747+314 & E1 & 277.3 &  23 &  12 &  59 &  49.4 &  14 &   4 &  58\\
         & E2 &   9.2 &  18 &   6 & 105 & & & &\\
         & C  &   8.0 &  11 &  -- &  -- & & & &\\
         & W2 &   3.1 &  26 &   7 &  53 & & & &\\
         & W1 & 274.6 &  19 &  15 &  46 &  39.8 &  10 &   6 &  46\\
\hline
0811+360 & E  & 206.0 &  14 &  10 & 160 &  46.0 &  11 &   3 & 174\\
         & W  & 204.7 &  21 &   9 & 118 &  36.6 &  10 &   2 & 143\\
\hline
0853+291 & C  &   6.1 &  10 &   5 &  72 & & & &\\
         & M  &  48.5 &  13 &  12 & 112 &  11.0 &   1 &  -- &  --\\
         & S  &   4.8 &  24 &  14 & 105 & & & &\\
\hline
0902+416 & C  &  51.9 &  10 &   8 & 170 &  17.7 &   2 &  -- &  --\\
         & S  & 144.5 &  31 &  22 & 172 &   4.2 &   7 &   5 &  56\\
\hline
1123+340 & E  &  46.0 &  33 &  12 &  32 & & & &\\
         & N1 &  90.7$\dagger$ &  28 &   8 &  60 &  17.4 &  12 &   5 &  57\\
         & N2 &       &     &     &     &  11.1 &  12 &   5 &  56\\
         & N3 & 111.4 &   8 &   5 &  75 &  50.8 &   3 &   2 &  92\\
         & W1 & 388.1 &  14 &  13 &  92 &  60.3 &  11 &   8 & 113\\
         & W2 & 164.6 &  15 &  12 &  43 &  20.0 &  13 &   3 &  76\\
         & W3 &   9.4 &   7 &   1 & 112 &   5.9 &   9 &   4 & 109\\
\hline
1232+295 & E1 &  11.2 &  26 &  10 & 116 & & & &\\
         & E2 &  43.7 &  38 &  11 & 114 &   1.6 &   4 &   3 & 107\\
         & E3 &  37.1 &  28 &   9 & 112 &  10.3 &  12 &  -- &  --\\
         & E4 &  29.0 &  26 &  10 & 130 & & & &\\
         & E5 &   9.6 &  26 &  15 & 168 & & & &\\
         & W1 &  11.2 &  35 &  11 &  42 & & & &\\
         & W2 &  31.1 &  29 &  11 &  22 & & & &\\
\hline
1242+364 & N2 & 247.5 &  26 &  13 & 108 &  11.3 &  14 &   2 & 113\\
         & N1 &  60.1 &  17 &   7 &  26 & & & &\\
         & C  &  16.7 &   4 &  -- &  -- &  14.3 &   2 &  -- &  --\\
         & S  &  68.8 &  15 &  12 & 160 &   9.0 &   5 &   3 & 133\\
\hline
1251+308 & N2 &  50.2 &  17 &   5 & 109 &  18.1 &   4 &  -- &  --\\
         & N1 &   6.1 &  23 &  12 & 141 & & & &\\
         & S  &  64.6 &  33 &  20 & 143 & & & &\\
\hline
1343+386 & N  &  69.0 &   8 &   4 & 178 &  46.5 &   2 &   1 &  14\\
         & C  &  15.1 &  12 &   6 & 164 & & & &\\
         & S  & 559.1 &   9 &   7 & 121 & 136.7 &   4 &   4 &  46\\
\hline
1441+409 & E  & 119.6 &  12 &   9 &  26 &  12.2 &   5 &   1 & 145\\
         & C  & 165.7 &  15 &   3 &  71 &  46.1 &  15 &   2 &  70\\
         & W  & 405.5 &  11 &   7 &  64 &  95.7 &   6 &   4 &  54\\
\hline
1619+378 & L  & 120.7 &  23 &  19 & 168 &   5.5 &   7 &   4 &  85\\
         & C  & 118.5 &  20 &   3 &  55 &  41.4 &  10 &   1 &  58\\
\hline
1632+391 & C  &  12.1 &   7 &  -- &  -- & & & &\\
         & M  &  26.0 &  15 &  12 & 162 & & & &\\
         & W  &  28.3 &  25 &  22 &  38 & & & &\\
\hline
1656+391 & E  &  96.2 &  22 &  13 & 121 & & & &\\
         & W  & 265.5 &  22 &  16 & 121 &  15.6 &  18 &  10 &  91\\
\hline
1709+303 & E  & 324.6 &  31 &  19 & 136 &  18.1 &   7 &   2 & 169\\
         & W  & 165.9 &  20 &  10 & 128 &  31.5 &   5 &   3 & 117\\
\hline
1717+315 & E  &  19.5 &  26 &  13 &  48 & & & &\\
         & W  &  75.1 &  31 &  20 &  93 & & & &\\
\hline
\end{tabular}
\end{flushleft}
\label{t_VLBI}
\end{table*}

\clearpage

\clearpage
\clearpage
\clearpage
\clearpage
\clearpage
\clearpage
\begin{figure*}
\begin{flushleft}
\centering
\includegraphics[scale=0.5]{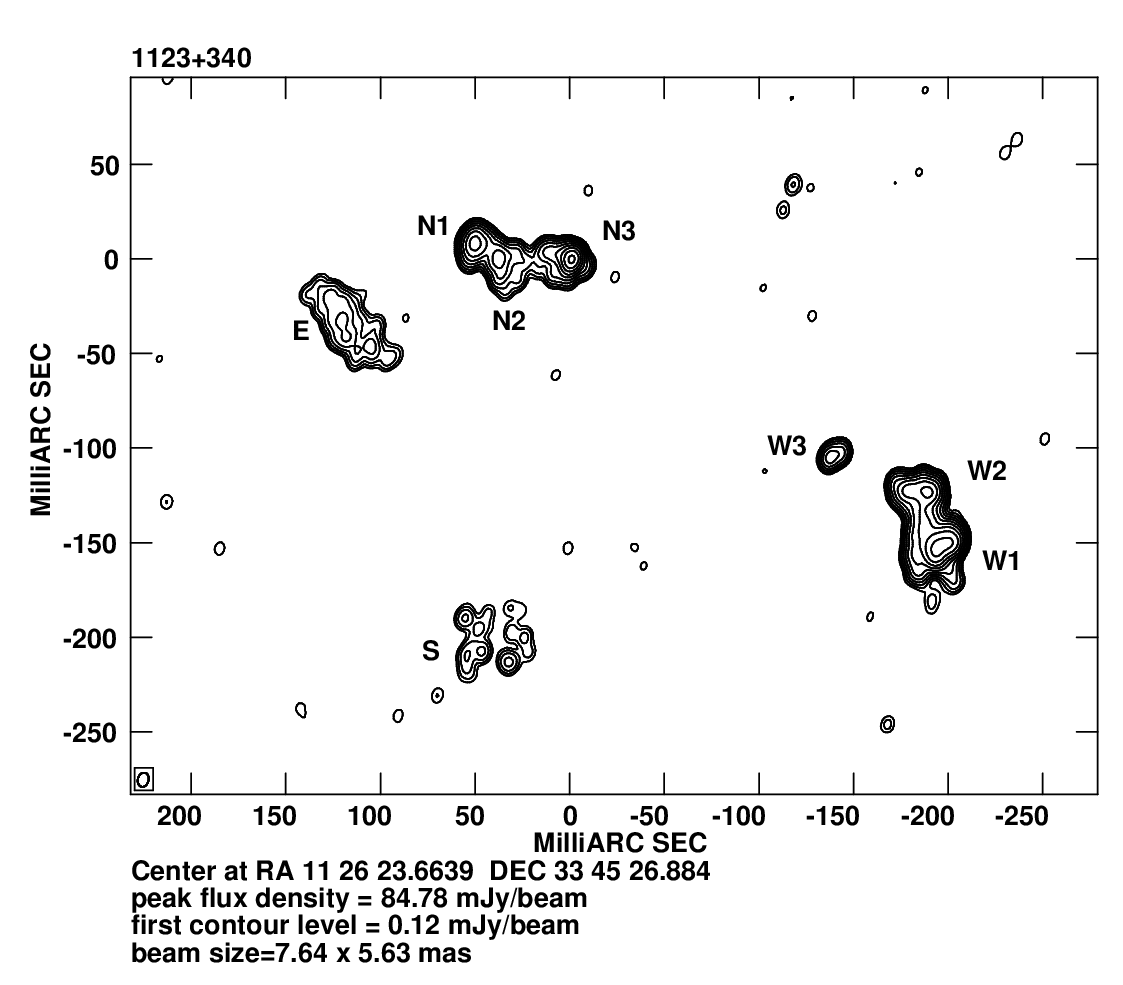}
\includegraphics[scale=0.5]{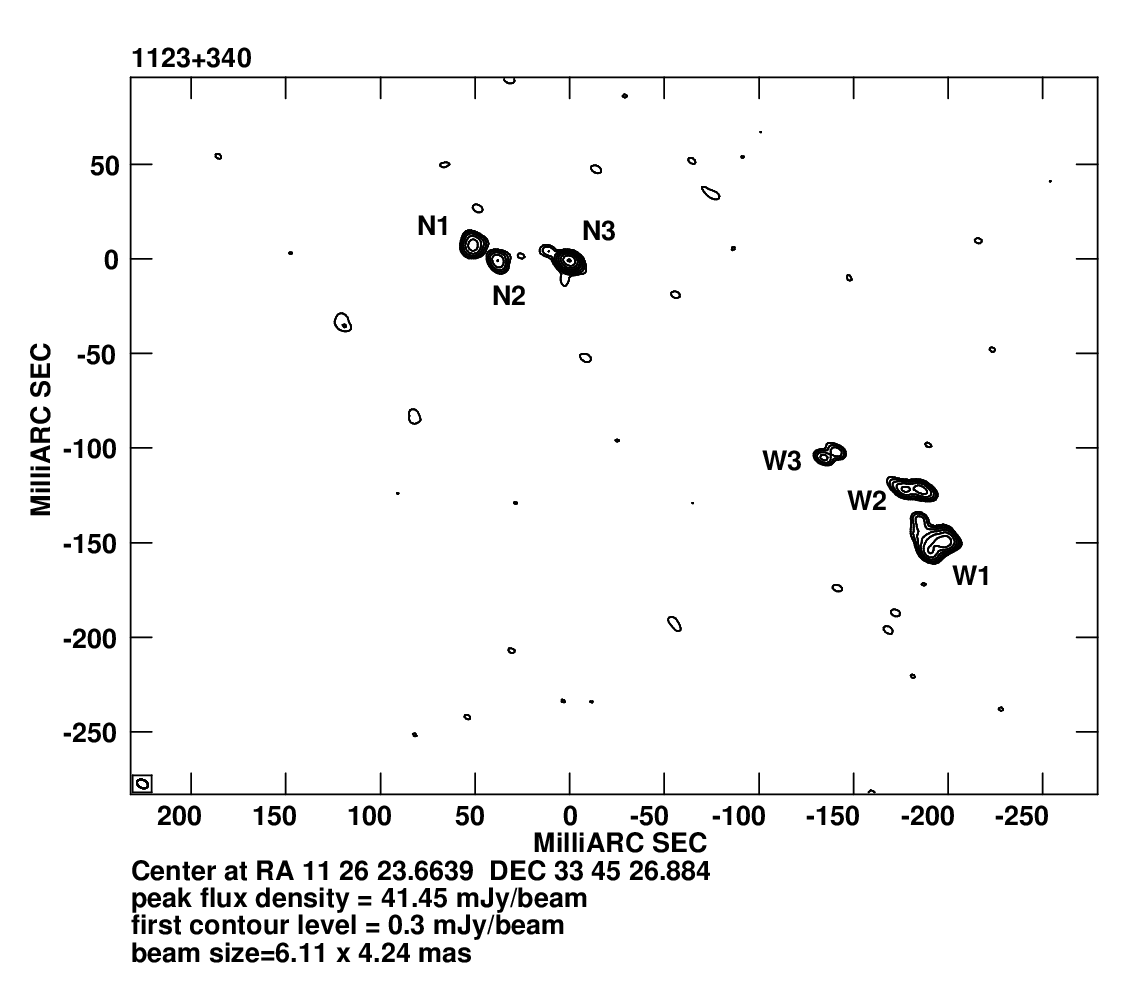}
{\small\bf\\
Fig.\,\thefigure{}.}~(continued)
1123+340 --- VLBA map at 1.7\,GHz (upper panel) and EVN map at 5\,GHz (lower panel).
\end{flushleft}
\end{figure*}
\clearpage
\begin{figure*}
\begin{flushleft}
\centering
\includegraphics[scale=0.5]{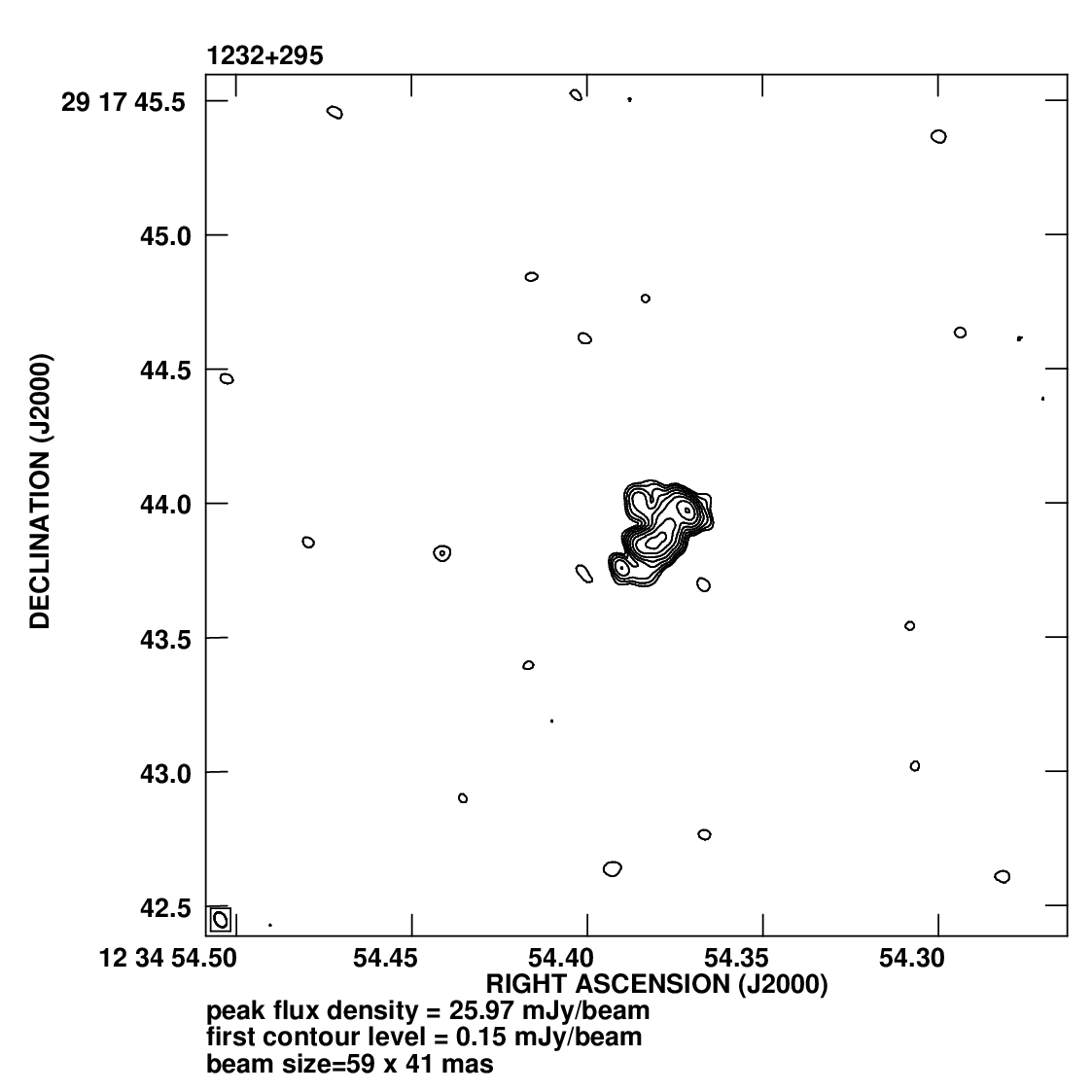}
\end{flushleft}
\begin{flushleft}
\centering
\includegraphics[scale=0.4]{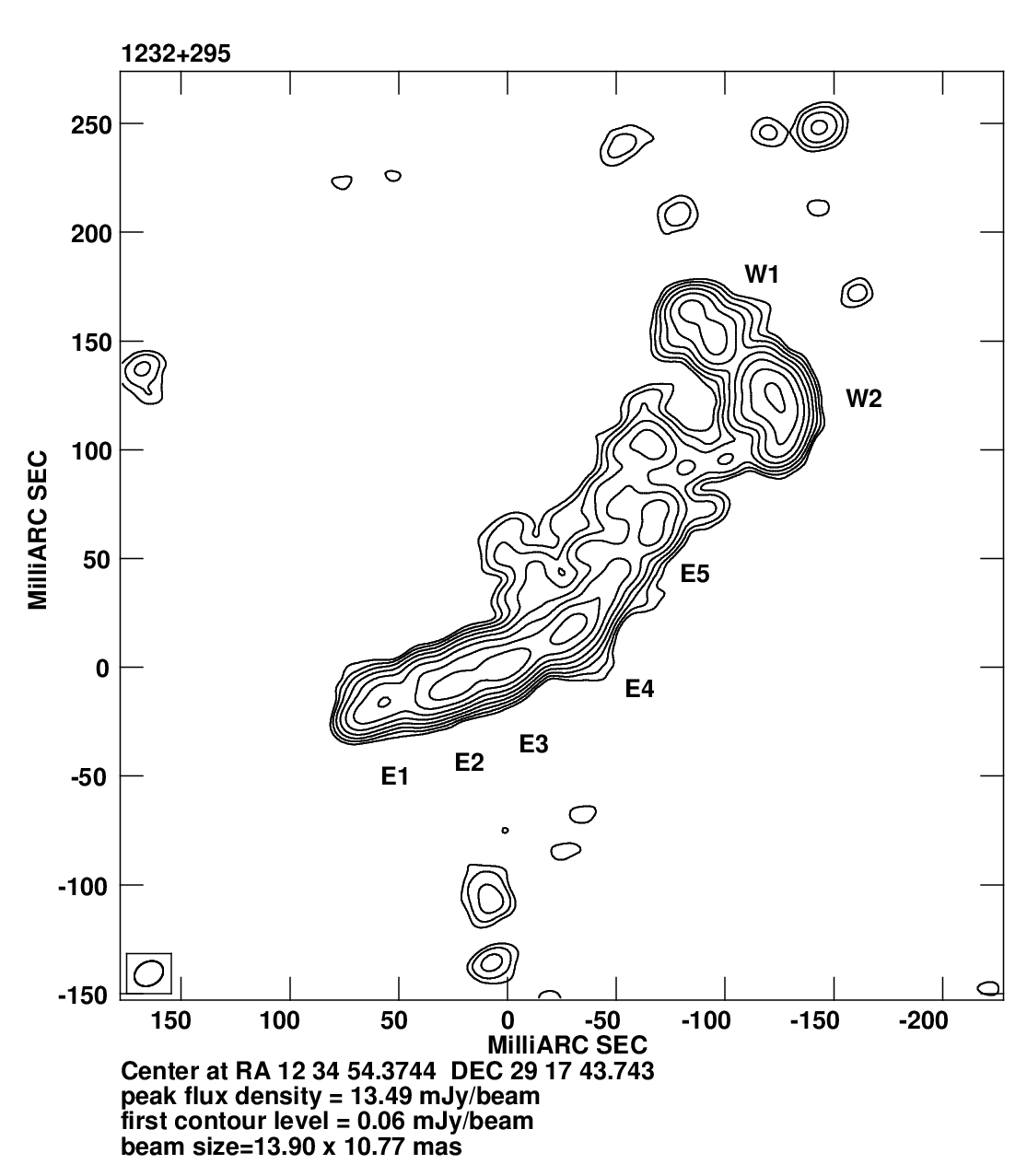}
\includegraphics[scale=0.4]{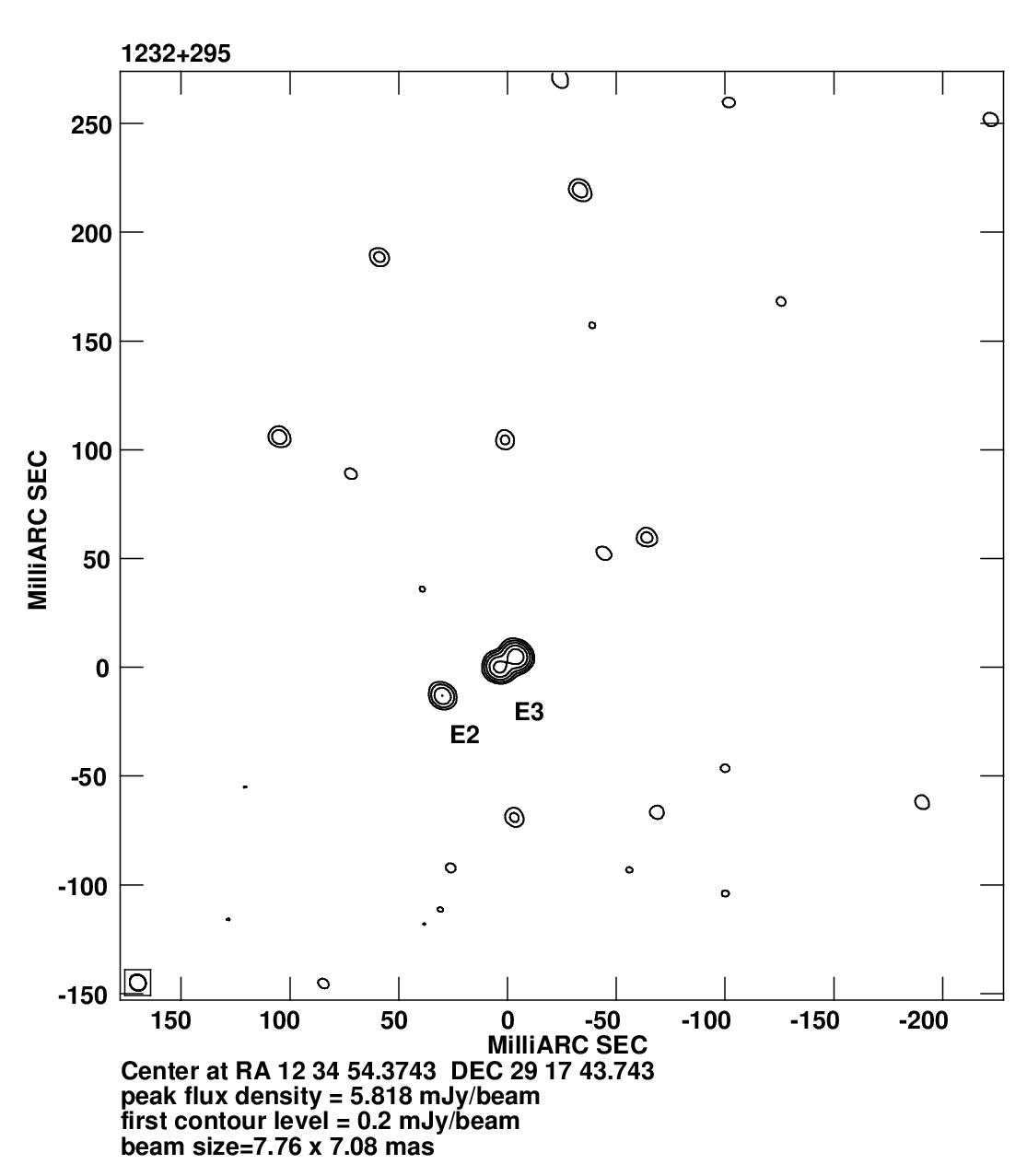}
\end{flushleft}
\caption{1232+295 -- MERLIN map at 5\,GHz (upper panel), VLBA map at 1.7\,GHz
(lower left panel) and EVN map at 5\,GHz (lower right panel).
}
\label{fig:1232+295}
\end{figure*}
\clearpage
\begin{figure*}
\begin{flushleft}
\centering
\includegraphics[scale=0.5]{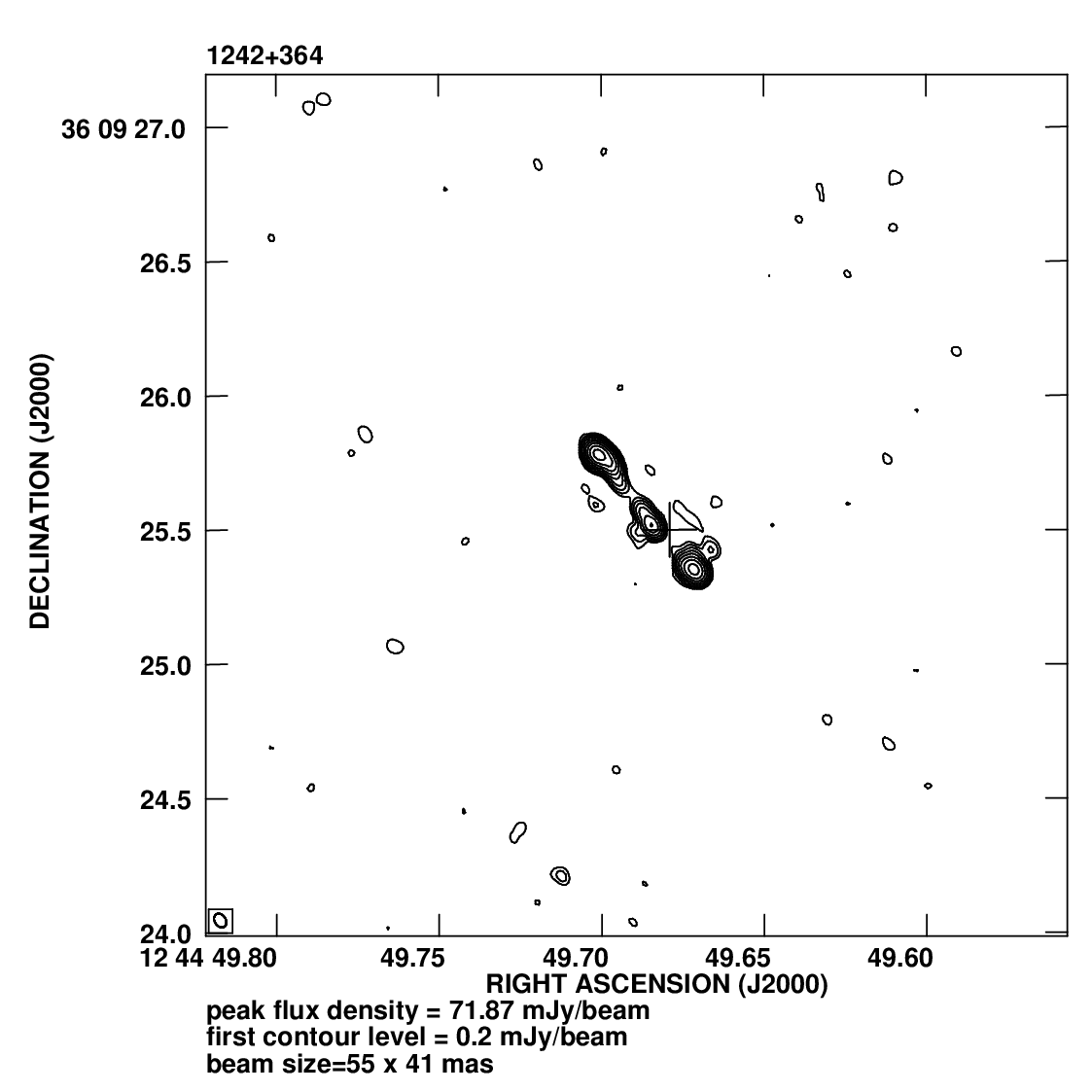}
\end{flushleft}
\begin{flushleft}
\centering
\includegraphics[scale=0.5]{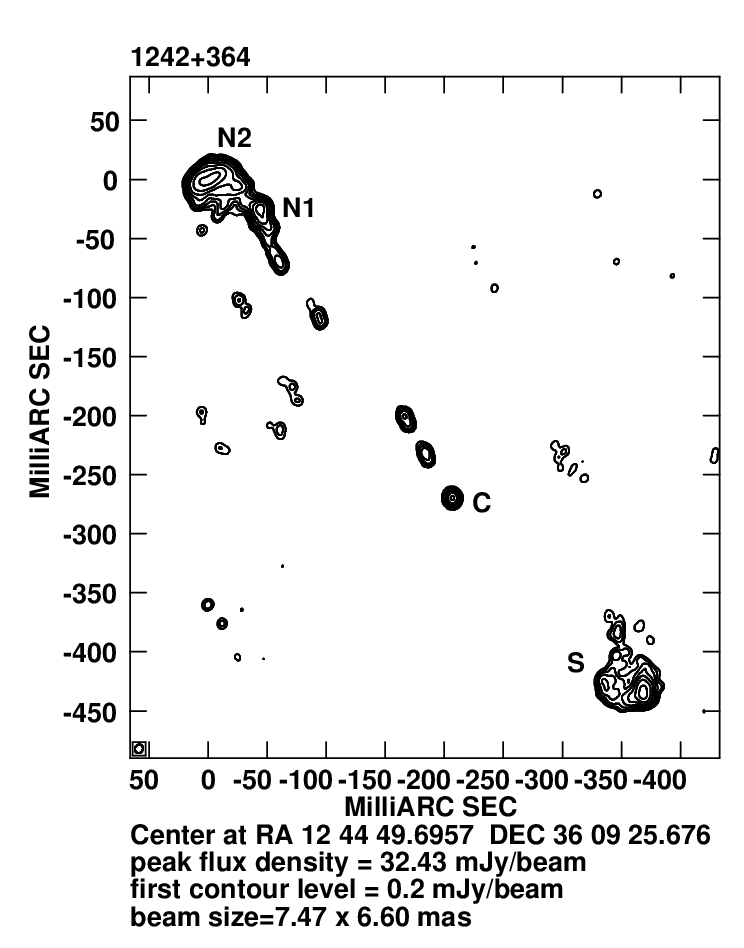}
\includegraphics[scale=0.5]{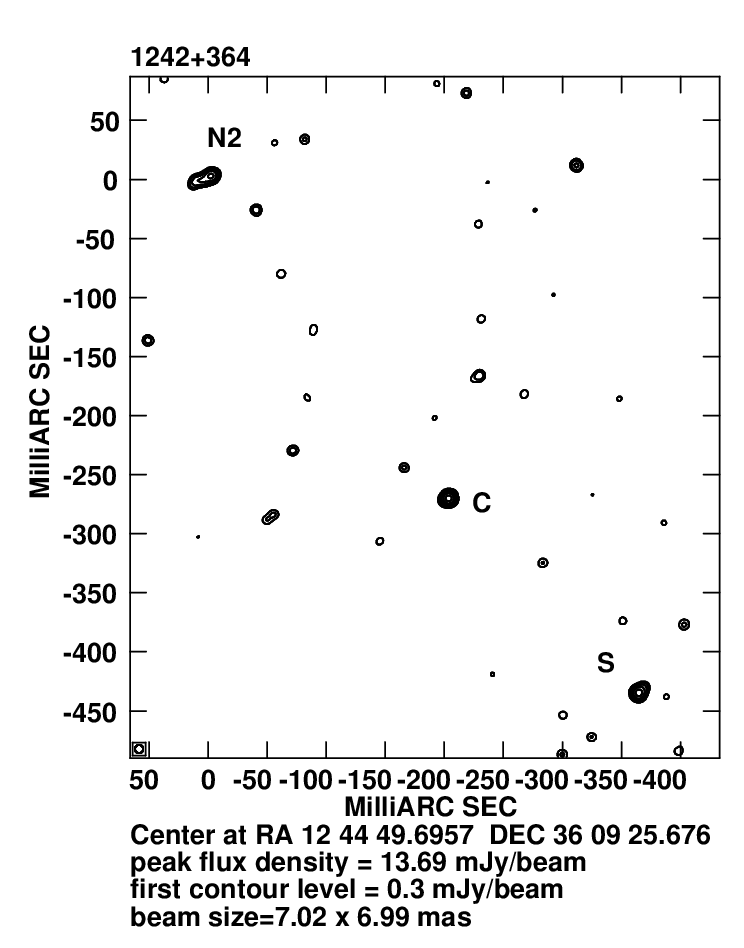}
\end{flushleft}
\caption{1242+364 -- MERLIN map at 5\,GHz (upper panel), VLBA map at 1.7\,GHz
(lower left panel) and EVN map at 5\,GHz (lower right panel).
The cross on the MERLIN
map indicates the position of an optical object in SDSS/DR4.}
\label{fig:1242+364}
\end{figure*}
\clearpage
\begin{figure*}
\begin{flushleft}
\centering
\includegraphics[scale=0.5]{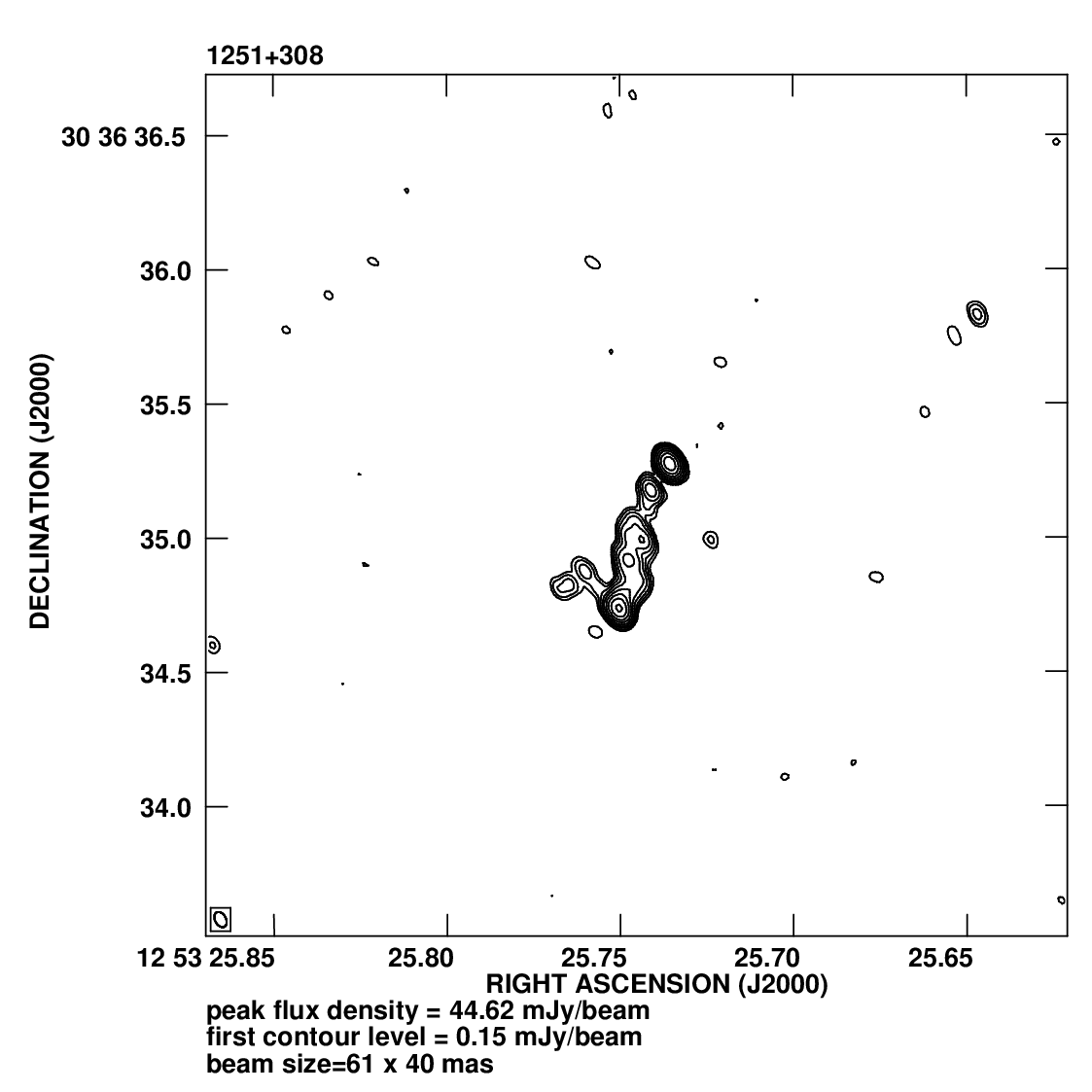}
\end{flushleft}
\begin{flushleft}
\centering
\includegraphics[scale=0.5]{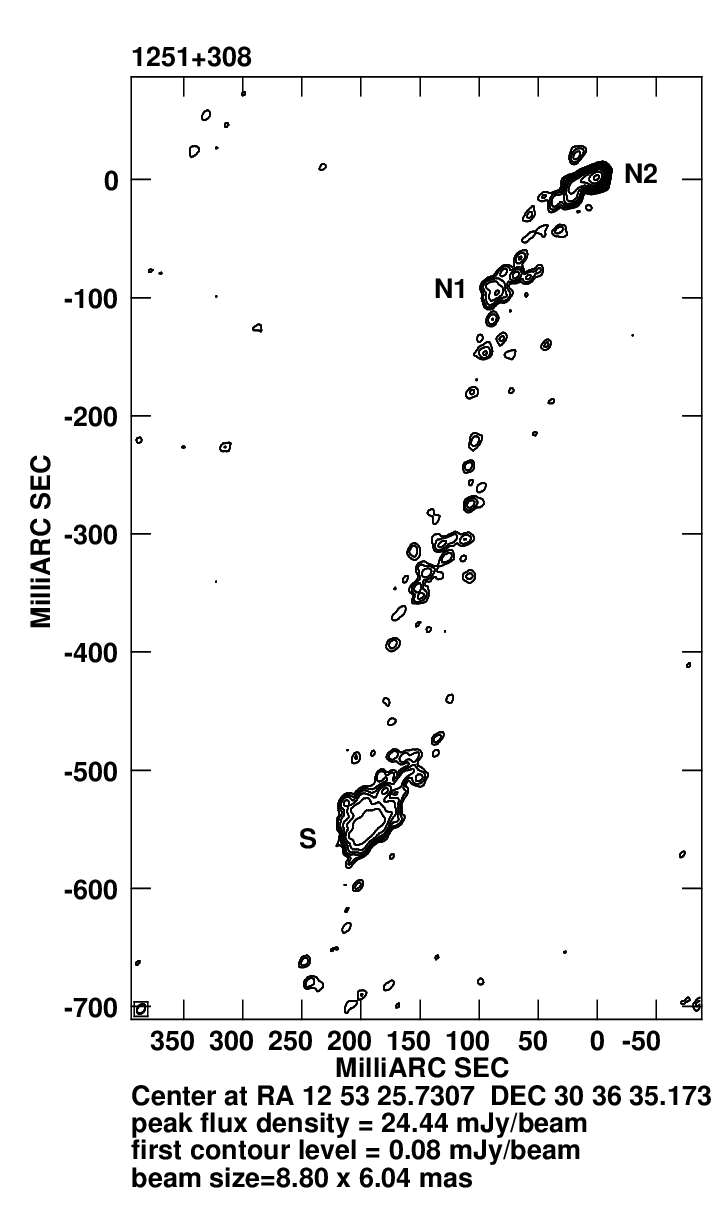}
\includegraphics[scale=0.5]{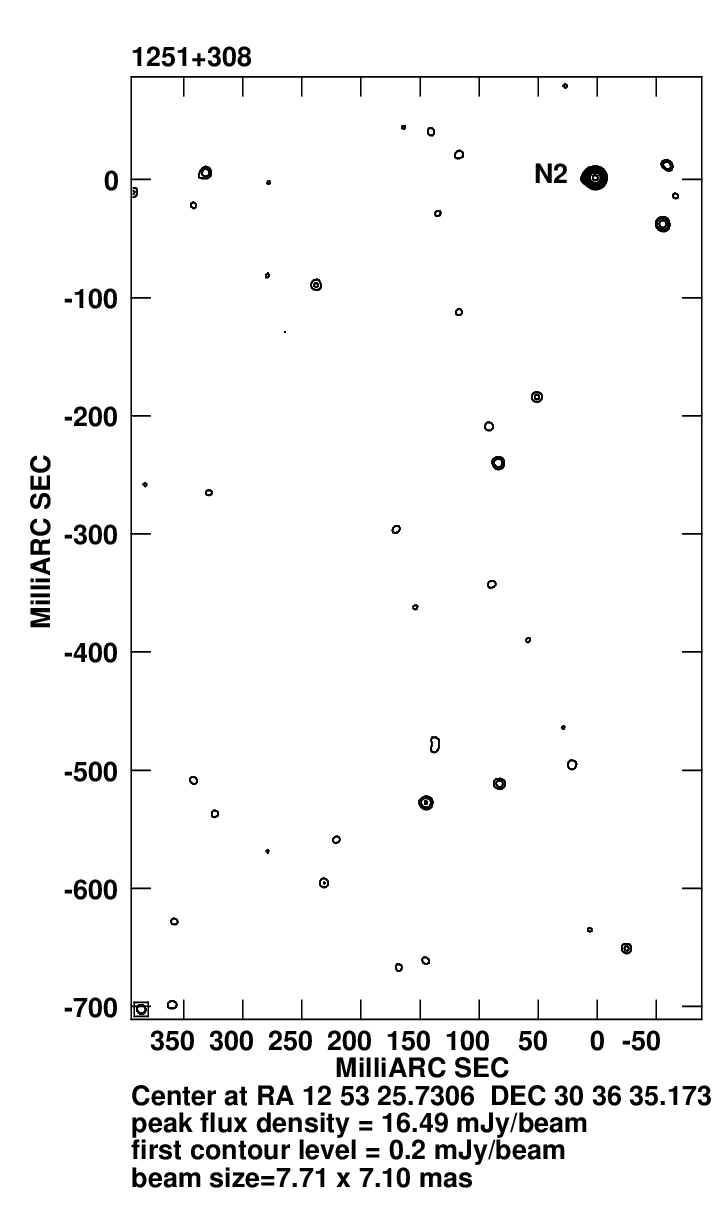}
\end{flushleft}
\caption{1251+308 -- MERLIN map at 5\,GHz (upper panel), VLBA map at 1.7\,GHz
(lower left panel) and EVN map at 5\,GHz (lower right panel).
}
\label{fig:1251+308}
\end{figure*}
\clearpage
\begin{figure*}
\begin{flushleft}
\centering
\includegraphics[scale=0.55]{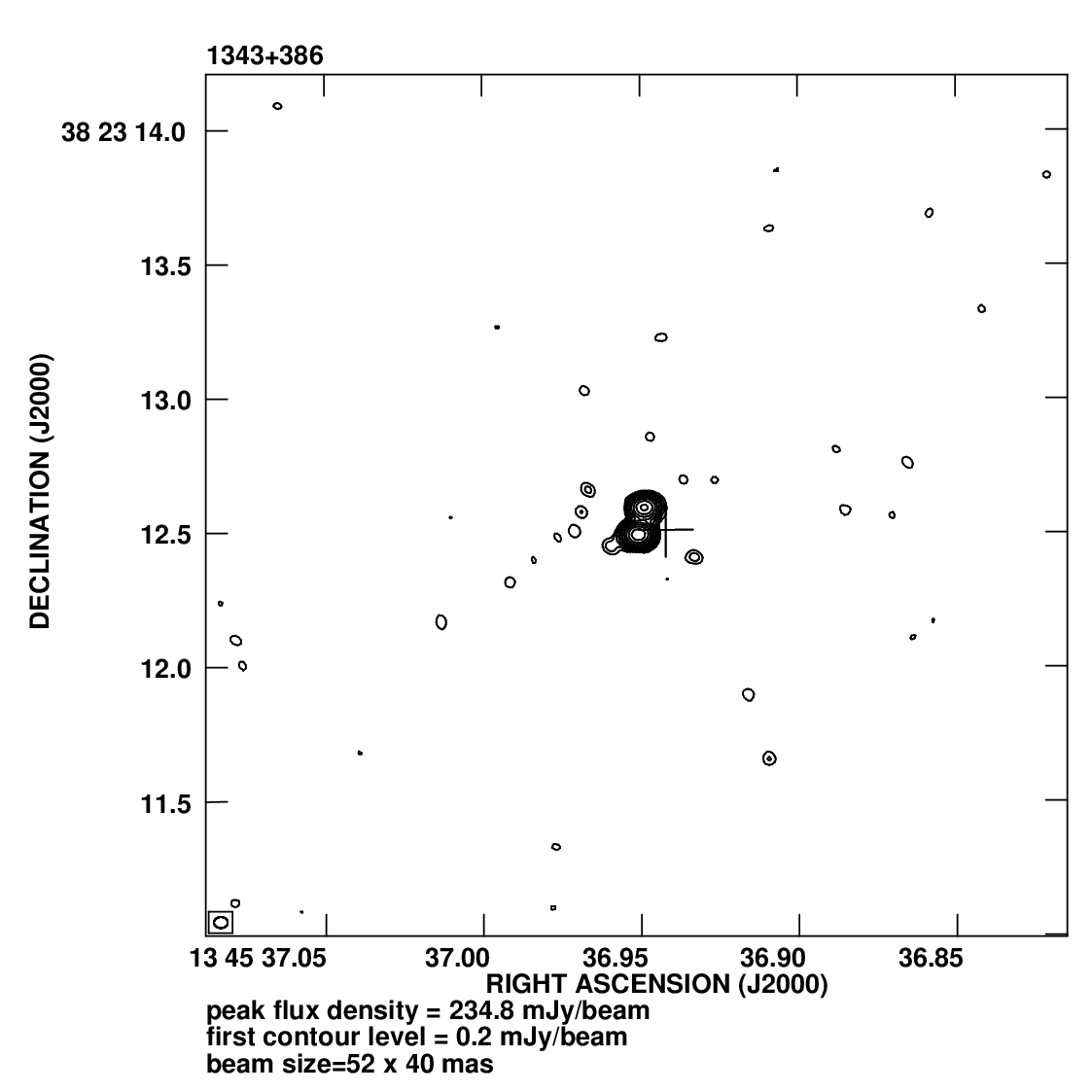}
\end{flushleft}
\begin{flushleft}
\centering
\includegraphics[scale=0.55]{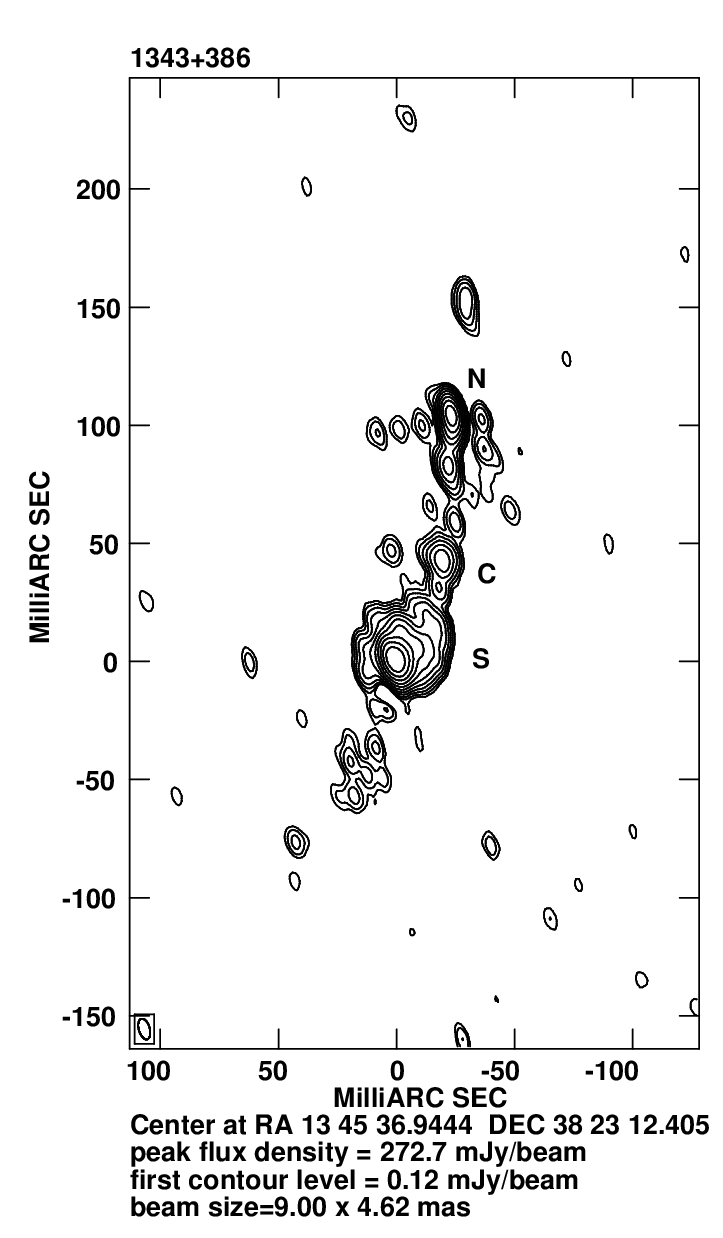}
\includegraphics[scale=0.55]{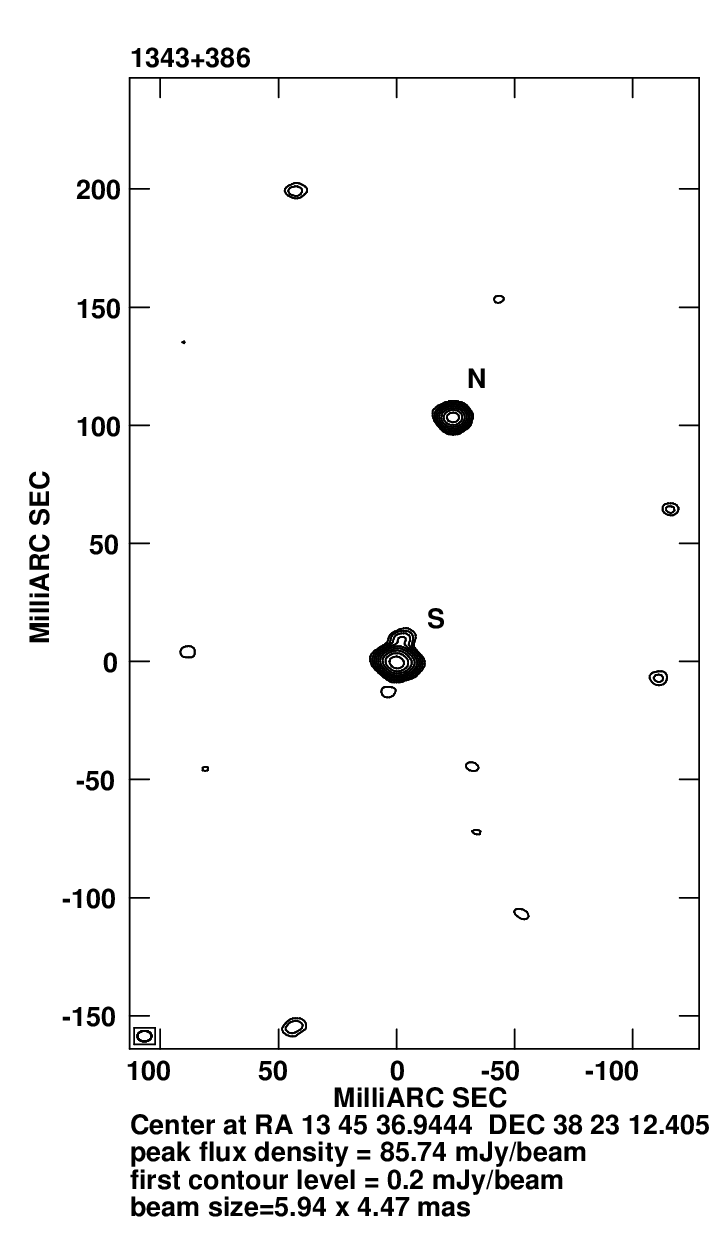}
\end{flushleft}
\caption{1343+386 -- MERLIN map at 5\,GHz (upper panel), VLBA map at 1.7\,GHz
(lower left panel) and EVN map at 5\,GHz (lower right panel).
The cross on the MERLIN
map indicates the position of an optical object in SDSS/DR4.}
\label{fig:1343+386}
\end{figure*}
\clearpage
\begin{figure*}
\begin{flushleft}
\centering
\includegraphics[scale=0.5]{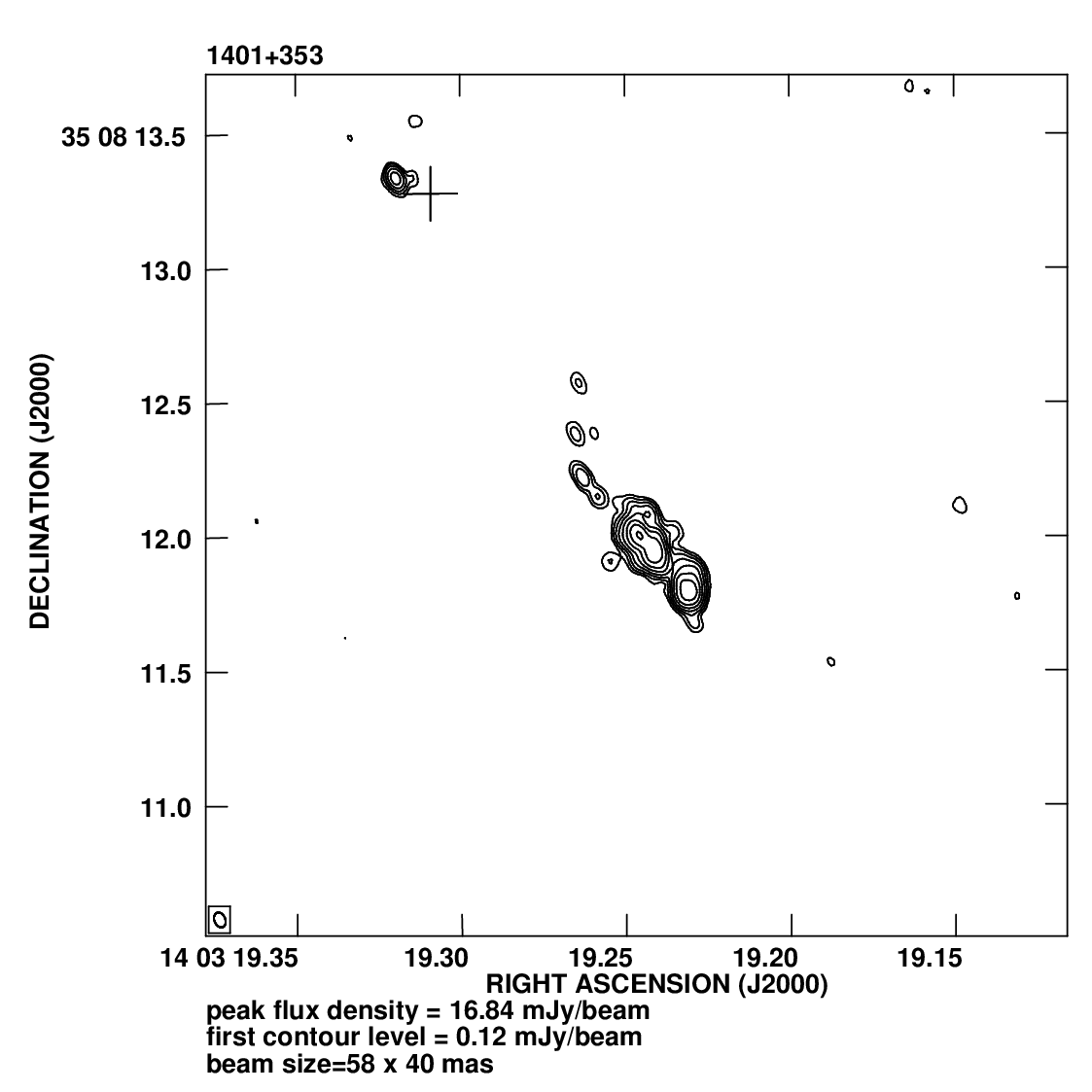}
\end{flushleft}
\caption{MERLIN map of 1401+353 at 5\,GHz.
The cross on the MERLIN
map indicates the position of an optical object in SDSS/DR4.}
\label{fig:1401+353}
\end{figure*}
\clearpage
\begin{figure*}
\begin{flushleft}
\centering
\includegraphics[scale=0.5]{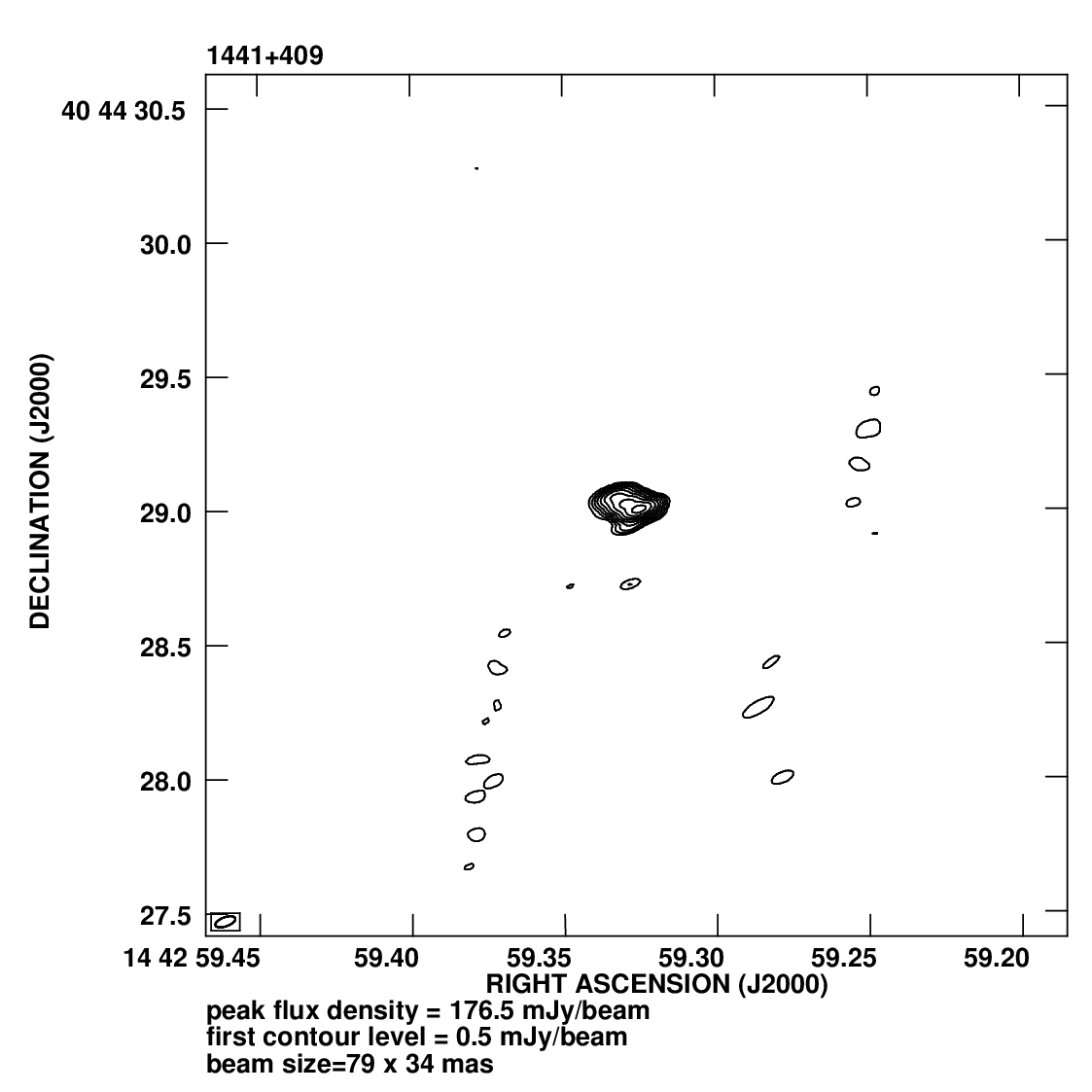}
\end{flushleft}
\begin{flushleft}
\centering
\includegraphics[scale=0.5]{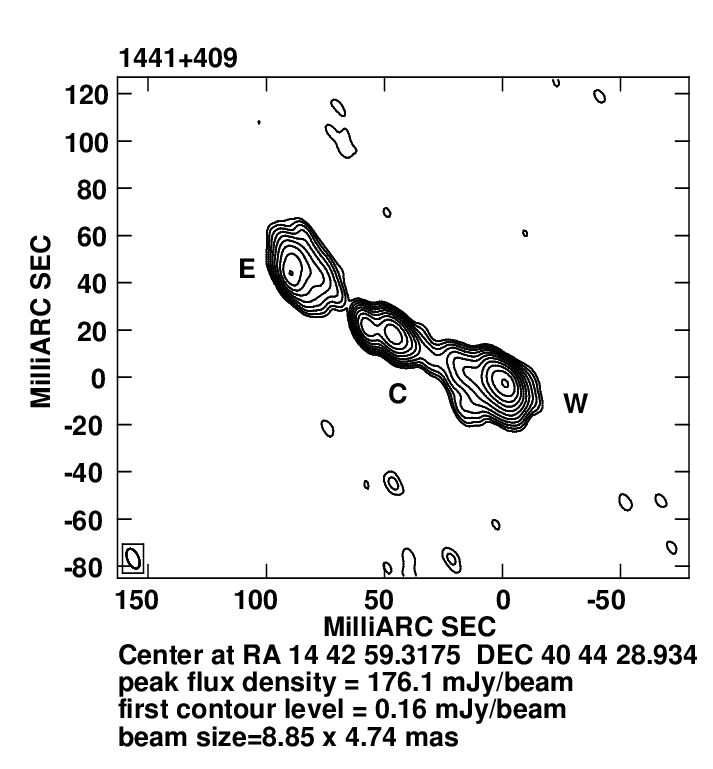}
\includegraphics[scale=0.5]{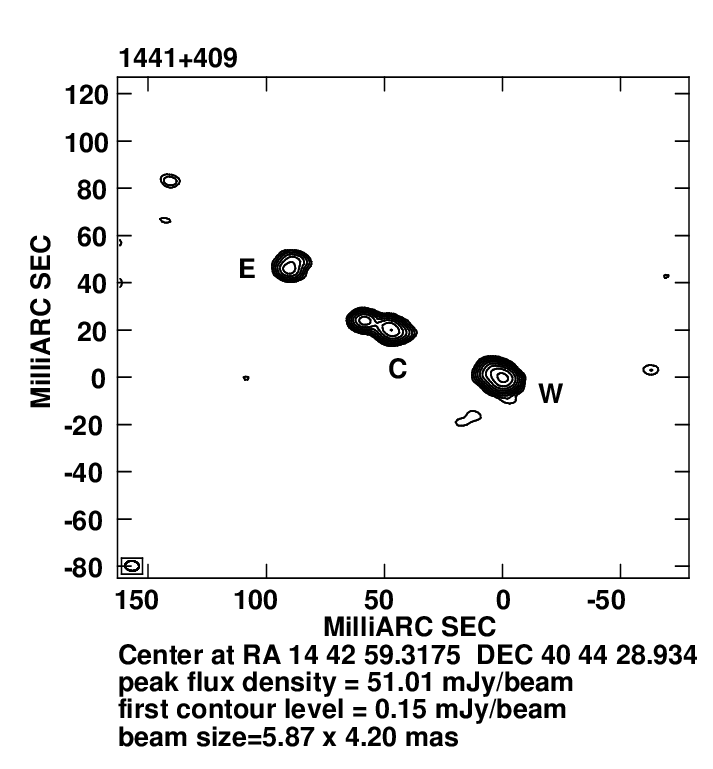}
\end{flushleft}
\caption{1441+409 -- MERLIN map at 5\,GHz (upper panel), VLBA map at 1.7\,GHz
(lower left panel) and EVN map at 5\,GHz (lower right panel).
}
\label{fig:1441+409}
\end{figure*}
\clearpage
\begin{figure*}
\begin{flushleft}
\centering
\includegraphics[scale=0.5]{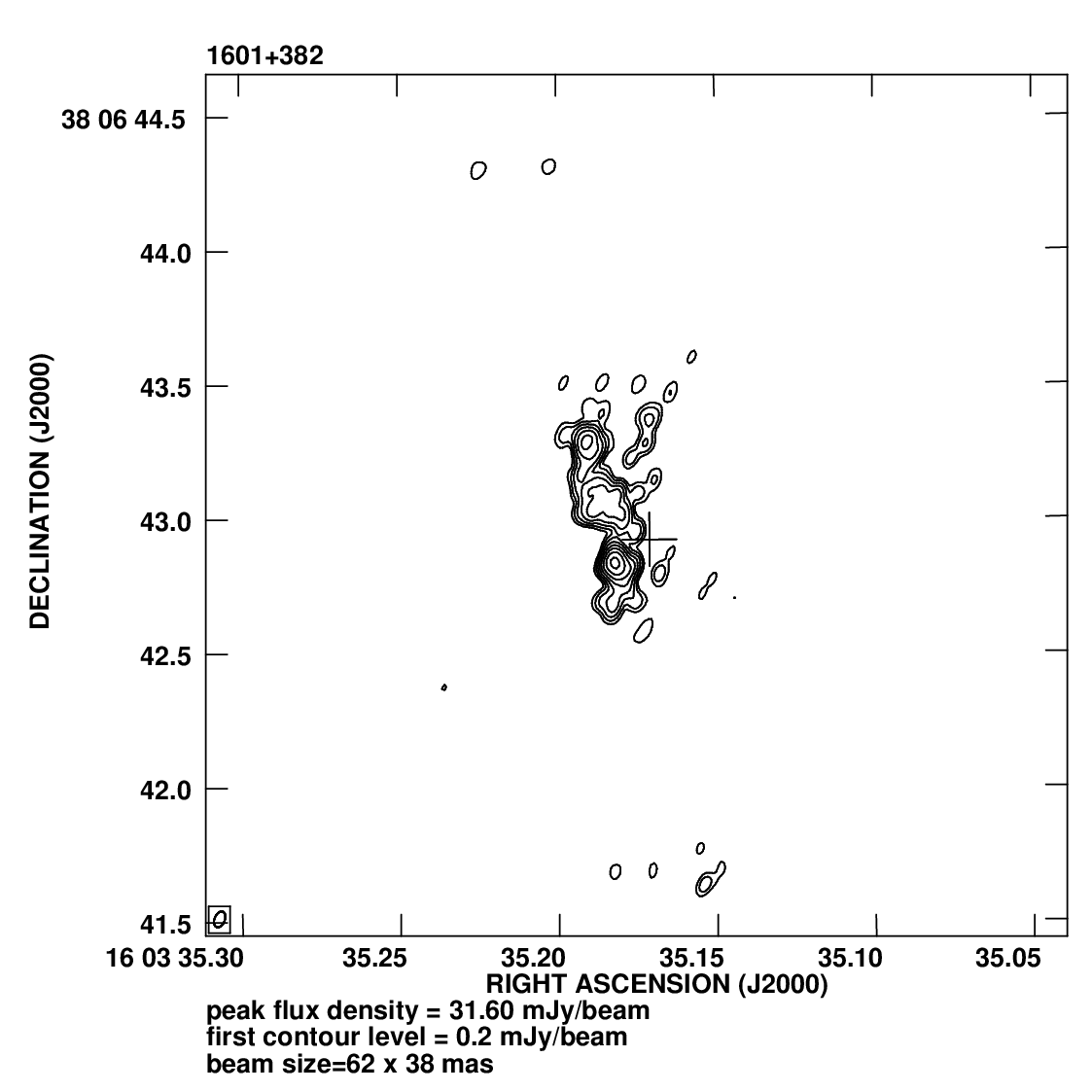}
\end{flushleft}
\caption{MERLIN map of 1601+382 at 5\,GHz.
The cross on the MERLIN map
indicates the position of an optical object in SDSS/DR4.}
\label{fig:1601+382}
\end{figure*}
\clearpage
\begin{figure*}
\begin{flushleft}
\centering
\includegraphics[scale=0.5]{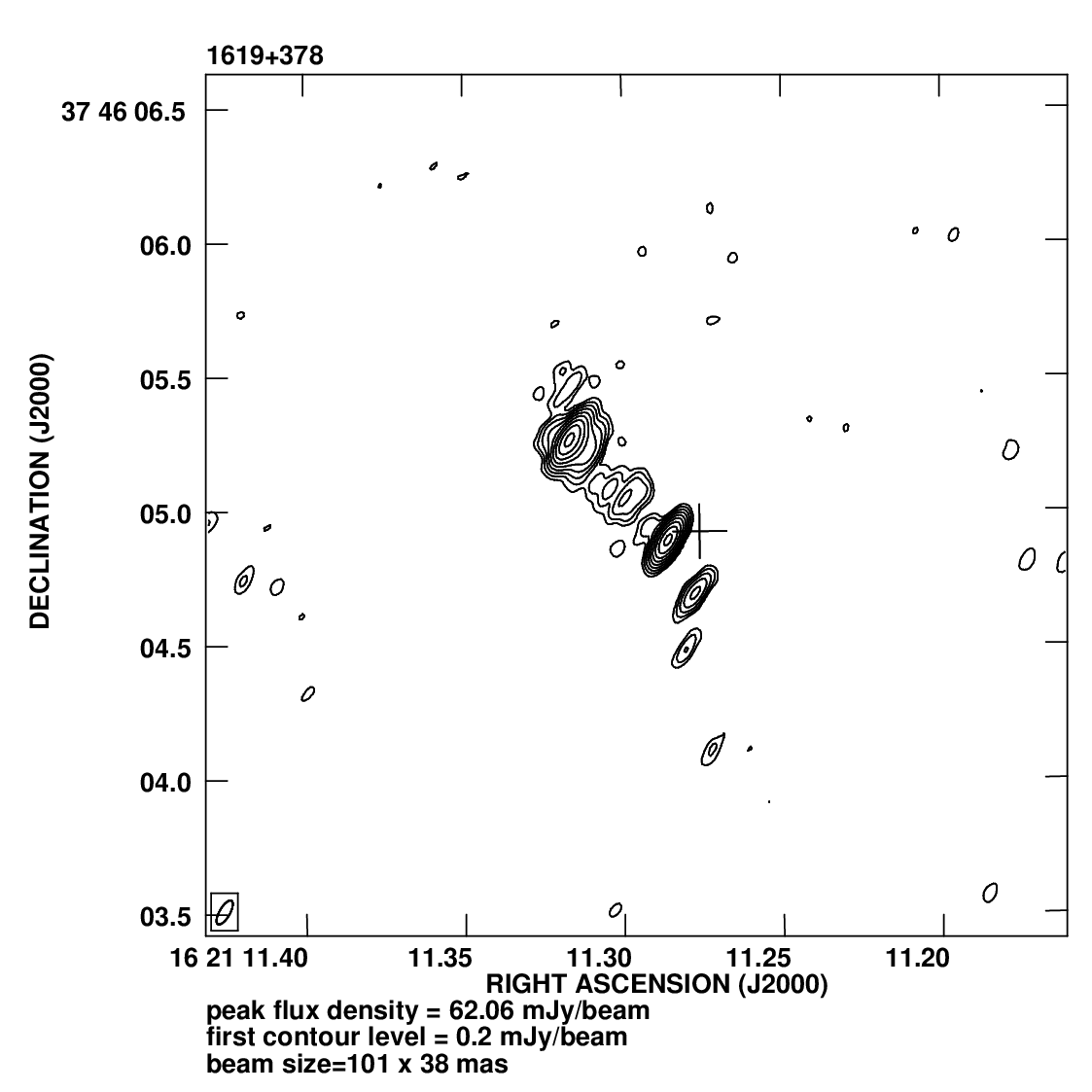}
\end{flushleft}
\begin{flushleft}
\centering
\includegraphics[scale=0.5]{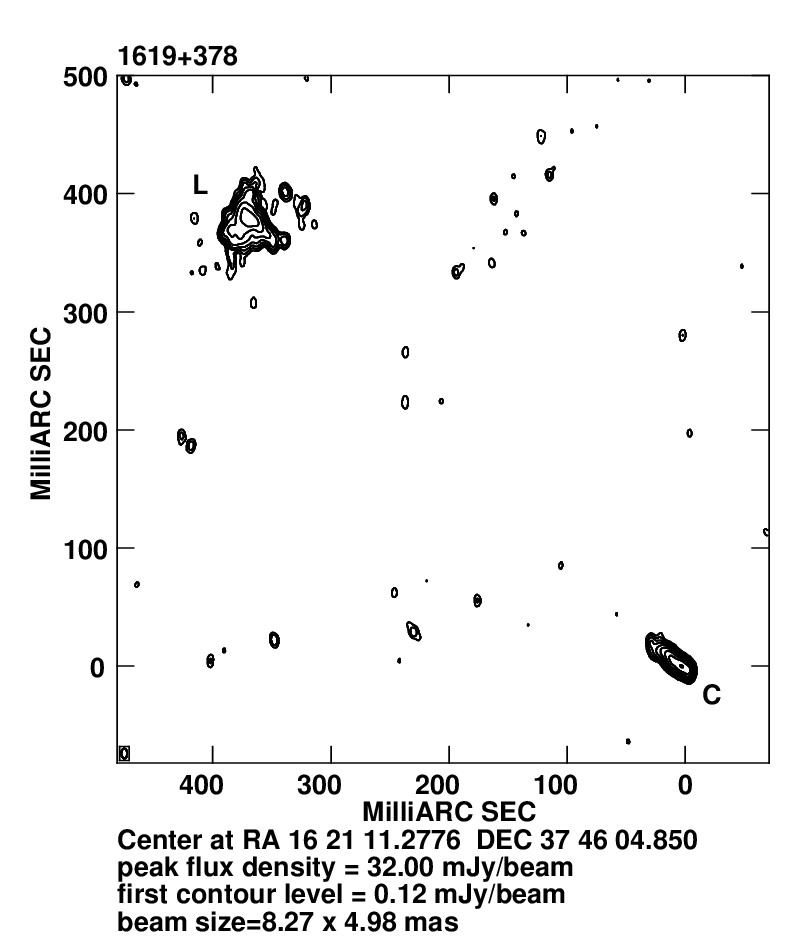}
\includegraphics[scale=0.5]{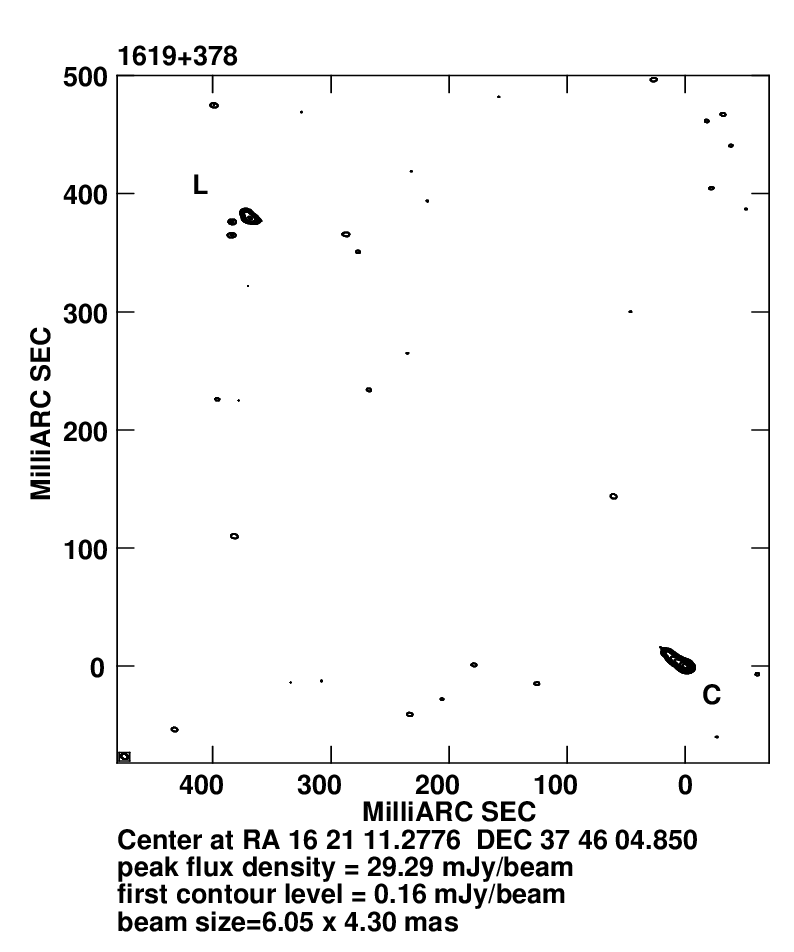}
\end{flushleft}
\caption{1619+378 -- MERLIN map at 5\,GHz (upper panel), VLBA map at 1.7\,GHz
(lower left panel) and EVN map at 5\,GHz (lower right panel).
The cross on the MERLIN
map indicates the position of an optical object in SDSS/DR4.}
\label{fig:1619+378}
\end{figure*}
\clearpage
\begin{figure*}
\begin{flushleft}
\centering
\includegraphics[scale=0.5]{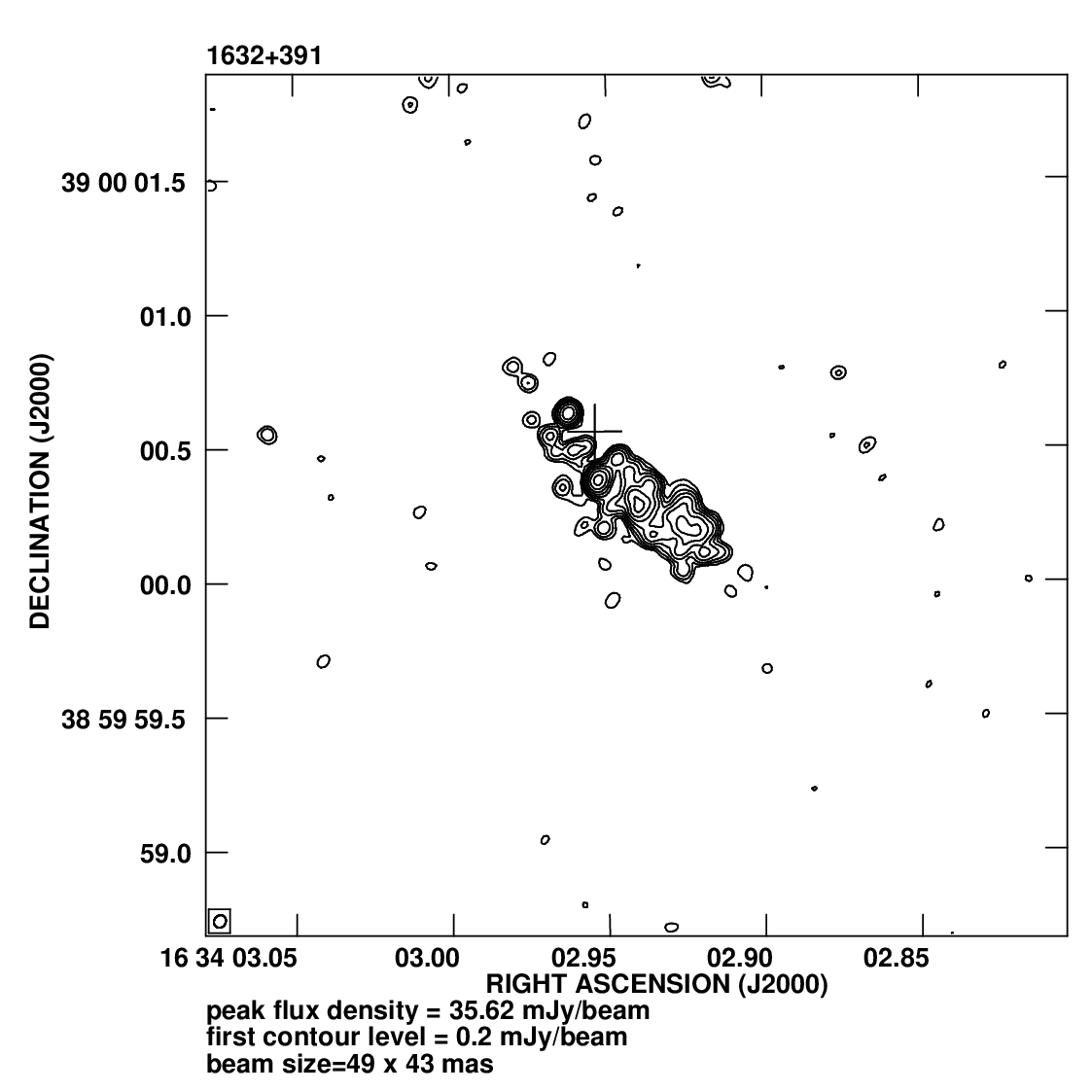}
\includegraphics[scale=0.5]{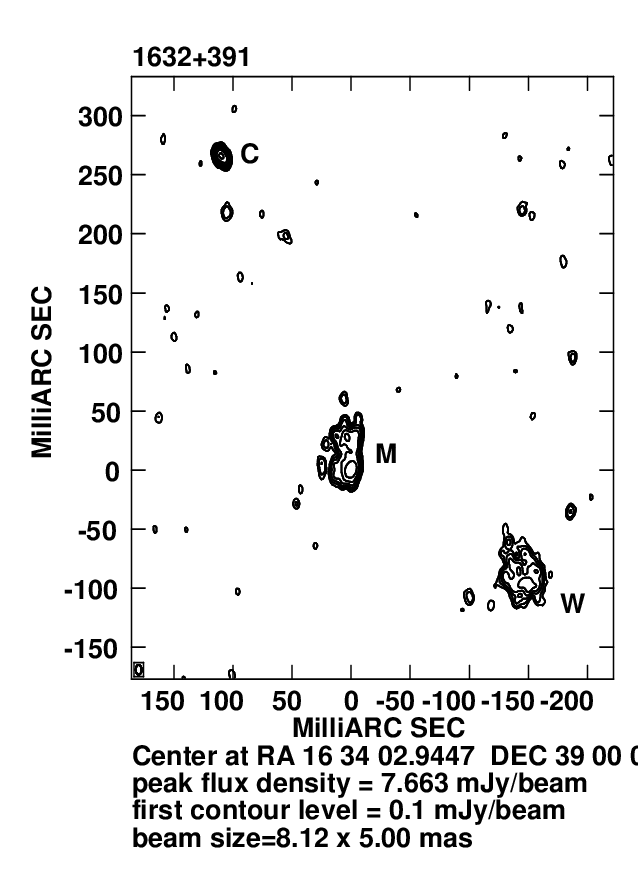}
\end{flushleft}
\caption{1632+391 -- MERLIN map at 5\,GHz (left panel) and VLBA map at 1.7\,GHz
(right panel).
The cross on the MERLIN map indicates the
position of an optical object in SDSS/DR4.}
\label{fig:1632+391}
\end{figure*}
\clearpage
\begin{figure*}
\begin{flushleft}
\centering
\includegraphics[scale=0.5]{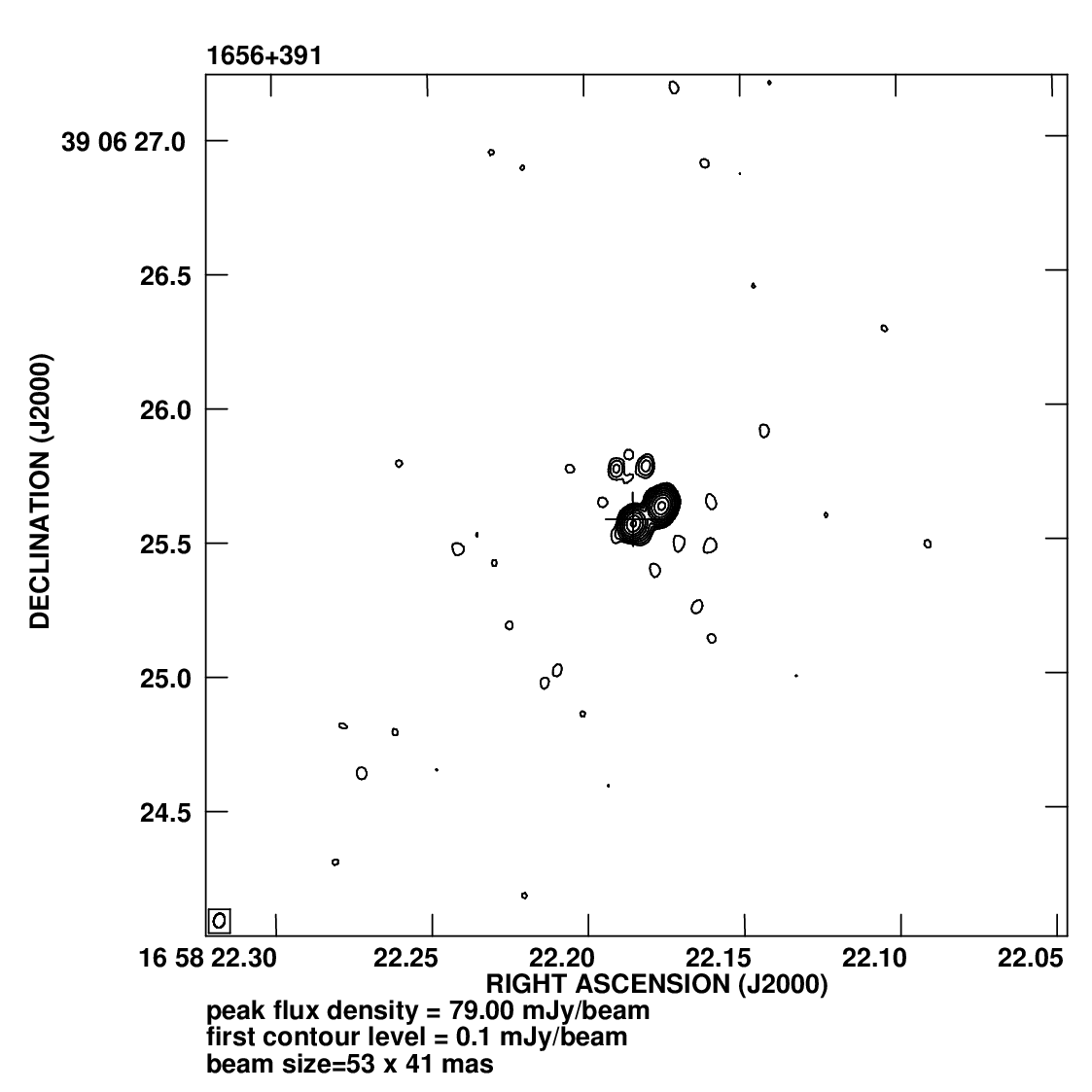}
\end{flushleft}
\begin{flushleft}
\centering
\includegraphics[scale=0.5]{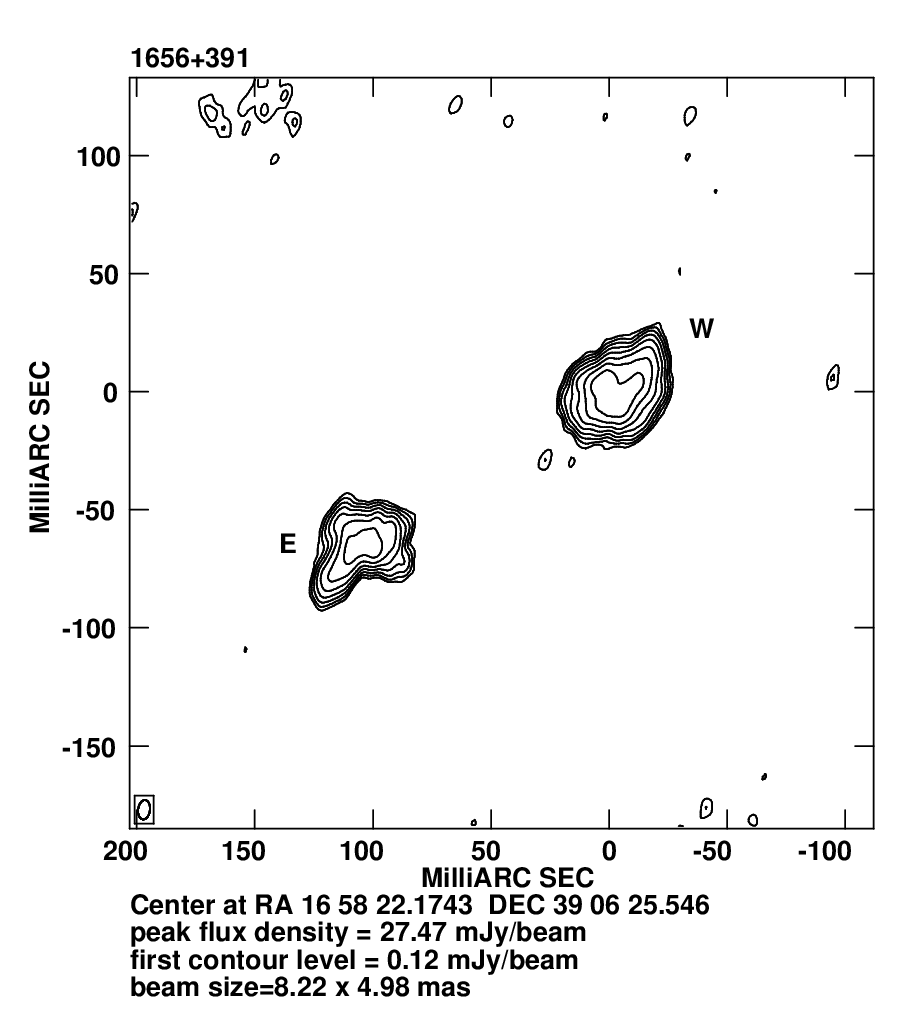}
\includegraphics[scale=0.5]{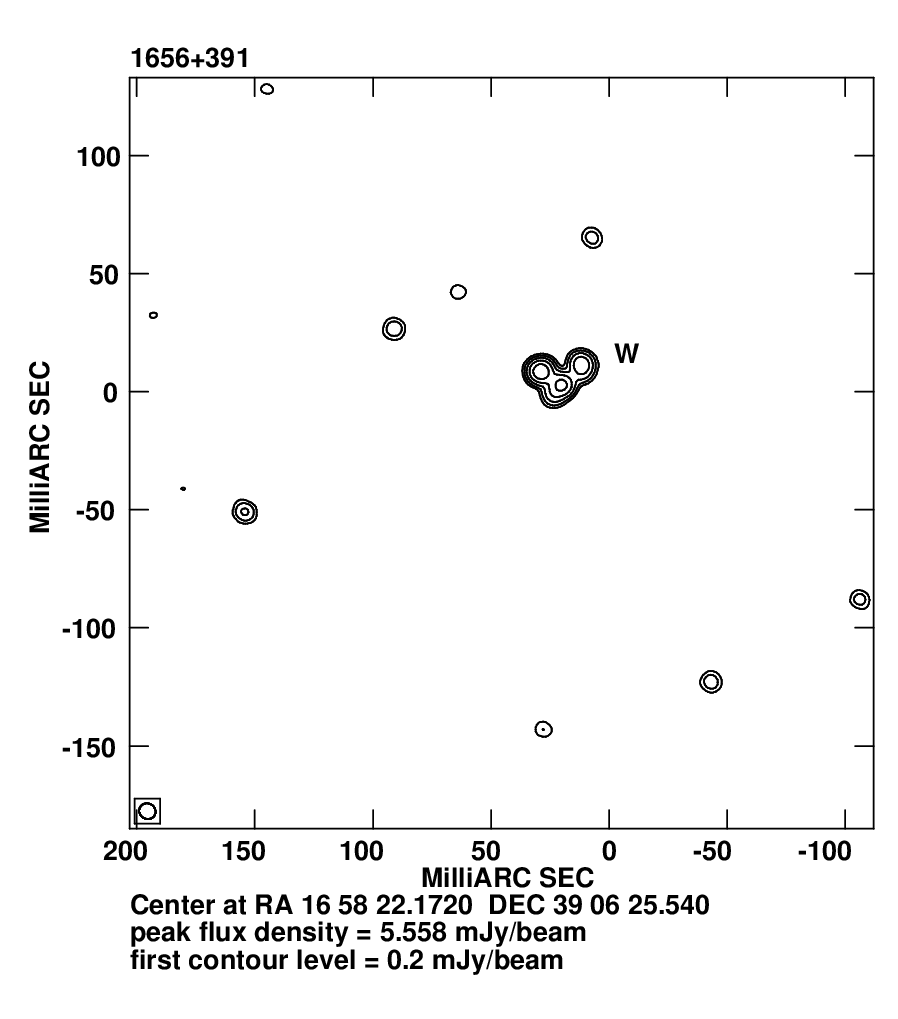}
\end{flushleft}
\caption{1656+391 -- MERLIN map at 5\,GHz (upper panel), VLBA map at 1.7\,GHz
(lower left panel) and EVN map at 5\,GHz (lower right panel).
The cross on the MERLIN
map indicates the position of an optical object in SDSS/DR4.}
\label{fig:1656+391}
\end{figure*}
\clearpage
\begin{figure*}
\begin{flushleft}
\centering
\includegraphics[scale=0.5]{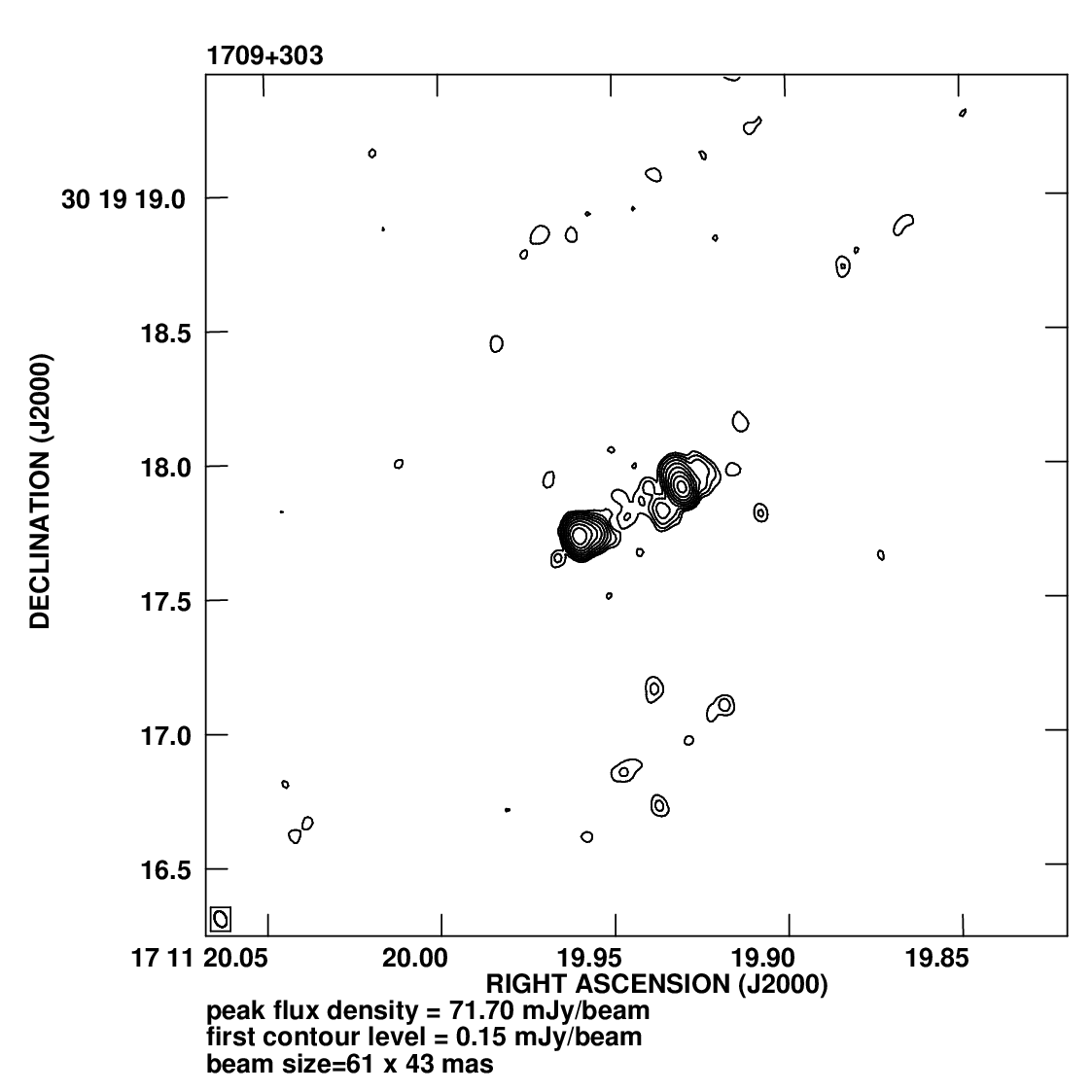}
\end{flushleft}
\begin{flushleft}
\centering
\includegraphics[scale=0.5]{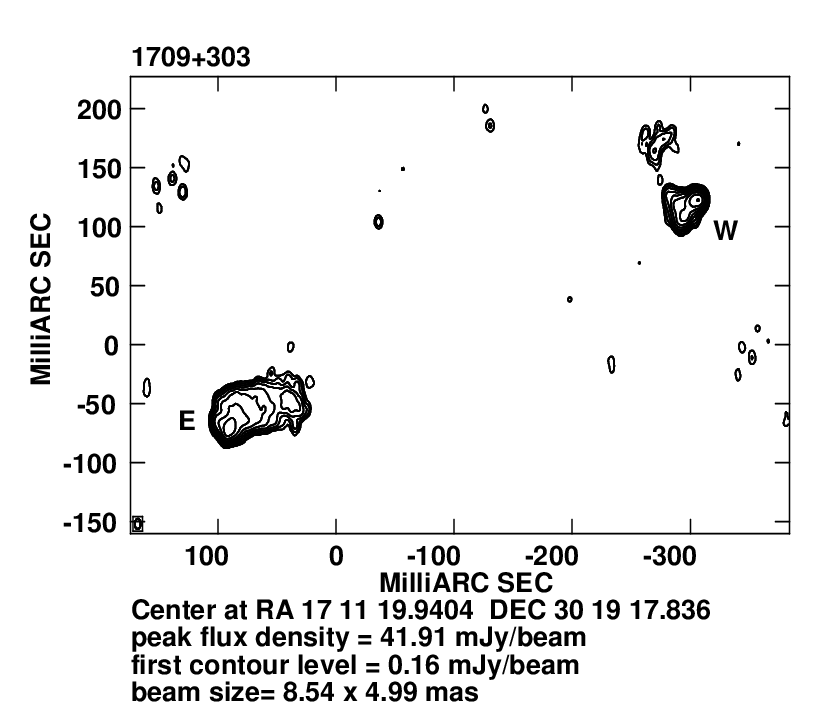}
\includegraphics[scale=0.5]{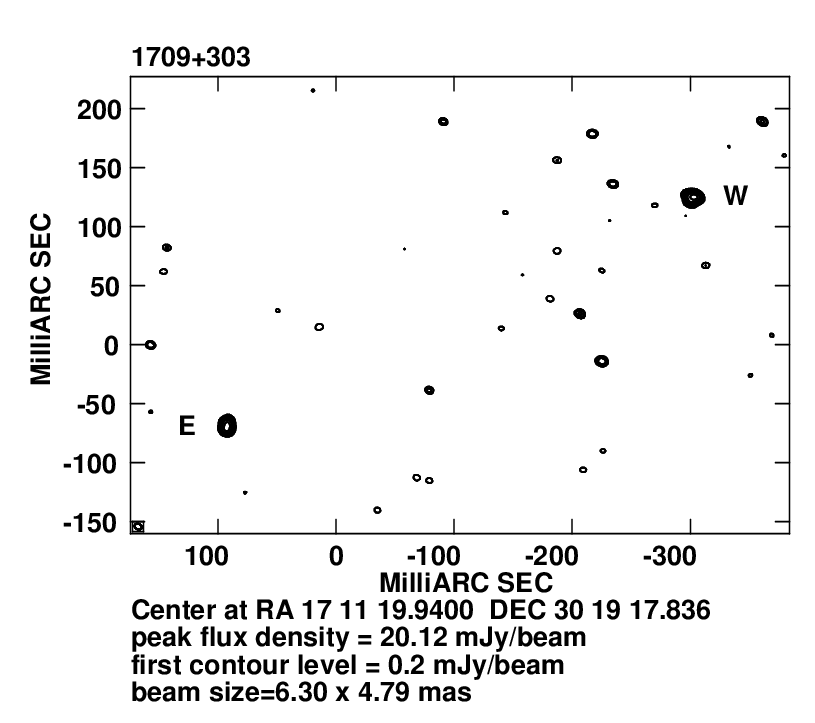}
\end{flushleft}
\caption{1709+303 -- MERLIN map at 5\,GHz (upper panel), VLBA map at 1.7\,GHz
(lower left panel) and EVN map at 5\,GHz (lower right panel).
}
\label{fig:1709+303}
\end{figure*}
\clearpage
\begin{figure*}
\begin{flushleft}
\centering
\includegraphics[scale=0.5]{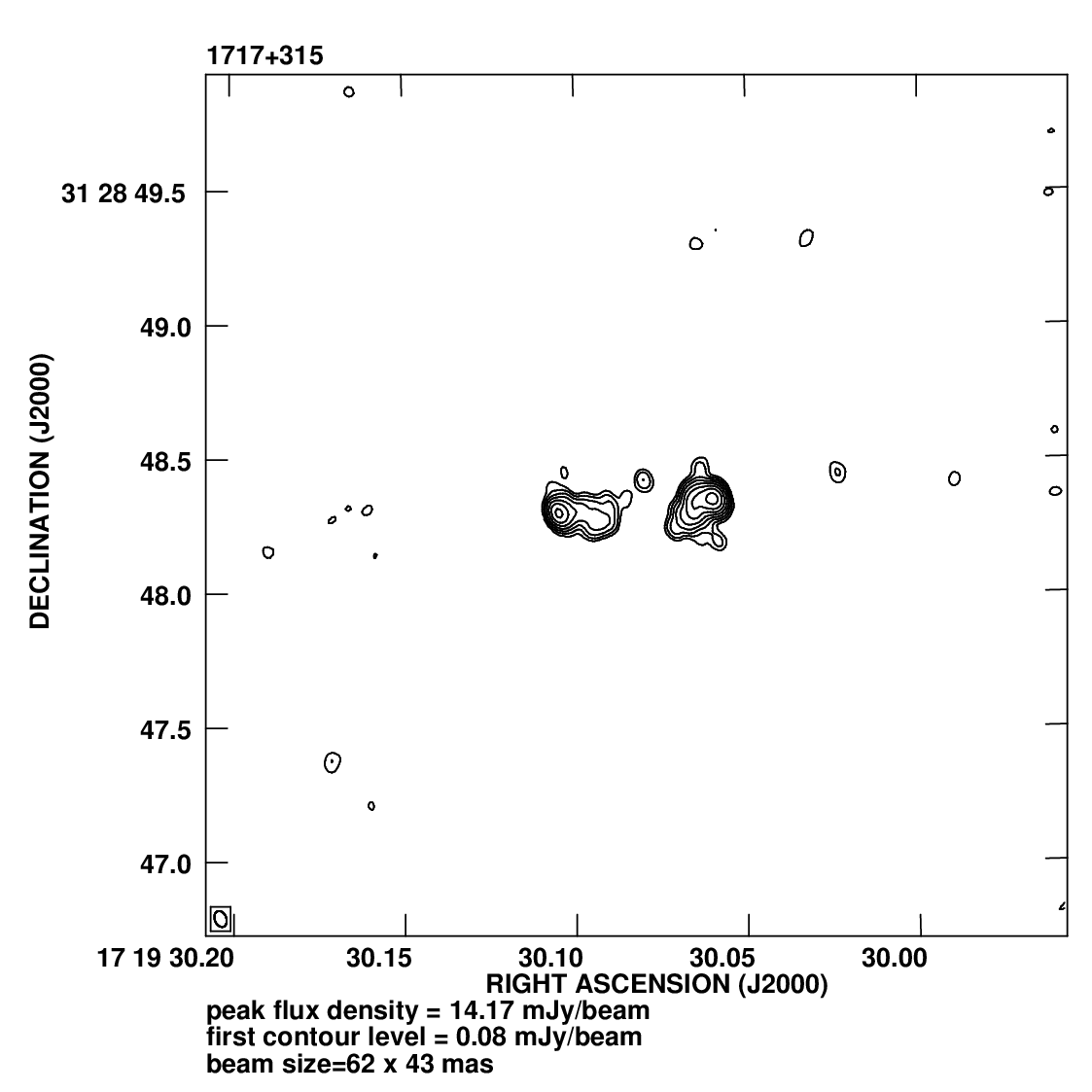}
\includegraphics[scale=0.5]{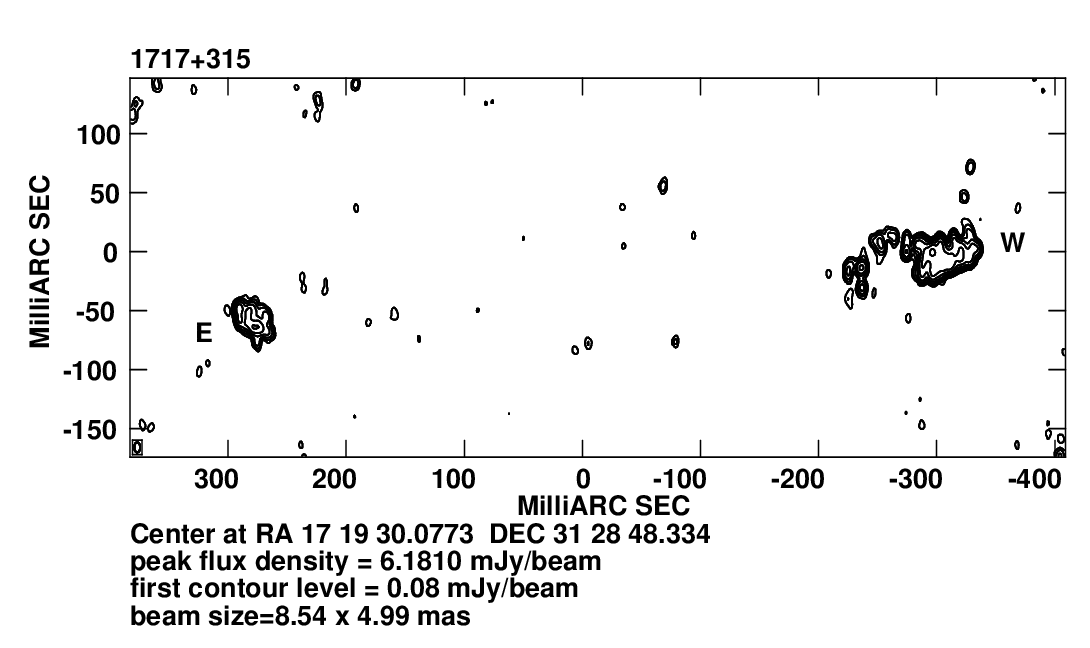}
\end{flushleft}
\caption{1717+315 -- MERLIN map at 5\,GHz (upper panel) and VLBA map at
1.7\,GHz (lower panel).}
\label{fig:1717+315}
\end{figure*}

\end{document}